\documentclass[10pt,journal]{IEEEtran}

\ifCLASSOPTIONcompsoc
  \usepackage[nocompress]{cite}
\else
  \usepackage{cite}
\fi
\ifCLASSINFOpdf
\else
\fi

\usepackage{fancyhdr}
\usepackage{graphicx}
\usepackage{dsfont}
\usepackage{cite}
\usepackage{subfigure}
\usepackage{caption}
\usepackage{bm}
\usepackage{bbm}
\usepackage{booktabs} 
\usepackage{amssymb}
\usepackage{amsmath}
\usepackage{amsthm}
\usepackage{algorithm}
\usepackage{algorithmic}
\usepackage{multirow}
\usepackage{array}
\usepackage{comment}
\usepackage{stfloats}
\usepackage{color}
\usepackage{setspace}

\newtheorem{theorem}{Theorem}
\newtheorem{lemma}{Lemma}

\newtheorem{proposition}{Proposition}

\newtheorem{observation}{Observation}
\newtheorem{example}{Example}

\usepackage{url}

\begin{document}

\title{Monetizing Mobile Data via Data Rewards}


\author{\IEEEauthorblockN{Haoran Yu,~\IEEEmembership{Member,~IEEE,} Ermin Wei,~\IEEEmembership{Member,~IEEE,} and Randall A. Berry,~\IEEEmembership{Fellow,~IEEE}}
\thanks{Manuscript received September 1, 2019; accepted November 9, 2019. This research was supported by the NSF grants AST-1343381, AST-1547328 and CNS-1701921. Some results in the paper were presented at IEEE INFOCOM, Paris, France, April 2019 \cite{yu2019business}. (\emph{Corresponding author: Haoran Yu.})}
\thanks{The authors are with the Department of Electrical and Computer Engineering, Northwestern University, Evanston, IL 60208, USA (email: yhrhawk@gmail.com; \{ermin.wei, rberry\}@northwestern.edu).}
}

\twocolumn

\newpage
\setcounter{page}{1}
\maketitle

\begin{abstract}
Most mobile network operators generate revenues by directly charging users for data plan subscriptions. Some operators now also offer users data rewards to incentivize them to watch mobile ads, which enables the operators to collect payments from advertisers and create new revenue streams. In this work, we analyze and compare two data rewarding schemes: a \emph{Subscription-Aware Rewarding (SAR) scheme} and a \emph{Subscription-Unaware Rewarding (SUR) scheme}. 
Under the SAR scheme, only the subscribers of the operators' data plans are eligible for the rewards; under the SUR scheme, all users are eligible for the rewards (e.g., the users who do not subscribe to the data plans can still get SIM cards and receive data rewards by watching ads). 
We model the interactions among an operator, users, and advertisers by a two-stage Stackelberg game, and characterize their equilibrium strategies under both the SAR and SUR schemes. 
We show that the SAR scheme can lead to more subscriptions and a higher operator revenue from the data market, while the SUR scheme can lead to better ad viewership and a higher operator revenue from the ad market. 
We further show that the operator's optimal choice between the two schemes is sensitive to the users' data consumption utility function and the operator's network capacity. We provide some counter-intuitive insights. For example, when each user has a logarithmic utility function, the operator should apply the SUR scheme (i.e., reward both subscribers and non-subscribers) if and only if it has a small network capacity.
\end{abstract}


\thispagestyle{empty}


\IEEEpeerreviewmaketitle

\section{Introduction}\label{sec:intro}
Despite the rapid growth of global mobile traffic, several leading analyst firms estimate that global mobile service revenue has nearly reached a saturation point. For example, Strategy Analytics forecasts that the global mobile service revenue will only increase by $3$\% between 2018 and 2021 \cite{Analytics2018}. As suggested in \cite{AnalyticsAD}, one promising approach for the mobile network operators to create new revenue streams is to offer \emph{mobile data rewards}: the network operators reward users with free mobile data every time the users watch mobile ads delivered by the operators, and the operators are paid by the corresponding advertisers.


The data rewarding paradigm leads to a ``win-win-win'' outcome \cite{AnalyticsAD}. First, the operators monetize their services based on the mobile advertising, the global revenue of which was estimated to reach \$80 billion at the end of 2017 \cite{AnalyticsAD}. Second, the advertisers gain \emph{incentivized advertising}, where the rewards incentivize the users to better engage with ads and the advertisers allow the users to have more control over their experiences (e.g., whether and when to watch ads). {According to surveys conducted by Forrester Consulting, IPG Media Lab, and Kiip, most mobile app users prefer to watch ads with rewards than to watch targeted ads \cite{emarketer}.} Third, the users earn free mobile data to satisfy their growing data demand. 

There has been an increasing number of businesses entering this space. Aquto and Unlockd are two leading companies that provide technical support for data rewarding (e.g., they develop mobile apps that display ads and track the amount of rewarded data). 
Aquto has collaborated with operators, such as Verizon and Telefonica \cite{Aquto}. Unlockd has collaborated with Tesco Mobile (in the United Kingdom), Boost Mobile (in the United States), Lebara Mobile (in Australia), and AXIS (in Indonesia) \cite{Unlockdjsac}. {Other examples of operators that have offered data rewards include DOCOMO, Optus, and ChungHwa Telecom \cite{DOCOMO,Optus}. Furthermore, AT\&T recently acquired AppNexus (a leading online advertising company) and will make a significant investment in the advertising business \cite{ATTadnew}.} Offering mobile data rewards could become a natural and effective approach to further monetize an operator's mobile service.


{We use an example in Table \ref{table:example} to show that offering data rewards might lead to a significant revenue improvement for an operator. Suppose that an operator rewards $0.5$MB of data per image ad.{\footnote{{To ensure that users carefully watch the ads, the operator can ask ad-related questions before giving the rewards \cite{qqensure}.}}} If a user watches $40$ image ads every day,{\footnote{{According to \cite{unlocktimes}, a mobile user unlocks its phone $80$ times per day on average. Then, watching $40$ image ads per day is similar to watching an image ad every two times the user unlocks its phone.}}} it can get $600$MB of data after $30$ days. When the CPM (cost per thousand impressions, also called cost per mille) is \$8.2 \cite{CPMReport}, the operator's corresponding ad revenue is \$9.84. In other words, the operator gets \$9.84 by rewarding $600$MB of data to the user. As a comparison, the conventional data pricing is less profitable to the operator. As shown in \cite{ultramobile}, operators only charge a user an extra \$4 when the user switches from a 1GB data plan to a 2GB data plan.}

Based on the eligibility of receiving rewards, there are two basic types of data rewarding schemes. In the \emph{Subscription-Aware Rewarding (SAR) scheme}, the operators only allow the users who subscribe to the operators' existing data plans (with monthly fees) to watch ads for rewards.{\footnote{{Some operators, such as AT\&T and Verizon, offer unlimited data plans \cite{unlimited}. However, when the actual data usage of an unlimited data plan's subscriber exceeds a threshold, the subscriber's network speed will be throttled. Hence, the unlimited data plans' subscribers may also earn free high-speed data by watching ads.}}} In the \emph{Subscription-Unaware Rewarding (SUR) scheme}, the operators reward all users for watching ads, regardless of whether the users subscribe to the data plans.{\footnote{The operators can offer free specialized SIM cards to the users who do not subscribe to the data plans. These users can top up the cards by watching ads, as shown in \cite{DOCOMO}.}} Intuitively, the SAR scheme leads to more subscriptions and the SUR scheme incentivizes more users to watch ads. The optimal design and comparison of the two schemes are crucial for realizing the full potential of the mobile data rewards, which motivates our work.


\subsection{Our Contributions}
\begin{figure}[t]
  \centering
  \includegraphics[scale=0.4]{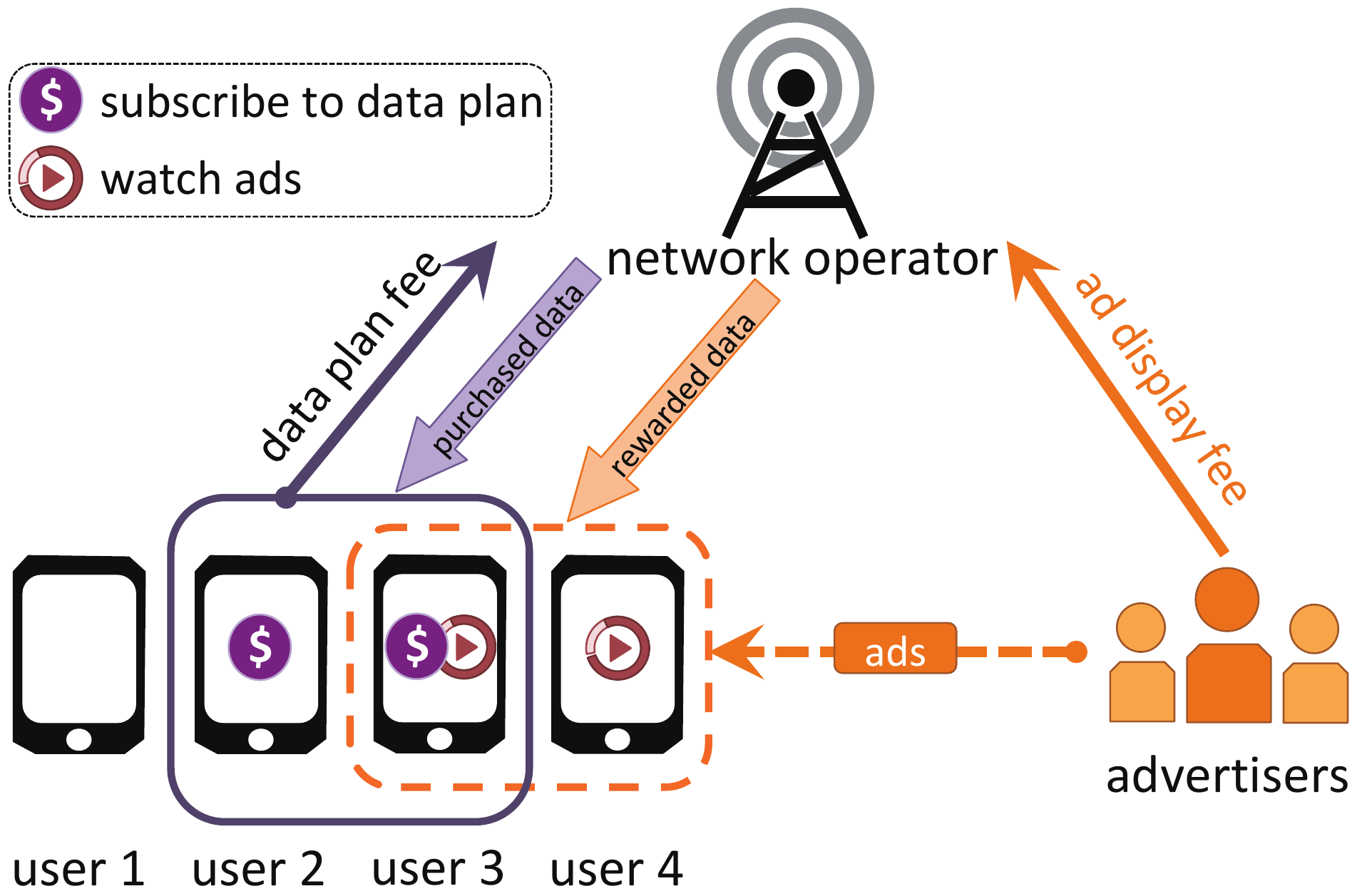}\\
  \caption{Data rewarding ecosystem (user $4$ is feasible under the \emph{SUR} scheme, but is infeasible under the \emph{SAR} scheme).}
  \label{fig:system}
\end{figure}

We illustrate the data rewarding ecosystem in Fig. \ref{fig:system}. The purple arrows indicate that an operator charges the users for data plan subscriptions. The orange arrows indicate that the operator rewards the users for watching ads and gets payments from the advertisers. 

{{
\begin{table*}[t]
\caption{{{Example of Data Rewards}}}\label{table:example}
\begin{tabular}{|p{3.2cm}|p{2.6cm} |p{2.8cm} |p{1.5cm} |p{5.2cm}|}
\hline
\multirow{2}{*}{Rewarding Plan} & \multicolumn{2}{l|}{{A User's Views and Reward (Per Month)}} &  \multicolumn{2}{l|}{{Calculation of Operator's Ad Revenue}}\\
\cline{2-5}
& {Views} & {Reward} & {CPM} &{Views/1000$\times$CPM$=$Ad Revenue}\\
\hline
{$0.5$MB per image ad} & {1200 image ads} & {600MB} & {\$8.2} &{1200/1000$\times$\$8.2=\$9.84}\\
\hline
\end{tabular}
\end{table*}
}}


We model the interactions among the operator, users, and advertisers by a two-stage Stackelberg game. In Stage I, the operator decides the unit data reward (i.e., the amount of data rewarded for watching one ad) for the users, and the ad price (i.e., the payment for purchasing one ad slot) for the advertisers. In Stage II, the users with different valuations for the mobile service make their data plan subscription and ad watching decisions. We consider a general data consumption utility function and a general distribution of user valuation. Meanwhile, the advertisers decide the number of ad slots to purchase, considering the advertising's \emph{wear-out effect} (i.e., an ad's effectiveness can decrease if it reaches a user who has watched the same ad for several times \cite{pechmann1988advertising,kirmani1997advertising}). 

We analyze the two-stage game for both the SAR and SUR schemes. In particular, we characterize the operator's optimal strategy that maximizes the total revenue from the data market and ad market. Our key findings in this work are as follows.

{\bf{I. Design of Unit Data Reward (Theorems \ref{theorem:subscriptionaware} and \ref{theorem:notuse}):}} \emph{Under both the SAR and SUR schemes, the operator should not always use up the available network capacity for data rewards.} Under the SAR scheme, increasing the unit data reward can lead to more data plan subscriptions and motivate more users to watch ads. However, it also allows a user to obtain a larger amount of data after watching a few ads. Hence, a user may watch fewer ads under a larger unit data reward. As a result, increasing the unit data reward may decrease the operator's revenue. Under the SUR scheme, (besides the above negative impact) increasing the unit data reward may lead to a loss in data plan subscriptions, and even generate a revenue that is lower than the revenue when the operator does not offer any data reward. In our work, we derive two sufficient conditions, under which the operator does and does not use up the capacity for data rewards, respectively.

{\bf{II. Design of Ad Price (Theorems \ref{theorem:SAR:price} and \ref{theorem:differentiation}):}} \emph{Given the unit data reward, the operator's optimal ad price is affected by the wear-out effect if and only if the wear-out effect is small.} If the wear-out effect is small, the operator should sell all ad slots and its optimal ad price should decrease with the wear-out effect; otherwise, the operator should not sell all ad slots and its optimal ad price will be independent of the wear-out effect. Moreover, \emph{under the SUR scheme, the operator can differentiate the ad slots generated by the subscribers and non-subscribers when selling the ad slots to the advertisers and displaying the ads to the users. We numerically show that this can improve the operator's total revenue by up to $20.3\%$}. Under the SUR scheme, both the subscribers and non-subscribers watch ads. Since the subscribers also obtain data from the data plan, the subscribers and non-subscribers may watch different numbers of ads. Because of the advertising's wear-out effect, each advertiser has a different willingness to purchase the ad slots generated by the subscribers and non-subscribers, and it is beneficial for the operator to differentiate these ad slots.


{\bf{III. Choice of Rewarding Scheme (Theorem \ref{theorem:comparison}; Observations \ref{observation:log}, \ref{observation:alpha}, and \ref{observation:exp}):}} \emph{The operator's choice between the SAR and SUR schemes is heavily affected by the users' data consumption utility function and network capacity.} When each user has a logarithmic utility function {or each user has a generalized $\alpha$-fair utility function} \cite{joe2018sponsoring}, if the network capacity is limited, the operator should apply the SUR scheme (i.e., reward both subscribers and non-subscribers); if the capacity is large, it should apply the SAR scheme (i.e., only reward the subscribers). When each user has an exponential utility: (i) under a large wear-out effect, the choice between the two schemes is similar to the logarithmic utility case; (ii) under a small wear-out effect, the operator should always apply the SUR scheme, regardless of the capacity.

Our comparison between the SAR and SUR schemes also provides insights for a more general problem, where the operator offers \emph{multiple} data plans and decides whether to only allow the subscribers of the \emph{expensive} data plans to earn rewards. Our analysis of the SAR and SUR schemes captures the key considerations of choosing these schemes (e.g., whether to motivate more subscriptions to the expensive data plans or incentivize more ad watching).


\subsection{Related Work}
\subsubsection{{Provision of Fee-Based and Ad-Based Services}}
There has been some work studying markets where providers offer both a fee-based service and an ad-based free service.
{For example, Riggins in \cite{riggins2002market} studied an online publisher that offers both the fee-based and ad-based versions of its website.} In \cite{yu2017public}, a Wi-Fi network provider allows users to either directly pay or watch ads to access the Wi-Fi network. In \cite{guo2017economic}, an app developer offers virtual items, and each app user will either pay or watch ads to obtain them in the equilibrium. In these studies, the fee-based and ad-based services are always \emph{substitutes}, and each user chooses between these two options. 
In our work, their relation is more complicated, since a user may subscribe to the data plan and meanwhile watch ads for more data. Under the SAR scheme, increasing the reward for watching ads can increase the number of subscribers, which shows the \emph{complementary} relation between the subscription and data rewards. Therefore, our work studies a novel structure, and derives new insights for the joint provision of fee-based and ad-based services. 
Furthermore, our work considers the operator's capacity for providing the service and the advertising's wear-out effect, which were not considered in \cite{yu2017public} and \cite{guo2017economic}.

\subsubsection{{Sponsored Mobile Data}}
As studied in \cite{andrews2013economic,lotfi2017economics,joe2018sponsoring,zhang2015sponsored}, sponsored data provides another way for operators to create new revenue streams: content providers sponsor the data usage of their content, and users can access the content free of charge. 
There are several key differences between sponsored data and data rewards as studied here. First, the users can consume sponsored data only for the content specified by the content providers, while they can use reward data to access any online content. Second, with sponsored data, the content providers benefit from the users' data consumption on the corresponding content. With data rewards, the advertisers aim to deliver ads effectively, and do not benefit from the users' data consumption.


\subsubsection{{Other Related References}}
Other related work includes \cite{bangera2017advertising,sen2017incentive,harishankaraccept}. Bangera \emph{et al.} in \cite{bangera2017advertising} conducted a survey, which shows that $76\%$ of the respondents are interested in watching ads in exchange for mobile data. Sen \emph{et al.} in \cite{sen2017incentive} conducted an experiment to study the effectiveness of monetary rewards in increasing ads' viewership. Both \cite{bangera2017advertising} and \cite{sen2017incentive} did not analyze the equilibrium strategies of the entities, such as operators, advertisers, and users. Harishankar \emph{et al.} in \cite{harishankaraccept} studied monetizing the operator's idle network capacity by providing users with supplemental discount offers, which are not related to advertising. 






\section{Model}\label{sec:model}
In this section, we model the strategies of the operator, users, and advertisers, and introduce the two-stage game. We use capital letters to denote parameters, and lower-case letters to denote decision variables or random variables. 

\subsection{Network Operator}
We consider a monopolistic operator, who offers a predetermined (monthly) flat-rate data plan $\left(F,Q\right)$ to users. Parameter $F>0$ denotes the subscription fee, and $Q>0$ denotes the data amount associated with a subscription.{\footnote{Compared with designing data rewards, the operator has less flexibility to adjust its data plan (e.g., subscribers may sign long-term contracts with the operator). Hence, we study the operator's reward design, \emph{given} its existing data plan. In our future work, we plan to extend our analysis by jointly optimizing the data plan and reward.}} To derive insights into the data reward design, we focus on a single-operator, single-data plan scenario, which has been widely considered in literature (e.g., \cite{joe2018sponsoring,zhang2015sponsored}). 

The operator decides two variables: (i) a unit data reward $\omega\in\left[0,\infty\right)$, which is the amount of data that a user receives for watching one ad; (ii) an ad price $p\in\left(0,\infty\right)$, which is the price that the operator charges the advertisers for buying one ad slot. 
Here, we consider a price-based mechanism, where the operator sells the ad slots in advance at a fixed price.{\footnote{The operator and advertisers usually have large-scale collaborations, e.g., an advertiser's ads are displayed around 300,000 times per promotion activity. In this case, the price-based mechanism facilitates the customization and communication process \cite{choi2017online}. The operator can also sell the slots via the real-time auction, especially when it has some user profiles and the advertisers want to target different user categories \cite{choi2017online}. We leave the study of heterogeneous advertisers and real-time auctions to future work.}}


\subsection{Users}
We consider a continuum of users, and denote the mass of users by $N$. Let $\theta$ denote a user's type, which parameterizes its valuation for mobile service. We assume that $\theta$ is a continuous random variable drawn from $\left[0,\theta_{\max}\right]$, and its probability density function $g\left(\theta\right)$ satisfies $g\left(\theta\right)>0$ for all $\theta\in\left[0,\theta_{\max}\right]$. 

Let $r\in\left\{0,1\right\}$ denote a user's data plan subscription decision, and $x\in\left[0,\infty\right)$ denote the number of ads that a user chooses to watch (during one month). We allow $x$ and the advertisers' purchasing decisions to be fractional \cite{yu2017public,bergemann2011targeting}. The amount of data that a user obtains from its subscription and ad watching is $Qr+\omega x$. We use $\theta u\left(Qr+\omega x\right)$ to capture a type-$\theta$ user's utility of using the mobile service. Here, $u\left(z\right),z\ge0,$ is the same for all users, and can be any strictly increasing, strictly concave, and twice differentiable function that satisfies $u\left(0\right)=0$ and $\lim_{z\rightarrow \infty} u'\left(z\right)=0$. The concavity of $u\left(z\right)$ captures the diminishing marginal return with respect to the data amount. Unless otherwise specified, our results are derived under a general $u\left(z\right)$ that satisfies these properties. To study the impact of $u\left(z\right)$'s shape, we will also consider {three} concrete choices of $u\left(z\right)$ used in the literature:
\begin{itemize}
\item \emph{Logarithmic function \cite{duan2015pricing,schmidt2009minimum}:} $u\left(z\right)=\ln\left(1+z\right)$;
\item {\emph{Generalized $\alpha$-fair function \cite{joe2018sponsoring}:} $u\left(z\right)=\frac{\left(z+\mu\right)^{1-\alpha}}{1-\alpha}-\frac{\mu^{1-\alpha}}{1-\alpha}, 0<\alpha<1,\mu\ge0$;}
\item \emph{Exponential function \cite{zhou2005utility}:} $u\left(z\right)=1-e^{-\gamma z},\gamma>0$.
\end{itemize}
One reason for considering these is that the logarithmic function {and generalized $\alpha$-fair function are} not upper bounded for $z\ge0$, while the exponential function is upper bounded. This difference will affect the optimal choice between the SAR and SUR schemes. For ease of exposition, we call $u\left(\cdot\right)$ a user's utility function (although the actual utility is $\theta u\left(\cdot\right)$).

A type-$\theta$ user's payoff is
\begin{align}
\Pi^{\rm user}\left(\theta,r,x,\omega\right)=\theta u\left(Qr+\omega x\right)-Fr-\Phi x,\label{equ:userpayoff}
\end{align}
where $F$ is the subscription fee, and $\Phi>0$ denotes a user's average disutility (e.g., inconvenience) of watching one ad. We assume that the total disutility of watching ads linearly increases with the number of watched ads \cite{anderson2015advertising,guo2017economic}.


In Sections \ref{subsec:stageII:user} and \ref{subsec:unaware:user}, we will analyze the users' optimal decisions $r^*\left(\theta,\omega\right)$ and $x^*\left(\theta,\omega\right)$. Next, we introduce two notations to capture the total number of ad slots created by users. {Let $N^{\rm ad}\left(\omega\right)$ denote the mass of users with $x^*\left(\theta,\omega\right)>0$ (i.e., who watch ads), and let $y$ be the value of $x^*\left(\theta,\omega\right)$ chosen by one of these $N^{\rm ad}\left(\omega\right)$ users. Because these $N^{\rm ad}\left(\omega\right)$ users may have different types $\theta$, they may have different values of $x^*\left(\theta,\omega\right)$, i.e., watch different numbers of ads. Therefore, $y$ is a \emph{random variable}. The distribution of $y$ gives the distribution of the number of ads watched by {a user \emph{given that the user watches ads}}.{\footnote{{In Example \ref{example:1} in Section \ref{subsec:advertiser}, we will compute the concrete distribution of $y$ given the assumption of $\theta$.} Moreover, the distribution of $y$ depends on the operator's decision $\omega$. For the simplicity of presentation, we omit this dependence in the notation.}} The expected total number of created ad slots is simply the expected total number of ads watched by the users, given by ${\mathbb E}\left[y\right]N^{\rm ad}\left(\omega\right)$.}

\subsection{Advertisers}\label{subsec:model:advertiser}
We consider $K$ homogeneous advertisers. When $N^{\rm ad}\left(\omega\right)>0$, we assume that to display the ads to a user, the operator randomly draws ads from all the ${\mathbb E}\left[y\right]N^{\rm ad}\left(\omega\right)$ ad slots without replacement.


Suppose an advertiser purchases $m\in\left[0,\infty\right)$ ad slots from the operator (in Sections \ref{subsec:aware:operator} and \ref{subsec:unaware:stageI}, the operator will choose its ad price $p$ to ensure that the total number of sold ad slots does not exceed ${\mathbb E}\left[y\right]N^{\rm ad}\left(\omega\right)$). If a user watches $y$ ads, on average, $\frac{my}{{\mathbb E}\left[y\right]N^{\rm ad}\left(\omega\right)}$ ads among the $y$ watched ads belong to this advertiser. 
We let $\psi\left(m,y,\omega\right)$ denote the overall effectiveness of the advertiser's advertising on the user (e.g., a large $\psi\left(m,y,\omega\right)$ implies that the user has a good impression of the advertiser's product). We model $\psi\left(m,y,\omega\right)$ by
\begin{align}
\psi\left(m,y,\omega\right) = B\frac{my}{{\mathbb E}\left[y\right]N^{\rm ad}\left(\omega\right)}-A\left(\frac{my}{{\mathbb E}\left[y\right]N^{\rm ad}\left(\omega\right)}\right)^2,\label{equ:wearout}
\end{align}
where $B>0$ and $A\ge0$ are parameters. 
{Eq. (\ref{equ:wearout}) means that $\psi\left(m,y,\omega\right)$ is quadratic in $\frac{my}{{\mathbb E}\left[y\right]N^{\rm ad}\left(\omega\right)}$.} This reflects the advertising's \emph{wear-out effect}: the advertising's effectiveness may first increase and then decrease with the number of ads delivered by this advertiser to the user. 
This is because too much repetition may lead the user to have a bad impression of the product. 
The wear-out effect has been widely observed in the literature \cite{pechmann1988advertising,kirmani1997advertising}. Some studies, such as \cite{anand2011advertising} and \cite{campbell2003brand}, explicitly considered a quadratic relation between the ad repetition and the advertising's effectiveness, which is similar to (\ref{equ:wearout}). Note that a larger $A$ in (\ref{equ:wearout}) reflects a stronger degree of wear-out effect.\footnote{{When advertising its product, the advertiser can make several different versions of ads, and fill the $m$ purchased ad slots with them. This can reduce $A$, as it mitigates the feeling of repetition from the perspective of the users.}}


We define an advertiser's utility as the expected total value of its advertising's effectiveness on all users. If a user does not see the advertiser's ads, the advertising's effectiveness on the user is zero. Therefore, an advertiser's utility is simply ${\mathbb E}_y\left[\psi\left(m,y,\omega\right)\right]N^{\rm ad}\left(\omega\right)$. Considering the advertiser's payment for purchasing $m$ ad slots, the advertiser's payoff is
{\begin{align}
\nonumber
& \Pi^{\rm ad}\left(m,\omega,p\right) ={\mathbb E}_y\left[\psi\left(m,y,\omega\right)\right]N^{\rm ad}\left(\omega\right)-mp\\
\nonumber
& \overset{(a)}{=}{\mathbb E}_y\left[ \frac{Bmy}{{\mathbb E}\left[y\right]N^{\rm ad}\left(\omega\right)}-\frac{A m^2y^2}{\left({\mathbb E}\left[y\right]N^{\rm ad}\left(\omega\right)\right)^2} \right]N^{\rm ad}\left(\omega\right)-mp\\
& = \left(B-p\right)m -\frac{A {\mathbb E}\left[y^2\right]}{\left({\mathbb E}\left[y\right]\right)^2 N^{\rm ad}\left(\omega\right)}m^2. \label{equ:adpayoff}
\end{align}
Note that ${\mathbb E}\left[y\right]N^{\rm ad}\left(\omega\right)$ and $\left({\mathbb E}\left[y\right]N^{\rm ad}\left(\omega\right)\right)^2$ in the denominators on the right-hand side of equality (a) are deterministic.}

When $N^{\rm ad}\left(\omega\right)=0$, we simply define $\Pi^{\rm ad}\left(m,\omega,p\right)\triangleq -mp$, and it is easy to see that the advertiser will not purchase any ad slot in this case. 


\subsection{Two-Stage Stackelberg Game}\label{subsec:gamestructure}
We model the interactions among the operator, users, and advertisers by a two-stage Stackelberg game. In Stage I, the operator decides the unit data reward $\omega$ and ad price $p$. In Stage II, each type-$\theta$ user chooses the subscription decision $r$ and the number of watched ads $x$, and each advertiser decides the number of purchased ad slots $m$.{\footnote{If we break Stage II into two stages and consider the sequential decision making of the advertisers and users, the game's outcome will not change. This is because given the operator's unit data reward, the users' decisions are not directly affected by the advertisers' decisions. Hence, the advertisers can anticipate the users' decisions, regardless of their decision sequence.}}

We assume that the users' maximum valuation $\theta_{\max}$ satisfies $\theta_{\max}>\frac{u'\left(0\right) F}{u'\left(Q\right)u\left(Q\right)}$. Similar assumptions about the range of users' attributes have been made in \cite{dewenter2012file,rasch2013piracy,yu2019pricing}. As shown in Sections \ref{sec:awarereward} and \ref{sec:unawarereward}, this assumption implies that the high-valuation users may both subscribe to the data plan and watch ads under a small reward $\omega$. 
In fact, we can easily see that the user equilibrium under $\theta_{\max}\le\frac{u'\left(0\right) F}{u'\left(Q\right)u\left(Q\right)}$ will be a special case of that under $\theta_{\max}>\frac{u'\left(0\right) F}{u'\left(Q\right)u\left(Q\right)}$. {{We summarize our paper's key notations in Appendix \ref{appendix:notationtable}.}}






\section{Subscription-Aware Rewarding}\label{sec:awarereward}
In this section, we analyze the two-stage game under the SAR scheme, i.e., the operator only allows the subscribers of the data plan to watch ads for rewards. {{Note that we do not study the scheme which only rewards the non-subscribers for watching ads. This scheme is less reasonable in practice, i.e., the subscribers should not have a lower priority of using the service than the non-subscribers.}}

%

\subsection{Users' Decisions in Stage II}\label{subsec:stageII:user}
Given $\omega$, a type-$\theta$ user solves the following problem:
\begin{align}
\max_{r\in\left\{0,1\right\},{~}x\in\left[0,\infty\right)} \Pi^{\rm user}\left(\theta,r,x,\omega\right),{~~}{~~}{~~}{~~}{\rm s.t.}{~~}{~}x=xr,\label{equ:user:problem}
\end{align}
where $\Pi^{\rm user}\left(\theta,r,x,\omega\right)$ is given in (\ref{equ:userpayoff}), and $x=xr$ implies that a user can watch ads ($x>0$) only if it subscribes ($r=1$). 

We use $\left(u'\right)^{-1}\left(\cdot\right)$ to denote the inverse function of $u'\left(\cdot\right)$. In Lemma \ref{lemma:theta0}, we introduce several thresholds of $\theta$, which will be used to characterize the users' decisions {{(due to space limits, we leave all proofs in our appendices)}}. 


\begin{lemma}\label{lemma:theta0}
Define $\theta_0\triangleq \frac{F}{u\left(Q\right)}$ and $\theta_1\triangleq \frac{\Phi}{\omega u'\left(Q\right)}$. When $\omega\in\left(\frac{\Phi u\left(Q\right)}{Fu'\left(Q\right)},\infty\right)$, there is a unique $\theta\in\left(\theta_1,\theta_0\right)$ that satisfies $\theta u\left(\left(u'\right)^{-1}\left(\frac{\Phi}{\omega \theta}\right)\right)-F-\frac{\Phi}{\omega}\left(\left(u'\right)^{-1}\left(\frac{\Phi}{\omega\theta}\right)-Q\right)=0$, and we denote it by $\theta_2$.
\end{lemma}
Although $\theta_1$, $\theta_2$ in Lemma \ref{lemma:theta0} (and $\theta_3$, $\theta_4$ in Lemma \ref{lemma:theta3}) are functions of $\omega$, we omit this dependence in the notation to simplify the presentation. Based on these thresholds, we characterize the users' decisions in the following proposition. 

\begin{proposition}\label{proposition:stageII:user}
Under the SAR scheme, the optimal decisions of a type-$\theta$ user ($\theta\in\left[0,\theta_{\max}\right]$) are as follows:{\footnote{Here, ${\mathbbm 1}_{\left\{\cdot\right\}}$ denotes the indicator function. It equals $1$ if the event in braces is true, and equals $0$ otherwise.}}

Case A: When $\omega\in\left[0,\frac{\Phi}{u'\left(Q\right)\theta_{\max}}\right]$,
\begin{align}
\nonumber
r^*\left(\theta,\omega\right)={\mathbbm 1}_{\left\{\theta\ge \theta_0\right\}},{~~}x^*\left(\theta,\omega\right)=0;
\end{align}

Case B: When $\omega\in \left(\frac{\Phi}{u'\left(Q\right)\theta_{\max}},\frac{\Phi u\left(Q\right)}{Fu'\left(Q\right)}\right]$,
\begin{align}
\nonumber
r^*\!\left(\theta,\omega\right)\!=\!{\mathbbm 1}_{\left\{\theta\ge \theta_0\right\}},x^*\!\left(\theta,\omega\right)\!=\!\frac{1}{\omega}\!\left(\!\left(u'\right)^{-1}\!\left(\frac{\Phi}{\omega\theta}\right)\!-\!Q\right) \!{\mathbbm 1}_{\left\{\theta\ge \theta_1\right\}};
\end{align}

Case C: When $\omega\in \left(\frac{\Phi u\left(Q\right)}{Fu'\left(Q\right)},\infty \right)$,
\begin{align}
\nonumber
r^*\!\left(\theta,\omega\right)\!=\!{\mathbbm 1}_{\left\{\theta\ge \theta_2\right\}},x^*\!\left(\theta,\omega\right)\!=\!\frac{1}{\omega}\!\left(\!\left(u'\right)^{-1}\!\left(\frac{\Phi}{\omega\theta}\right)\!-\!Q\right) \!{\mathbbm 1}_{\left\{\theta\ge \theta_2\right\}}.
\end{align}
\end{proposition}

\begin{figure}[t]
  \centering
  \includegraphics[scale=0.45]{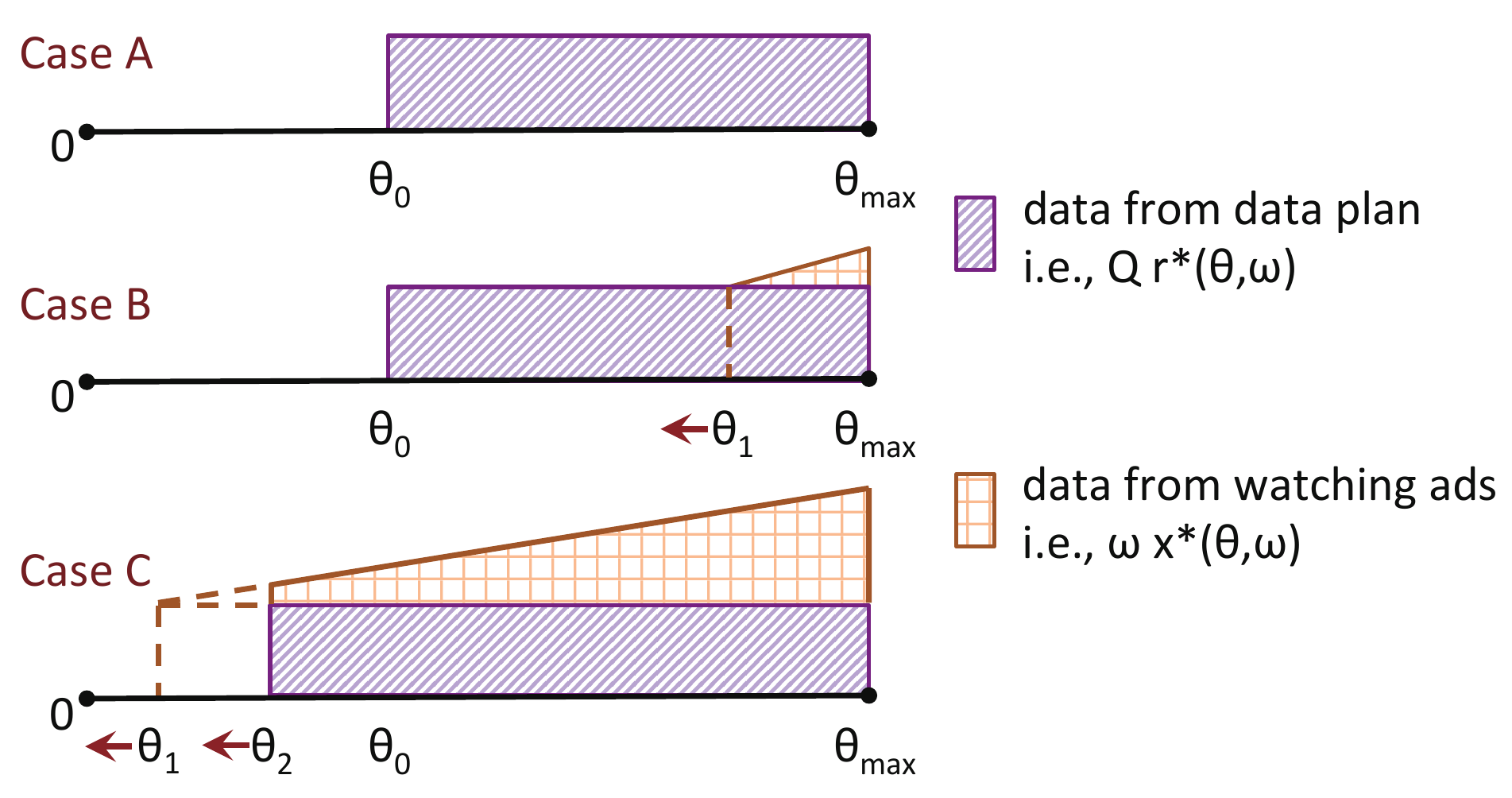}\\
  \caption{Illustration of data obtained under the SAR scheme (based on Proposition \ref{proposition:stageII:user}). For $u\left(z\right)=\ln\left(1+z\right)$, the amount of data obtained via watching ads (i.e., $\omega x^*\left(\theta,\omega\right)$) linearly increases with $\theta$ when $x^*\left(\theta,\omega\right)>0$. {{The red arrows indicate the change of $\theta_1$ and $\theta_2$ as $\omega$ increases.}}}
  \label{fig:user}
\end{figure}


In Fig. \ref{fig:user}, we illustrate the data that users with different valuations $\theta$ obtain from data plan subscriptions (i.e., $Qr^*\left(\theta,\omega\right)$) and watching ads (i.e., $\omega x^*\left(\theta,\omega\right)$). 

In Case A, only the users with $\theta\ge\theta_0$ subscribe, and no user watches ads because of the small unit data reward $\omega$.

In Case B, the users who subscribe are the same as those in Case A. Users with $\theta\ge\theta_1$ watch ads, and the threshold $\theta_1$ decreases (i.e., more users watch ads) as $\omega$ increases. Next, we focus on the users with $\theta\ge\theta_1$. We can show that the number of watched ads $x^*\left(\theta,\omega\right)$ increases with $\theta$ (note that $\left(u'\right)^{-1}\left(\cdot\right)$ is decreasing because of the strict concavity of $u\left(\cdot\right)$). In particular, the marginal increase of $x^*\left(\theta,\omega\right)$ with respect to $\theta$ is affected by the utility function $u\left(z\right)$:
\begin{itemize}
\item If $u\left(z\right)=\ln\left(1+z\right)$, we can show that $x^*\left(\theta,\omega\right)$ linearly increases with $\theta$ (as illustrated in Fig. \ref{fig:user});
\item {If $u\left(z\right)=\frac{\left(z+\mu\right)^{1-\alpha}}{1-\alpha}-\frac{\mu^{1-\alpha}}{1-\alpha}, 0<\alpha<1,\mu\ge0$, then $x^*\left(\theta,\omega\right)$ convexly increases with $\theta$;}
\item If $u\left(z\right)=1-e^{-\gamma z},\gamma>0$, then $x^*\left(\theta,\omega\right)$ concavely increases with $\theta$.
\end{itemize}

In Case C, more users subscribe compared with Cases A and B, i.e., the subscription threshold $\theta_2$ is smaller than $\theta_0$. This is because the unit reward $\omega$ is large and users with $\theta\in\left[\theta_2,\theta_0\right)$ subscribe to be eligible for the data rewards. 
{{In Appendix \ref{appendix:monotonicity:theta2}, we prove that $\theta_2$ decreases (i.e., more users subscribe) as $\omega$ increases.}} Moreover, each subscriber watches a positive number of ads, i.e., $x^*\left(\theta,\omega\right)>0$ for $\theta\ge \theta_2$.

Based on these results, we can see one key advantage of the SAR scheme: it leads to a large number of data plan subscriptions.


\subsection{Advertisers' Decisions in Stage II}\label{subsec:advertiser}
Given $p$ and $\omega$, each advertiser solves the following problem:\!
\begin{align}
\max_{m\in\left[0,\infty\right)} \Pi^{\rm ad}\left(m,\omega,p\right),\label{problem:advertiser}
\end{align}
where the payoff $\Pi^{\rm ad}\left(m,\omega,p\right)$ is given in (\ref{equ:adpayoff}). We characterize the optimal number of purchased ad slots in Proposition \ref{proposition:advertiser}.

\begin{proposition}\label{proposition:advertiser}
If $N^{\rm ad}\left(\omega\right)=0$ or $p\ge B$, then $m^*\left(\omega,p\right)=0$; otherwise,
\begin{align}
m^*\left(\omega,p\right)=\frac{B-p}{2A} \frac{\left({\mathbb E}\left[y\right]\right)^2}{{\mathbb E}\left[y^2\right]} N^{\rm ad}\left(\omega\right).\label{equ:mlowp}
\end{align}
\end{proposition}


Recall that the random variable $y$ denotes the value of $x^*\left(\theta,\omega\right)$ when $x^*\left(\theta,\omega\right)>0$, and $N^{\rm ad}\left(\omega\right)$ is the mass of users watching ads. In (\ref{equ:mlowp}), $m^*\left(\omega,p\right)$ decreases with the degree of wear-out effect $A$. Moreover, since ${\mathbb E}\left[y^2\right]=\left({\mathbb E}\left[y\right]\right)^2+{\rm Var}\left[y\right]$, we can see that $m^*\left(\omega,p\right)$ decreases with ${\rm Var}\left[y\right]$ (i.e., the variance of $y$). This implies that the advertisers prefer a low variation in the number of ads watched by each of the $N^{\rm ad}\left(\omega\right)$ users. 
{The reason is that the advertising's effectiveness is concave in $y$ given ${\mathbb E}\left[y\right]$ (as shown in (\ref{equ:wearout})).}

Given the concrete utility function $u\left(\cdot\right)$ and the distribution of $\theta$, we can derive $x^*\left(\theta,\omega\right)$ based on Proposition \ref{proposition:stageII:user}, and further compute ${\mathbb E}\left[y\right]$, ${\mathbb E}\left[y^2\right]$, and $N^{\rm ad}\left(\omega\right)$ in (\ref{equ:mlowp}). {We give an example as follows.
\begin{example}\label{example:1}
Suppose that $u\left(z\right)=\ln\left(1+z\right)$, $\theta$ is uniformly distributed in $\left[0,\theta_{\max}\right]$, and $\omega$ satisfies Case B in Proposition \ref{proposition:stageII:user}, i.e., $\omega\in \left(\frac{\left(1+Q\right)\Phi}{\theta_{\max}},\frac{\Phi}{F}\left(1+Q\right)\ln\left(1+Q\right) \right]$. Based on Proposition \ref{proposition:stageII:user}, the users' ad watching decisions are characterized as follows:
\begin{align}
x^*\left(\theta,\omega\right)=\frac{\theta-\theta_1}{\Phi} {\mathbbm 1}_{\left\{\theta\ge \theta_1\right\}},\label{equ:example:x}
\end{align}
where $\theta_1=\frac{\Phi\left(1+Q\right)}{\omega}$. Hence, only the users with $\theta\ge\theta_1$ watch ads. Since $\theta$ is uniformly distributed in $\left[0,\theta_{\max}\right]$, we can further compute $N^{\rm ad}\left(\omega\right)$ as follows:
\begin{align}
N^{\rm ad}\left(\omega\right)=\frac{\theta_{\max}-\theta_1}{\theta_{\max}} N.
\end{align}
According to (\ref{equ:example:x}) and the fact that $\theta\sim{\cal U}\left[0,\theta_{\max}\right]$, the number of ads watched by one of the $N^{\rm ad}\left(\omega\right)$ users is uniformly distributed in $\left[0,\frac{\theta_{\max}-\theta_1}{\Phi}\right]$. This implies that $y$ is uniformly distributed in $\left[0,\frac{\theta_{\max}-\theta_1}{\Phi}\right]$.{\footnote{{Strictly speaking, $x^*\left(\theta_1,\omega\right)=0$ and hence only the users with $\theta>\theta_1$ will watch ads. As a result, $y$ should be uniformly distributed in $\left(0,\frac{\theta_{\max}-\theta_1}{\Phi}\right]$. However, the probability that $\theta=\theta_1$ is zero due to the continuous distribution of $\theta$. Therefore, we can consider users with $\theta\ge\theta_1$ when counting $N^{\rm ad}\left(\omega\right)$ and treat $y$ as a variable uniformly distributed in $\left[0,\frac{\theta_{\max}-\theta_1}{\Phi}\right]$ without affecting our analysis.}}} Then, we can compute ${\mathbb E}\left[y\right]$ and ${\mathbb E}\left[y^2\right]$ as follows:
\begin{align}
{\mathbb E}\left[y\right]=\frac{1}{2}\frac{\theta_{\max}-\theta_1}{\Phi},{\mathbb E}\left[y^2\right]=\frac{1}{3} \left(\frac{\theta_{\max}-\theta_1}{\Phi}\right)^2.
\end{align}
Based on Proposition \ref{proposition:advertiser}, we can derive $m^*\left(\omega,p\right)$ as follows:
\begin{align}
m^*\left(\omega,p\right)=\frac{3}{8}\frac{\max\left\{B-p,0\right\}}{A}  \frac{\theta_{\max}-\theta_1}{\theta_{\max}} N.
\end{align}
\end{example}
}

{{In Appendix \ref{appendix:details:SAR}, we compute $m^*\left(\omega,p\right)$ for other values of $\omega$ (i.e., $\omega$ that satisfies Cases A or C) under the logarithmic utility function and uniformly distributed user types.}}

\subsection{Operator's Decisions in Stage I}\label{subsec:aware:operator}
The operator obtains revenue from both the mobile data market and ad market. In the mobile data market, each user who subscribes to the data plan should pay $F$ to the operator. The operator's corresponding revenue is 
\begin{align}
R^{\rm data}\left(\omega\right) = NF\int_{0}^{\theta_{\max}} r^*\left(\theta,\omega\right) g\left(\theta\right) d\theta.\label{equ:define:Rdata}
\end{align}
In the ad market, each advertiser pays $p$ for each purchased ad slot. The operator's corresponding revenue is 
\begin{align}
R^{\rm ad}\left(\omega,p\right) = K m^*\left(\omega,p\right) p.\label{equ:define:Rad}
\end{align}

Let $D\left(\omega\right)$ denote the total data demand, i.e., the total amount of mobile data that users request (by subscription and watching ads) under reward $\omega$. We can compute $D\left(\omega\right)$ as
\begin{align}
D\left(\omega\right) = N\int_{0}^{\theta_{\max}} \left(Q r^*\left(\theta,\omega\right)+\omega x^*\left(\theta,\omega\right)\right)  g\left(\theta\right) d\theta,\label{equ:define:Demand}
\end{align}
where $Q r^*\left(\theta,\omega\right)$ and $\omega x^*\left(\theta,\omega\right)$ are illustrated in Fig. \ref{fig:user}.

Based on $R^{\rm data}\left(\omega\right)$, $R^{\rm ad}\left(\omega,p\right)$, and $D\left(\omega\right)$, we formulate the operator's problem as follows:
\begin{align}
& \max_{\omega\ge0,p>0} R^{\rm total}\left(\omega,p\right) \triangleq R^{\rm data}\left(\omega\right)+R^{\rm ad}\left(\omega,p\right)\label{equ:obj}\\
& {~~}{~~}{\rm s.t.}{~~} D\left(\omega\right) \le C,\label{equ:capacity:data}\\
& {~~}{~~}{~~}{~~}{~~}K m^*\left(\omega,p\right)\le {\mathbb E}\left[y\right] N^{\rm ad}\left(\omega\right).\label{equ:capacity:ad}
\end{align}
Here, $R^{\rm total}\left(\omega,p\right)$ is the operator's total revenue. 
Constraint (\ref{equ:capacity:data}) implies that the total data demand $D\left(\omega\right)$ cannot exceed a capacity $C$ \cite{joe2018sponsoring,zhang2015sponsored}. To ensure that choosing $\omega=0$ (i.e., no data reward) is feasible to the problem, we assume that $C \ge D\left(0\right)$. Here, $D\left(0\right)$ is the data demand when the operator only offers the data plan without any data reward. Constraint (\ref{equ:capacity:ad}) implies that the total number of sold ad slots (i.e., $K m^*\left(\omega,p\right)$) should not exceed the number of available ad slots. When the operator does not sell all ad slots, it can fill the unsold slots with content like public news {{and pictures}} to guarantee the fairness among the users choosing to watch ads {{(e.g., Optus displayed wallpapers to users when there were unsold ad slots \cite{OptusFill})}}. 



To solve (\ref{equ:obj})-(\ref{equ:capacity:ad}), we first analyze $p^*\left(\omega\right)$, which is the optimal ad price under a given $\omega$. Then, we substitute $p=p^*\left(\omega\right)$ into $R^{\rm total}\left(\omega,p\right)$, and analyze the optimal unit data reward $\omega^*$. We characterize $p^*\left(\omega\right)$ in the following theorem.

\begin{theorem}\label{theorem:SAR:price}
If $\omega\in\left[0,\frac{\Phi}{u'\left(Q\right)\theta_{\max}}\right]$, any positive price is optimal; 
if $\omega\in\left(\frac{\Phi}{u'\left(Q\right)\theta_{\max}},\infty\right)$, 
\begin{align}
p^*\left(\omega\right)=\max\left\{\frac{B}{2},B-\frac{2A {\mathbb E}\left[y^2\right]}{K {\mathbb E}\left[y\right]}\right\}.\label{equ:aware:optimalp}
\end{align}
\end{theorem}
Note that the random variable $y$ is the value of $x^*\left(\theta,\omega\right)$ when $\!x^*\left(\theta,\omega\right)\!>\!0$. Hence, both ${\mathbb E}\left[y^2\right]\!$ and ${\mathbb E}\left[y\right]$ depend on $\omega$. 

If $\omega\in\left[0,\frac{\Phi}{u'\left(Q\right)\theta_{\max}}\right]$, no user watches ads (based on Proposition \ref{proposition:stageII:user}). In this case, the advertisers do not purchase ad slots, regardless of the ad price $p$. 
If $\omega\in\left(\frac{\Phi}{u'\left(Q\right)\theta_{\max}},\infty\right)$, Eq. (\ref{equ:aware:optimalp}) implies that $p^*\left(\omega\right)$ decreases with $A$ (the degree of wear-out effect) when $A$ is small, but does not change with $A$ when $A$ is large. When $A<\frac{BK{\mathbb E}\left[y\right]}{4 {\mathbb E}\left[y^2\right]}$, the wear-out effect is small, and the advertisers have high willingness to purchase ad slots. Hence, the operator chooses $p^*\left(\omega\right)=B-\frac{2A {\mathbb E}\left[y^2\right]}{K {\mathbb E}\left[y\right]}$ to sell all the ad slots (which leads to $K m^*\left(\omega,p^*\left(\omega\right)\right)= {\mathbb E}\left[y\right] N^{\rm ad}\left(\omega\right)$). When $A\ge \frac{BK{\mathbb E}\left[y\right]}{4 {\mathbb E}\left[y^2\right]}$, the large wear-out effect decreases the advertisers' willingness to purchase slots. The operator will not sell all slots, and will choose $p^*\left(\omega\right)=\frac{B}{2}$, which is independent of $A$.


Next, we analyze $\omega^*$, which maximizes $R^{\rm total}\left(\omega,p^*\left(\omega\right)\right)$, subject to $D\left(\omega\right) \le C$. We first introduce Proposition \ref{proposition:monotonicity}.
\begin{proposition}\label{proposition:monotonicity}
\!\!Given $C\!\ge\! D\left(0\right)$, there is a unique $\omega\!\in\!\!\left[\frac{\Phi}{u'\left(Q\right)\theta_{\max}},\!\infty\!\right)\!$ such that $D\left(\omega\right)=C$. We denote this $\omega$ by $D^{-1}\left(C\right)$. Moreover, $D^{-1}\left(C\right)$ strictly increases with $C$.
\end{proposition}
Based on Proposition \ref{proposition:monotonicity}, we can rewrite $D\left(\omega\right) \le C$ as $\omega\le D^{-1}\left(C\right)$. From numerical experiments, $R^{\rm total}\left(\omega,p^*\left(\omega\right)\right)$ is either always increasing or unimodal in $\omega\in\left[0,\infty\right)$. Hence, we can easily search for $\omega^*$ in the interval $\left[0,D^{-1}\left(C\right)\right]$ (e.g., when $R^{\rm total}\left(\omega,p^*\left(\omega\right)\right)$ is unimodal, we can apply the Golden Section method \cite{bertsekas1999nonlinear}). 
Next, we study when the operator will choose $\omega$ to be $D^{-1}\left(C\right)$, i.e., use up the network capacity for data rewards. In Theorem \ref{theorem:subscriptionaware}, we show a sufficient condition under which $\omega^*=D^{-1}\left(C\right)$.


\begin{theorem}\label{theorem:subscriptionaware}
Under the SAR scheme, if both $\frac{\left({\mathbb E}\left[y\right]\right)^2}{{\mathbb E}\left[y^2\right]} N^{\rm ad}\left(\omega\right)$ and ${\mathbb E}\left[y\right] N^{\rm ad}\left(\omega\right)$ increase with $\omega$ for $\omega\in\left(\frac{\Phi}{u'\left(Q\right)\theta_{\max}},\infty\right)$, the operator's optimal unit data reward is given by $\omega^*=D^{-1}\left(C\right)$.
\end{theorem}

We explain this sufficient condition by discussing the unit data reward $\omega$'s influence on $R^{\rm data}\left(\omega\right)$ and $R^{\rm ad}\left(\omega,p^*\left(\omega\right)\right)$. First, increasing $\omega$ can increase $R^{\rm data}\left(\omega\right)$, because more users subscribe. Second, increasing $\omega$ has the following impacts on $R^{\rm ad}\left(\omega,p^*\left(\omega\right)\right)$: (i) (\emph{positive impact}) It increases $N^{\rm ad}\left(\omega\right)$, i.e., more users watch ads; (ii) (\emph{negative impact}) It may decrease ${\mathbb E}\left[y\right]$. Under a larger $\omega$, a user can obtain a larger amount of data after watching a few ads. Then, the user's willingness to watch more ads may decrease because of the concavity of the utility function; (iii) (\emph{negative impact}) It may increase ${\rm Var}\left[y\right]$. Under a larger $\omega$, more users with different valuations $\theta$ watch ads, which can increase the variance of $y$. As discussed in Section \ref{subsec:advertiser}, increasing ${\rm Var}\left[y\right]$ decreases the advertisers' willingness to purchase ad slots. Under a general utility function $u\left(\cdot\right)$ and a general distribution of $\theta$, it is challenging to analyze the net effect of the above impacts. Theorem \ref{theorem:subscriptionaware} implies that when both $\frac{\left({\mathbb E}\left[y\right]\right)^2}{{\mathbb E}\left[y^2\right]} N^{\rm ad}\left(\omega\right)$ and ${\mathbb E}\left[y\right] N^{\rm ad}\left(\omega\right)$ increase with $\omega$, the positive impact dominates the negative impacts. In this case, the operator should set $\omega$ as large as possible without violating the capacity constraint (\ref{equ:capacity:data}) under the SAR scheme. 

A widely considered setting is that each user has a logarithmic utility function (e.g., \cite{duan2015pricing,schmidt2009minimum}) and a uniformly distributed type (e.g., \cite{rasch2013piracy,yu2017public}). We can verify that this setting satisfies the sufficient condition in Theorem \ref{theorem:subscriptionaware}, and hence we have the following proposition.
\begin{proposition}\label{proposition:uniform:useup}
When $u\left(z\right)=\ln\left(1+z\right)$ and $\theta\sim{\cal U}\left[0,\theta_{\max}\right]$, the operator's optimal unit data reward is given by $\omega^*=D^{-1}\left(C\right)$.
\end{proposition}

{{When each user has an exponential utility function (i.e., $u\left(z\right)=1-e^{-\gamma z}$), ${\mathbb E}\left[y\right] N^{\rm ad}\left(\omega\right)$ may decrease with $\omega$ and $\omega^*$ can be smaller than $D^{-1}\left(C\right)$ (i.e., the operator does not use up the capacity for rewards). We show an example in Appendix \ref{appendix:numericalexample}.}}






\section{Subscription-Unaware Rewarding}\label{sec:unawarereward}
In this section, we consider the SUR scheme, i.e., both the subscribers and non-subscribers can watch ads for rewards. 

\subsection{Users' Decisions in Stage II}\label{subsec:unaware:user}
Since the users can watch ads without subscription, each type-$\theta$ user simply chooses $r$ and $x$ to maximize its payoff without the constraint $x=xr$, as in (\ref{equ:user:problem}) in Section \ref{subsec:stageII:user}.


In Lemma \ref{lemma:theta3}, we introduce two new thresholds of $\theta$.
\begin{lemma}\label{lemma:theta3}
Define $\theta_3\triangleq \frac{\Phi}{\omega u'\left(0\right) }$. When $\omega\in\left(\frac{\Phi u\left(Q\right)}{F u'\left(0\right)},\frac{\Phi Q}{F}\right)$, there is a unique $\theta\in\left(\theta_3,\theta_1\right)$ that satisfies $\theta u\left( \left(u'\right)^{-1}\left(\frac{\Phi}{\omega\theta}\right) \right)- \frac{\Phi}{\omega} \left(u'\right)^{-1}\left(\frac{\Phi}{\omega\theta}\right)=\theta u\left(Q\right)-F$, and we denote it by $\theta_4$.
\end{lemma}


{{Recall that $\left(u'\right)^{-1}\left(\cdot\right)$ denotes the inverse function of $u'\left(\cdot\right)$.}} Based on the thresholds introduced in Lemmas \ref{lemma:theta0} and \ref{lemma:theta3}, we characterize the users' decisions in the following proposition (we use symbol $\hat {~}$ to indicate that the results are obtained under the SUR scheme). 

\begin{figure}[t]
  \centering
  \includegraphics[scale=0.45]{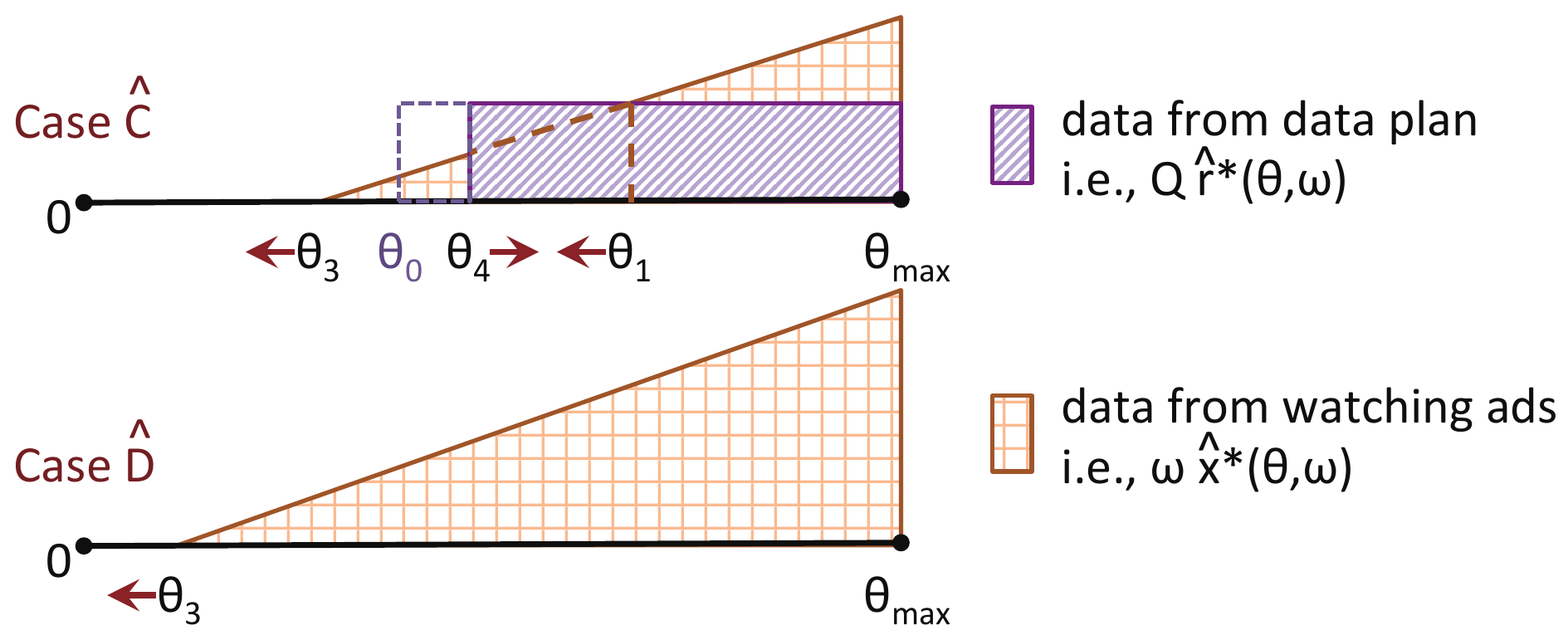}\\
  \caption{Illustration of data obtained under the SUR scheme (based on Proposition \ref{proposition:unaware:user}). For $u\left(z\right)=\ln\left(1+z\right)$, the amount of data obtained via watching ads (i.e., $\omega {\hat x}^*\left(\theta,\omega\right)$) linearly increases with $\theta$ when ${\hat x}^*\left(\theta,\omega\right)>0$. {{The red arrows indicate the change of $\theta_1$, $\theta_3$, and $\theta_4$ as $\omega$ increases.}}}
  \label{fig:user:unaware}
\end{figure}

\begin{proposition}\label{proposition:unaware:user}
Under the SUR scheme, the optimal decisions of a type-$\theta$ user ($\theta\in\left[0,\theta_{\max}\right]$) are as follows:

Case $\hat A$: When $\omega\in\left[0,\frac{\Phi}{u'\left(Q\right)\theta_{\max}}\right]$,
\begin{align}
\nonumber
{\hat r}^*\left(\theta,\omega\right)={\mathbbm 1}_{\left\{\theta\ge \theta_0\right\}},{~~}{~~}{\hat x}^*\left(\theta,\omega\right)=0;
\end{align}

Case $\hat B$: When $\omega\in\left(\frac{\Phi}{u'\left(Q\right)\theta_{\max}},\frac{\Phi u\left(Q\right)}{F u'\left(0\right)}\right]$,
\begin{align}
\nonumber
{\hat r}^*\!\left(\theta,\omega\right)\!=\!\!{\mathbbm 1}_{\left\{\theta\ge \theta_0\right\}},\!{\hat x}^*\!\left(\theta,\omega\right)\!\!=\!\!\frac{1}{\omega}\!\left(\!\left(u'\right)^{-1}\!\!\left(\frac{\Phi}{\omega\theta}\right)\!-\!Q\!\right) \!\!{\mathbbm 1}_{\!\left\{\theta\ge \theta_1\!\right\}};\!
\end{align}

Case $\hat C$: When $\omega\in\left(\frac{\Phi u\left(Q\right)}{F u'\left(0\right)},\frac{\Phi Q}{F}\right)$,
\begin{align}
\nonumber
& {\hat r}^*\!\left(\theta,\!\omega\right)\!\!=\!\!{\mathbbm 1}_{\left\{\theta\ge \theta_4\right\}},\\
\nonumber
& {\hat x}^*\!\!\left(\theta,\!\omega\right)\!\!=\!\!\frac{1}{\omega}\!\!\left(u'\right)^{-1}\!\!\left(\!\frac{\Phi}{\omega\theta}\!\right) \!\!{\mathbbm 1}_{\left\{\theta_3 \le \theta< \theta_4\right\}}\!\!+\!\!\frac{1}{\omega}\!\!\left(\!\!\left(u'\right)^{-1}\!\!\left(\!\frac{\Phi}{\omega\theta}\!\right)\!\!-\!Q\!\!\right) \!\!{\mathbbm 1}_{\!\left\{\!\theta\ge \theta_1\!\right\}}\!;
\end{align}

Case $\hat D$: When $\omega\in\left[\frac{\Phi Q}{F},\infty\right)$,
\begin{align}
\nonumber
{\hat r}^*\left(\theta,\omega\right)=0,{~~}{\hat x}^*\left(\theta,\omega\right)=\frac{1}{\omega}\left(u'\right)^{-1}\left(\frac{\Phi}{\omega\theta}\right){\mathbbm 1}_{\left\{\theta \ge \theta_3 \right\}}.
\end{align}

\end{proposition}
The users' optimal decisions in Cases $\hat A$ and $\hat B$ are the same as those in Cases $A$ and $B$ (under the SAR scheme), respectively. 
Hence, in Fig. \ref{fig:user:unaware}, we only illustrate the data obtained by users via subscription (i.e., $Q{\hat r}^*\left(\theta,\omega\right)$) and watching ads (i.e., $\omega {\hat x}^*\left(\theta,\omega\right)$) in Cases $\hat C$ and $\hat D$.

  
In Case $\hat C$, two segments of users watch ads: users with valuations $\theta\ge \theta_1$ watch ads and subscribe; users with valuations $\theta_3\le\theta<\theta_4$ watch ads without subscription. We characterize the properties of $\theta_4$ in the following lemma.
\begin{lemma}\label{lemma:theta4}
When $\omega\in\left(\frac{\Phi u\left(Q\right)}{F u'\left(0\right)},\frac{\Phi Q}{F}\right)$ (i.e., Case $\hat C$), (i) $\theta_4$ is greater than $\theta_0$, and (ii) $\theta_4$ increases with $\omega$.
\end{lemma}

In Case $\hat B$, the subscription threshold is $\theta_0$. Hence, result (i) of Lemma \ref{lemma:theta4} implies that some low-valuation users who subscribe in Case $\hat B$ become non-subscribers in Case $\hat C$. This is because these low-valuation users' marginal benefit of consuming data decreases after earning the data rewards, and it is no longer beneficial for them to subscribe to the data plan in Case $\hat C$. Result (ii) of Lemma \ref{lemma:theta4} shows that more subscribers become non-subscribers as the unit reward increases.

In Case $\hat D$, since $\omega$ is large, all users simply watch ads to earn the rewards, without paying for the subscription. 

\begin{figure*}[t]
  \centering
  \subfigure[Function ${\hat D}\left(\omega\right)$.]{
    \includegraphics[scale=0.4]{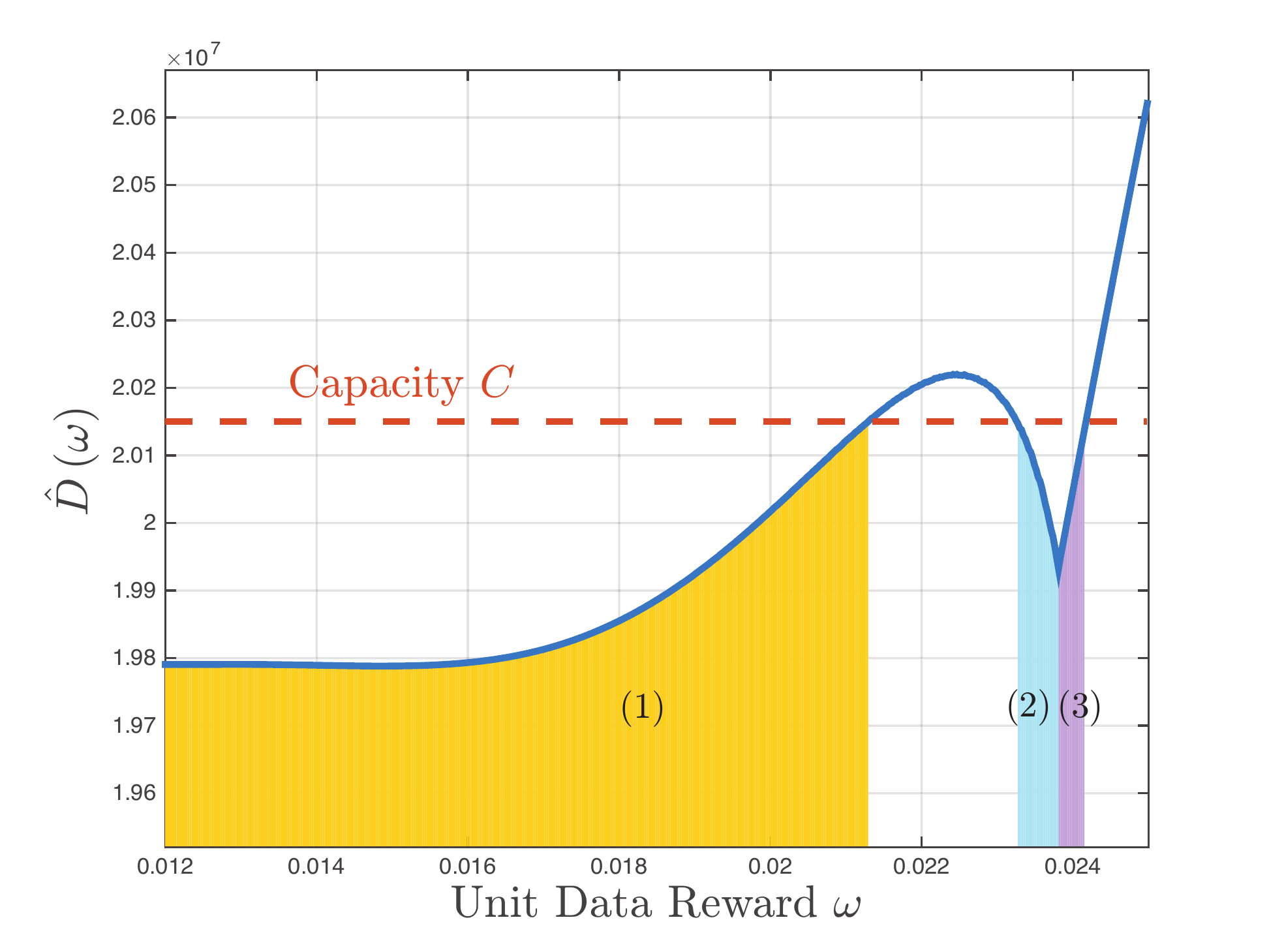}\label{fig:sur:dfunction}}
  \subfigure[Function ${\hat R}^{\rm total}\left(\omega,{\hat p}^*\left(\omega\right)\right)$.]{
    \includegraphics[scale=0.4]{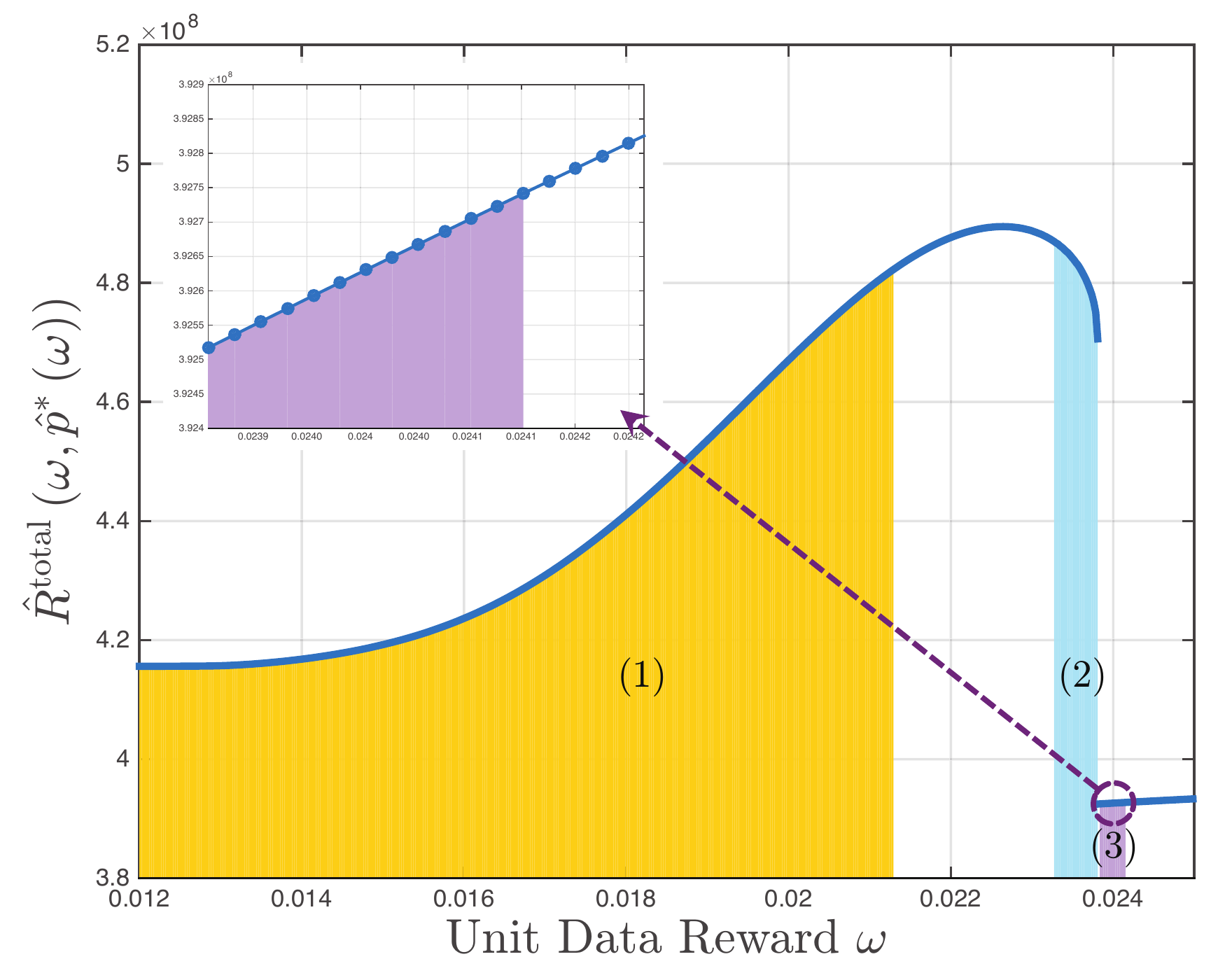}\label{fig:sur:rfunction}}
  \caption{{{Examples of ${\hat D}\left(\omega\right)$ and ${\hat R}^{\rm total}\left(\omega,{\hat p}^*\left(\omega\right)\right)$: {We assume that $u\left(z\right)=1-e^{-0.7z}$ and obtain the distribution of $\theta$ by truncating the normal distribution ${\cal N}\left(125,30\right)$ to interval $\left[0,250\right]$. We choose $N=10^7$, $F=42$, $Q=2$, $\Phi=0.5$, $K=13$, $A=1$, $B=5$, and $C=2.015\times10^7$.}}}}
\end{figure*}

\subsection{Advertisers' Decisions in Stage II}
Compared with the SAR scheme, the SUR scheme changes each advertiser's optimal decision through changing the mass of users watching ads and the distribution of the number of ads watched by each of these users. 

Given ${\hat r}^*\left(\theta,\omega\right)$ and ${\hat x}^*\left(\theta,\omega\right)$ in Proposition \ref{proposition:unaware:user}, we can compute ${\hat N}^{\rm ad}\left(\omega\right)$ (i.e., the mass of users watching ads) and the distribution of ${\hat y}$ (i.e., ${\hat x}^*\left(\theta,\omega\right)$'s value when ${\hat x}^*\left(\theta,\omega\right)>0$). 
To compute ${\hat m}^*\left(\omega,p\right)$, we can simply replace $N^{\rm ad}\left(\omega\right)$, ${\mathbb E}\left[y\right]$, and ${\mathbb E}\left[y^2\right]$ in Proposition \ref{proposition:advertiser} by ${\hat N}^{\rm ad}\left(\omega\right)$, ${\mathbb E}\left[{\hat y}\right]$, and ${\mathbb E}\left[{\hat y}^2\right]$. 

\subsection{Operator's Decisions in Stage I}\label{subsec:unaware:stageI}
Based on ${\hat r}^*\left(\theta,\omega\right)$, ${\hat x}^*\left(\theta,\omega\right)$, and ${\hat m}^*\left(\omega,p\right)$, we can compute ${\hat R}^{\rm data}\left(\omega\right)$, ${\hat R}^{\rm ad}\left(\omega,p\right)$, and ${\hat D}\left(\omega\right)$ in a similar manner as in (\ref{equ:define:Rdata})-(\ref{equ:define:Demand}). The operator's problem in Stage I is then given by:
\begin{align}
& \max_{\omega\ge0,p>0} {\hat R}^{\rm total}\left(\omega,p\right) \triangleq {\hat R}^{\rm data}\left(\omega\right)+{\hat R}^{\rm ad}\left(\omega,p\right)\label{equ:unaware:obj}\\
& {~~}{~~}{\rm s.t.}{~~} {\hat D}\left(\omega\right) \le C,{~~}K {\hat m}^*\left(\omega,p\right)\le {\hat N}^{\rm ad}\left(\omega\right) {\mathbb E}\left[{\hat y}\right],\label{equ:unaware:constraint}
\end{align}
which is similar to problem (\ref{equ:obj})-(\ref{equ:capacity:ad}). 

{{To solve (\ref{equ:unaware:obj})-(\ref{equ:unaware:constraint}), we first compute ${\hat p}^*\left(\omega\right)$, i.e., the optimal ad price under a given $\omega$. The analysis of ${\hat p}^*\left(\omega\right)$ is similar to that of $p^*\left(\omega\right)$ in Theorem \ref{theorem:SAR:price} under the SAR scheme. We can prove that if $\omega\in\left[0,\frac{\Phi}{u'\left(Q\right)\theta_{\max}}\right]$, no user watches ads and hence any positive ad price is optimal; otherwise, we have ${\hat p}^*\left(\omega\right)=\max\left\{\frac{B}{2},B-\frac{2A {\mathbb E}\left[{\hat y}^2\right]}{K {\mathbb E}\left[{\hat y}\right]}\right\}$.}}

Then, we compute ${\hat \omega}^*$ by maximizing ${\hat R}^{\rm total}\left(\omega,{\hat p}^*\left(\omega\right)\right)$, subject to ${\hat D}\left(\omega\right) \le C$. The computation of ${\hat \omega}^*$ is different from that of ${\omega}^*$ under the SAR scheme, because {(i) ${\hat D}\left(\omega\right)$ can be decreasing in $\omega\in\left(\frac{\Phi u\left(Q\right)}{F u'\left(0\right)},\frac{\Phi Q}{F}\right)$, and (ii) ${\hat R}^{\rm total}\left(\omega,{\hat p}^*\left(\omega\right)\right)$ is discontinuous at $\omega=\frac{\Phi Q}{F}$. Specifically, when $\omega\in\left(\frac{\Phi u\left(Q\right)}{F u'\left(0\right)},\frac{\Phi Q}{F}\right)$, increasing $\omega$ reduces the number of data plan subscribers, which may decrease ${\hat D}\left(\omega\right)$. Moreover, when $\omega$ increases to $\frac{\Phi Q}{F}$, all data plan subscribers quit their subscriptions and the distribution of users' ad watching times also changes. This leads to the discontinuity of ${\hat R}^{\rm total}\left(\omega,{\hat p}^*\left(\omega\right)\right)$ at $\omega=\frac{\Phi Q}{F}$. We illustrate examples of ${\hat D}\left(\omega\right)$ and ${\hat R}^{\rm total}\left(\omega,{\hat p}^*\left(\omega\right)\right)$ in Fig. \ref{fig:sur:dfunction} and Fig. \ref{fig:sur:rfunction}, respectively.}

{We can compute ${\hat \omega}^*$ as follows. First, we search for $\omega$'s feasible region, where ${\hat D}\left(\omega\right) \le C$. We can numerically show that $\omega$'s feasible region consists of at most three intervals. Then, we can show that ${\hat R}^{\rm total}\left(\omega,{\hat p}^*\left(\omega\right)\right)$ is either monotone or unimodal in each interval.\footnote{{For example, in Fig. \ref{fig:sur:dfunction}, $\omega$'s feasible region consists of the yellow, blue, and purple intervals (denoted by intervals (1), (2), and (3)). In Fig. \ref{fig:sur:rfunction}, ${\hat R}^{\rm total}\left(\omega,{\hat p}^*\left(\omega\right)\right)$ is increasing when $\omega$ is in the yellow or purple intervals, and is decreasing when $\omega$ is in the blue interval.}} Hence, we can determine ${\hat \omega}^*$ by comparing the local optimal unit data rewards found in these intervals.}


Under the SAR scheme, the operator always uses up the capacity for rewards if $u\left(z\right)=\ln\left(1+z\right)$ and $\theta\sim{\cal U}\left[0,\theta_{\max}\right]$. Under the SUR scheme, this does not hold, and a large $\omega$ may even generate a total revenue that is lower than the revenue when the operator does not offer any reward. This is because a large $\omega$ may reduce the number of subscribers (as shown in Case $\hat C$) and hence decrease ${\hat R}^{\rm data}\left(\omega\right)$. 
Next, we characterize a sufficient condition under which the network capacity is not used up for rewards (given general $u\left(z\right)$ and $g\left(\theta\right)$).
\begin{theorem}\label{theorem:notuse}
Under the SUR scheme, when the network capacity $C>N\left(u'\right)^{-1}\left(\frac{F}{\theta_{\max}Q}\right)$ and the degree of wear-out effect $A>\frac{B^2K}{8F \int_{\theta_0}^{\theta_{\max}}g\left(\theta\right) d\theta}$, we have ${\hat D}\left({\hat \omega}^*\right)<C$. 
\end{theorem}
When the operator has a large capacity and the wear-out effect is large, using up the capacity for rewards will significantly decrease ${\hat R}^{\rm data}\left(\omega\right)$ and will not significantly increase ${\hat R}^{\rm ad}\left(\omega,{\hat p}^*\left(\omega\right)\right)$. Hence, we have ${\hat D}\left({\hat \omega}^*\right)<C$ in this situation. {{We can show that both thresholds $N\left(u'\right)^{-1}\left(\frac{F}{\theta_{\max}Q}\right)$ and $\frac{B^2K}{8F \int_{\theta_0}^{\theta_{\max}}g\left(\theta\right) d\theta}$ decrease with $F$ (i.e., the subscription fee). Intuitively, if the data plan is expensive, the operator should not use up the capacity for rewards under the SUR scheme.}}

\subsection{Extension: Differentiation of Ad Slots}\label{subsec:differentiation}
In the above analysis, we assume that the operator does not differentiate the ad slots generated by the users. It sells all ad slots to the advertisers at the same price, and randomly draws ads from all ad slots when a user watches ads. Under the SUR scheme, the ad slots can be generated by both the subscribers and non-subscribers. In this section, we consider the differentiation of these two types of ad slots,{\footnote{Besides the subscription decision $r$, a user decides $x$, e.g., the number of ads to watch within a month. Different from $r$, the operator does not precisely know the user's decision of $x$ until the end of the month. If the operator can estimate $x$'s range based on the user's historical behavior, it can classify users into different categories and differentiate the corresponding ad slots similarly.}} which affects both the pricing and ad display rule. The operator can sell these two types of ad slots at different prices. When a subscriber or non-subscriber watches ads, the operator draws ads only from the corresponding type of ad slots (e.g., if an advertiser only purchases the ad slots generated by the subscribers, its ads will only be seen by the subscribers).

Given $\omega$, we use ${\hat N}_{\rm I}^{\rm ad}\left(\omega\right)$ and ${\hat N}_{\rm II}^{\rm ad}\left(\omega\right)$ to denote the number of the subscribers that watch ads and the number of the non-subscribers that watch ads, respectively. Let random variables ${\hat y}_{\rm I}$ and ${\hat y}_{\rm II}$ denote the numbers of ads watched by one of these subscribers and one of these non-subscribers, respectively. Similar to Proposition \ref{proposition:advertiser}, we have the following results:
\begin{itemize}
\item For the ad slots generated by the subscribers, the operator can set a price $p_{\rm I}>0$. If ${\hat N}_{\rm I}^{\rm ad}\left(\omega\right)>0$, the number of these slots purchased by each advertiser is ${\hat m}_{\rm I}^*\left(\omega,p_{\rm I}\right)=\frac{\max\left\{B-p_{\rm I},0\right\}}{2A} \frac{\left({\mathbb E}\left[{\hat y}_{\rm I}\right]\right)^2}{{\mathbb E}\left[{\hat y}_{\rm I}^2\right]} {\hat N}_{\rm I}^{\rm ad}\left(\omega\right)$; otherwise, ${\hat m}_{\rm I}^*\left(\omega,p_{\rm I}\right)=0$;
\item For the slots generated by the non-subscribers, the operator can set ${p}_{\rm II}>0$. If ${\hat N}_{\rm II}^{\rm ad}\left(\omega\right)>0$, the number of these slots purchased by each advertiser is ${\hat m}_{\rm II}^*\left(\omega,p_{\rm II}\right)=\!\frac{\max\left\{B-p_{\rm II},0\right\}}{2A} \frac{\left({\mathbb E}\left[{\hat y}_{\rm II}\right]\right)^2}{{\mathbb E}\left[{\hat y}_{\rm II}^2\right]} \!{\hat N}_{\rm II}^{\rm ad}\!\left(\omega\right)$; otherwise, ${\hat m}_{\rm II}^*\left(\omega,p_{\rm II}\right)\!=0$.
\end{itemize}

The operator's problem with differentiation is given by:
\begin{align}
& \!\!\!\!\!\max_{\omega\ge0,p_{\rm I},p_{\rm II}>0} \!{\hat R}^{\rm data}\!\left(\omega\right)\!+\!K {\hat m}_{\rm I}^*\!\left(\omega,p_{\rm I}\right) p_{\rm I}\!+\!K {\hat m}_{\rm II}^*\!\left(\omega,p_{\rm II}\right) {p}_{\rm II}\label{equ:diff:obj}\\
& {~~}{\rm s.t.}{~~}{~~}{~~} {\hat D}\left(\omega\right) \le C,\label{equ:diff:capacity}\\
& {~~}{~~}{~~}{~~}{~~}{~~} K {\hat m}_{\rm I}^*\left(\omega,p_{\rm I}\right) \le {\mathbb E}\left[{\hat y}_{\rm I}\right] {\hat N}_{\rm I}^{\rm ad}\left(\omega\right),\label{equ:diff:adc1}\\
& {~~}{~~}{~~}{~~}{~~}{~~} K {\hat m}_{\rm II}^*\left(\omega,p_{\rm II}\right) \le {\mathbb E}\left[{\hat y}_{\rm II}\right] {\hat N}_{\rm II}^{\rm ad}\left(\omega\right).\label{equ:diff:adc2}
\end{align}
Constraint (\ref{equ:diff:adc1}) means that the total number of sold ad slots that correspond to the subscribers should not exceed the number of ad slots generated by the subscribers. Constraint (\ref{equ:diff:adc2}) can be explained similarly for the non-subscribers. In fact, only when $\omega$ satisfies Case $\hat C$ in Proposition \ref{proposition:unaware:user}, both the subscribers and non-subscribers watch ads (i.e., ${\hat N}_{\rm I}^{\rm ad}\left(\omega\right),{\hat N}_{\rm II}^{\rm ad}\left(\omega\right)>0$), and problem (\ref{equ:diff:obj})-(\ref{equ:diff:adc2}) is different from problem (\ref{equ:unaware:obj})-(\ref{equ:unaware:constraint}) (i.e., the problem without differentiation). In the remaining cases, problem (\ref{equ:diff:obj})-(\ref{equ:diff:adc2}) reduces to problem (\ref{equ:unaware:obj})-(\ref{equ:unaware:constraint}). 

We define $\Pi^{\rm SUR}\triangleq {\hat R}^{\rm total}\left({\hat \omega}^*,{\hat p}^*\left({\hat \omega}^*\right)\right)$, which is the optimal objective value of problem (\ref{equ:unaware:obj})-(\ref{equ:unaware:constraint}). Let $\Pi^{\rm SURD}$ denote the optimal objective value of problem (\ref{equ:diff:obj})-(\ref{equ:diff:adc2}), i.e., $\Pi^{\rm SURD}$ is the operator's optimal total revenue under the SUR scheme with differentiation. We compare $\Pi^{\rm SUR}$ and $\Pi^{\rm SURD}$ in the following theorem.

\begin{theorem}\label{theorem:differentiation}
We always have $\Pi^{\rm SURD}\ge\Pi^{\rm SUR}$.
\end{theorem}
Hence, differentiation does not decrease the operator's optimal total revenue (given general $u\left(z\right)$ and $g\left(\theta\right)$). In general, it is easy to show that allowing a seller to sell items at different prices does not decrease its revenue. However, the differentiation here affects the ad display rule as well as the pricing, so it is non-trivial to prove Theorem \ref{theorem:differentiation}. For example, one conjecture is that given any $\left(\omega,p\right)$ which is feasible to (\ref{equ:unaware:obj})-(\ref{equ:unaware:constraint}), the operator can choose the same $\omega$ and set $p_{\rm I}=p_{\rm II}=p$ in (\ref{equ:diff:obj})-(\ref{equ:diff:adc2}) to ensure that the value of objective (\ref{equ:diff:obj}) is no smaller than that of (\ref{equ:unaware:obj}). In fact, the conjecture does not hold, because $\left(\omega,p_{\rm I},p_{\rm II}\right)$ \emph{may be infeasible for (\ref{equ:diff:obj})-(\ref{equ:diff:adc2})}.


Intuitively, if the optimal unit data reward satisfies Case $\hat C$ and the distributions of ${\hat y}_{\rm I}$ and ${\hat y}_{\rm II}$ are significantly different, the gap between $\Pi^{\rm SURD}$ and $\Pi^{\rm SUR}$ will be large. In the next section, we will show this gap numerically.

\begin{figure*}[t]
  \centering
  \subfigure[Logarithmic Utility.]{
    \includegraphics[scale=0.25]{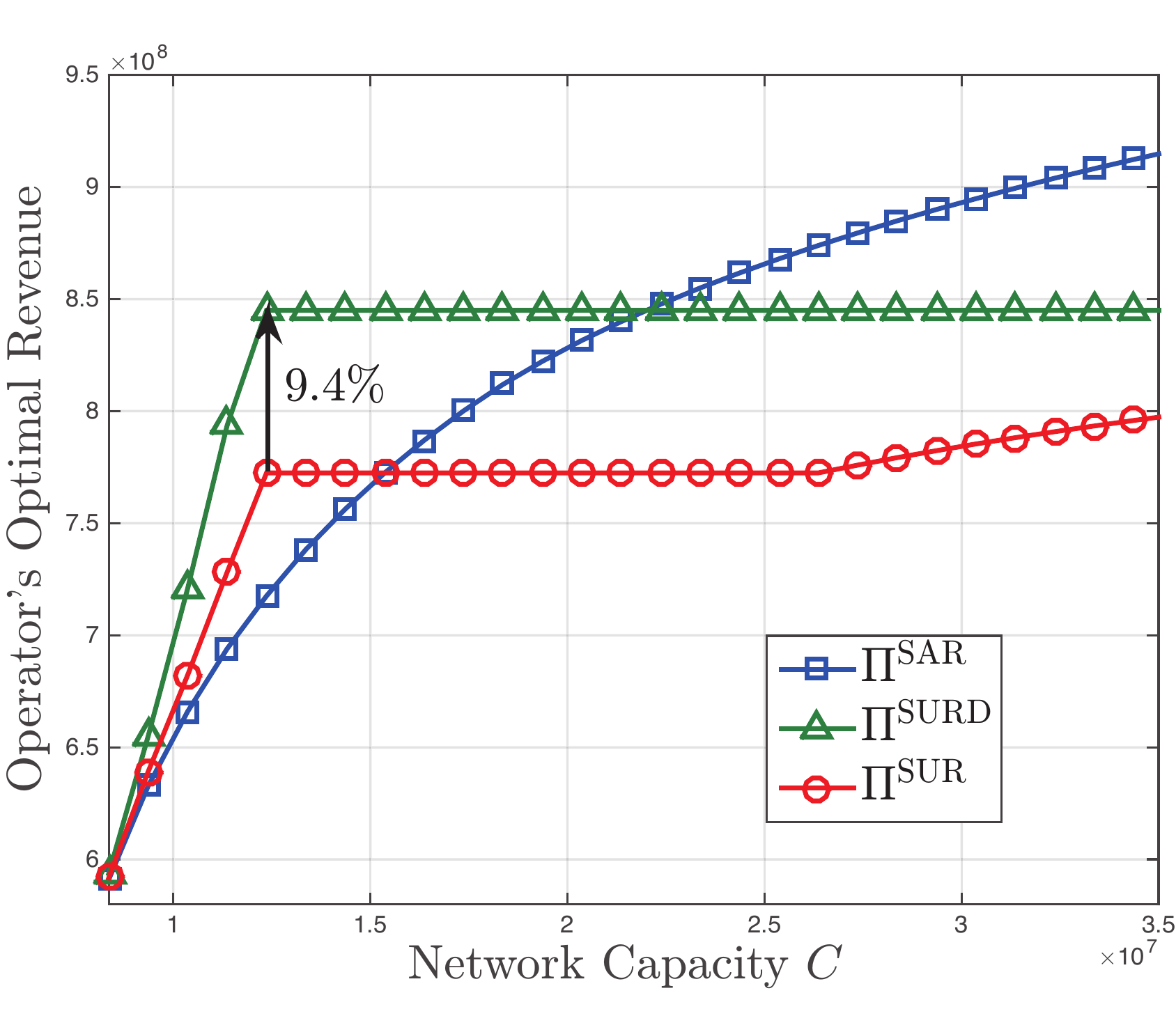}\label{fig:insimu:a}}
  \subfigure[Generalized $\alpha$-Fair Utility.]{
    \includegraphics[scale=0.25]{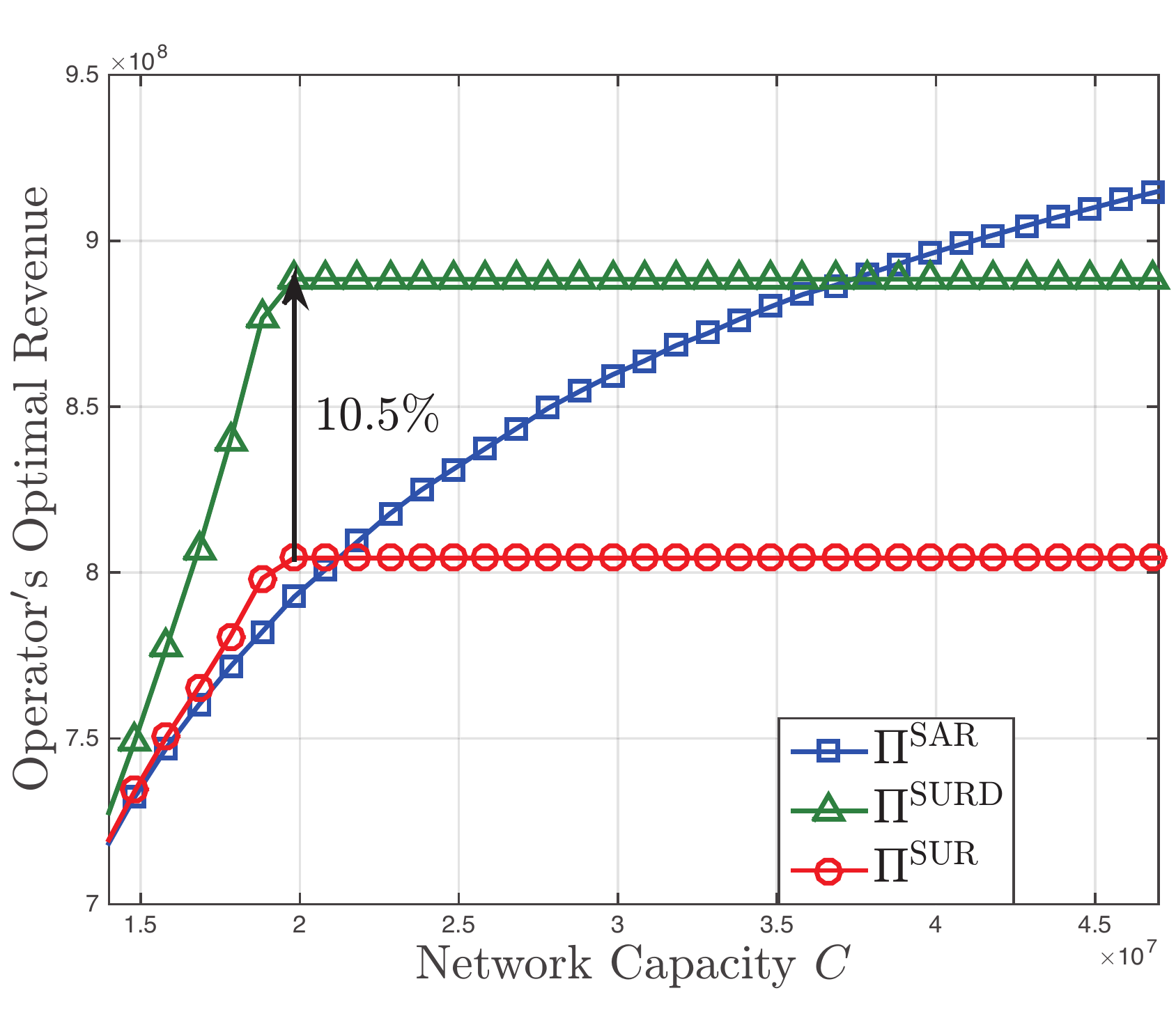}\label{fig:newsimu:b}}
    \subfigure[Exponential Utility (Large $A$).]{
    \includegraphics[scale=0.25]{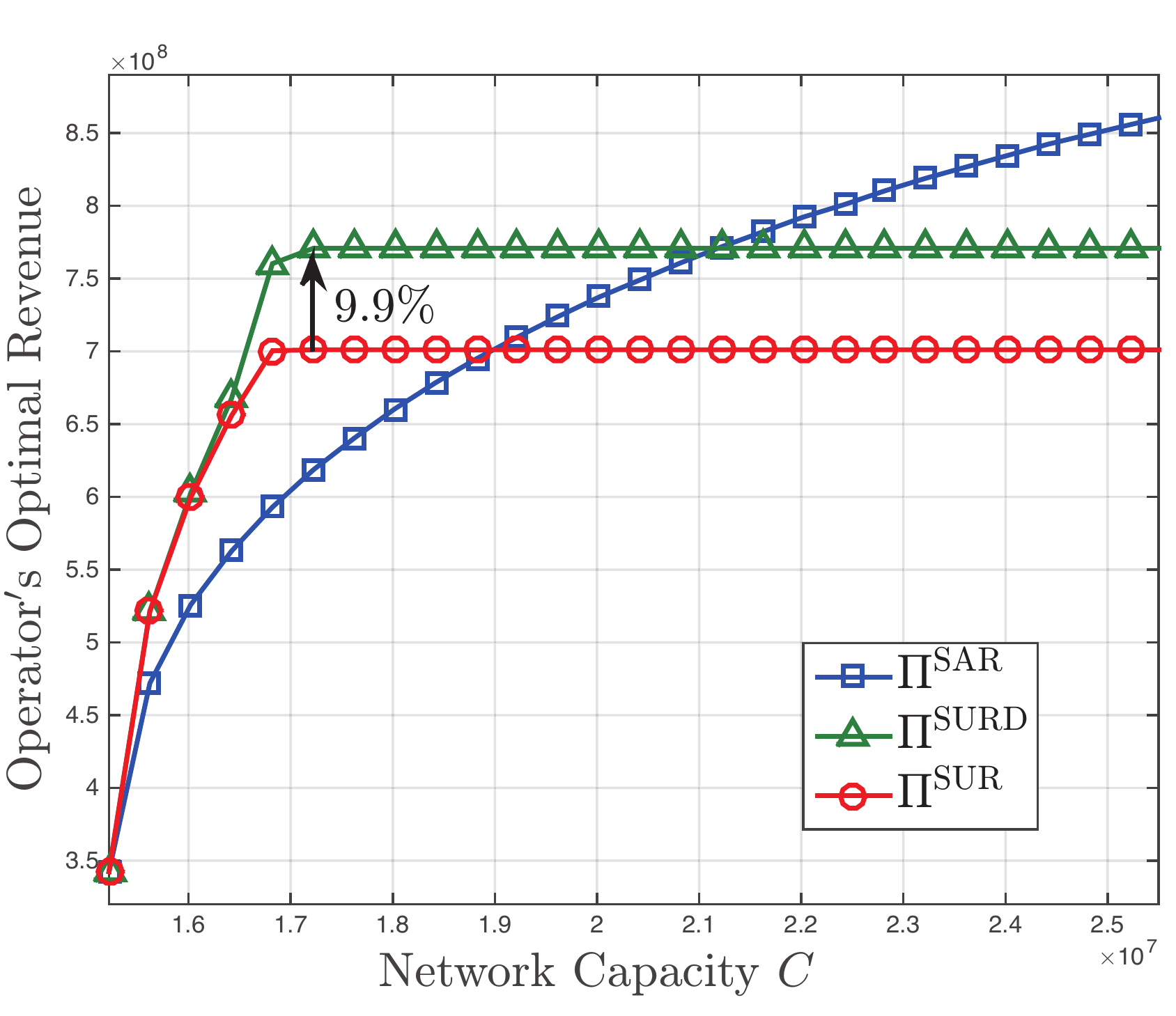}\label{fig:newsimu:c}}
  \subfigure[Exponential Utility (Small $A$).]{
    \includegraphics[scale=0.25]{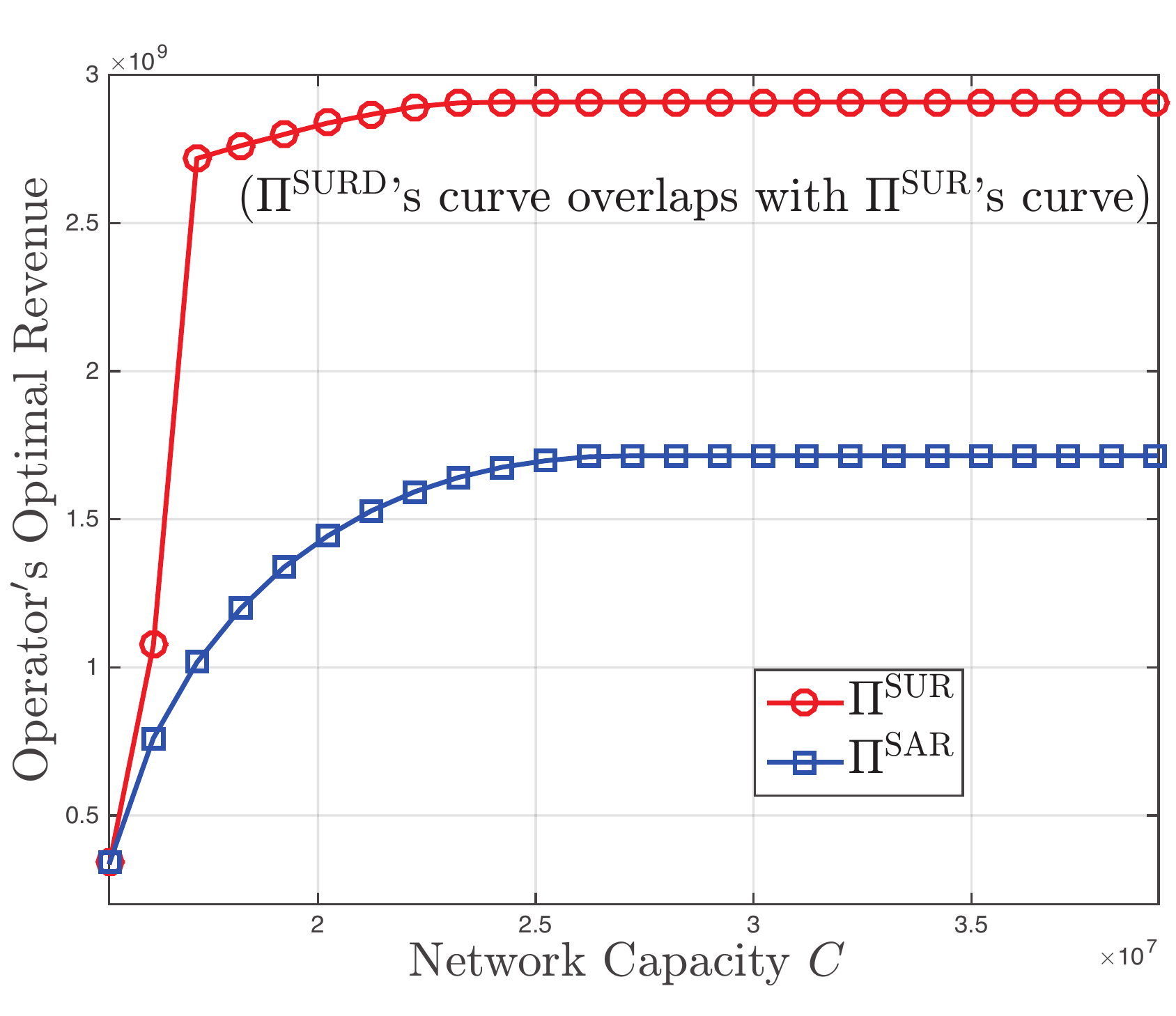}\label{fig:newsimu:d}}
  \caption{$\Pi^{\rm SAR}$, $\Pi^{\rm SUR}$, and $\Pi^{\rm SURD}$ Under Different Network Capacity (Uniformly Distributed $\theta$).}
  \label{fig:insimu:uni}
\end{figure*}




\section{Comparison Between Rewarding Schemes}\label{sec:comparison}
We define $\Pi^{\rm SAR}\triangleq{R}^{\rm total}\left(\omega^*,p^*\left(\omega^*\right)\right)$, which is the operator's optimal total revenue under the SAR scheme. In this section, we compare $\Pi^{\rm SAR}$, $\Pi^{\rm SUR}$, and $\Pi^{\rm SURD}$. 
{Since the comparison is challenging under a general user type distribution and a general utility function, we focus on specific user type distributions and utility functions. In Sections \ref{subsec:uni} and \ref{subsec:trun}, we consider uniformly distributed user types and truncated normally distributed user types, respectively.}

\subsection{{Uniformly Distributed User Types}}\label{subsec:uni}
{In this section, we assume that each user's type $\theta$ follows a uniform distribution. We will consider logarithmic utility, generalized $\alpha$-fair utility, and exponential utility.}

\subsubsection{Logarithmic Utility Function}

We assume that $u\left(z\right)=\ln\left(1+z\right)$. Theorem \ref{theorem:comparison} characterizes the analytical comparison between different schemes as $C\rightarrow \infty$.

\begin{theorem}\label{theorem:comparison}
When $\theta\sim{\cal U}\left[0,\theta_{\max}\right]$ and $u\left(z\right)=\ln\left(1+z\right)$, if network capacity $C\rightarrow \infty$, then $\Pi^{\rm SAR}>\Pi^{\rm SURD}\ge\Pi^{\rm SUR}$.
\end{theorem}

Theorem \ref{theorem:comparison} implies that if the operator has sufficiently large network capacity, it should only reward the subscribers for watching ads. Intuitively, this allows the operator to motivate all users to subscribe and watch ads via high data rewards. It maximizes the operator's revenue from both the data market and the ad market.

Under a finite network capacity $C$, none of $\Pi^{\rm SAR}$, $\Pi^{\rm SUR}$, or $\Pi^{\rm SURD}$ has a closed-form expression, and their analytical comparison is challenging. {Next, we compare them numerically. In the numerical experiment, we choose $N=10^7$, $F=30$, $Q=0.8$, $\theta\sim{\cal U}\left[0,155\right]$, $\Phi=0.3$, $K=23$, $A=0.6$, and $B=5$. Here, we consider an area with 10 million users. In Appendix \ref{appendix:morepara}, we consider different parameter settings (e.g., different values of $N$), and the key observations summarized in this section still hold under those settings.} 

In Fig. \ref{fig:insimu:a}, we plot $\Pi^{\rm SAR}$, $\Pi^{\rm SUR}$, and $\Pi^{\rm SURD}$ against $C$. We can see that only $\Pi^{\rm SAR}$ strictly increases with $C$. As shown in Proposition \ref{proposition:uniform:useup}, when each user has a logarithmic utility and a uniformly distributed type, the operator always uses up the capacity for rewards under the SAR scheme. Hence, the operator can always benefit from $C$'s increase in this situation. 

\begin{figure*}[t]
  \centering
  \subfigure[Logarithmic Utility and Uniformly Distributed $\theta$.]{
    \includegraphics[scale=0.35]{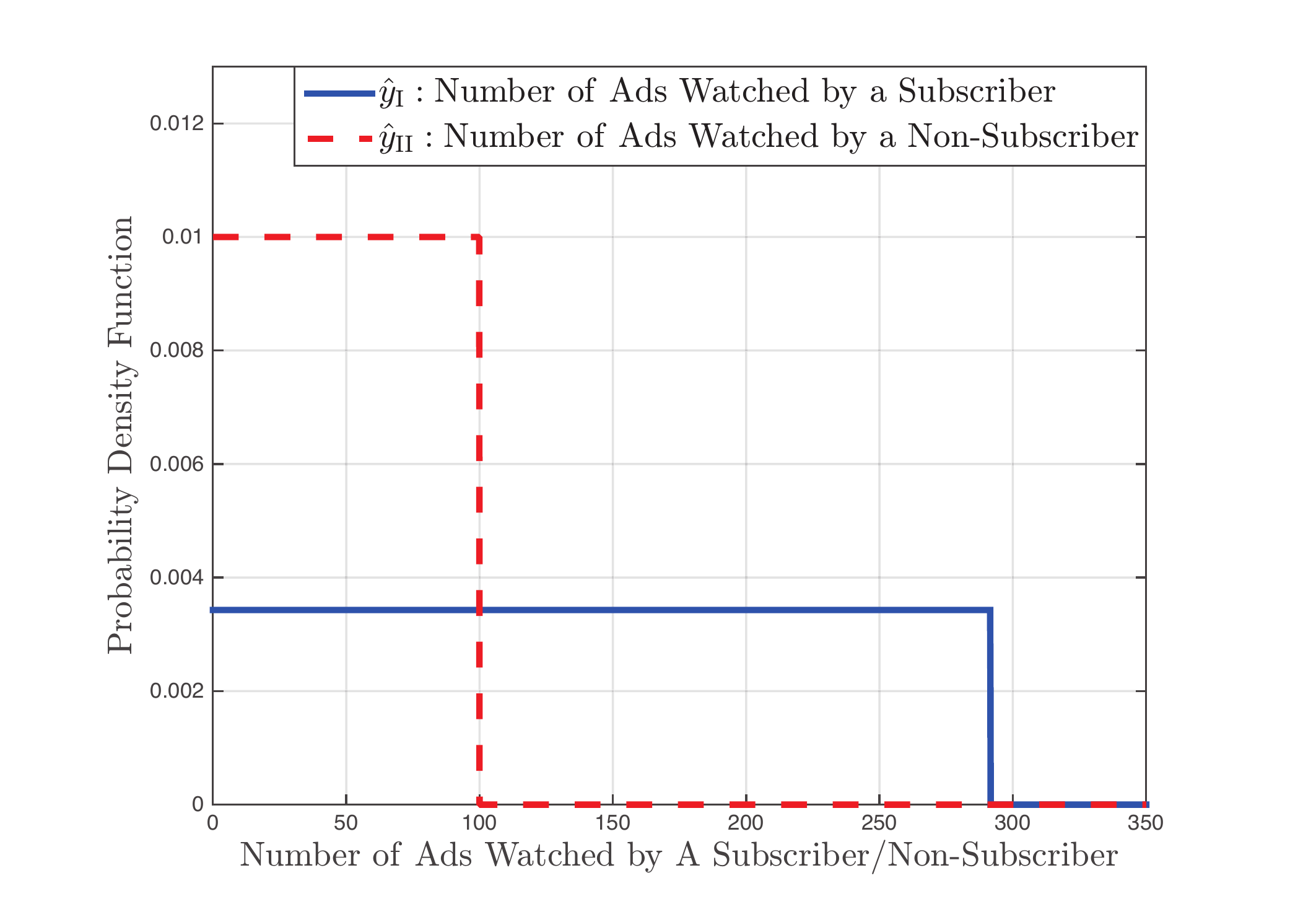}\label{fig:differentiation:uni}}
  \subfigure[\!Exponential Utility and Truncated Normally \!Distributed \!$\theta$.]{
    \includegraphics[scale=0.35]{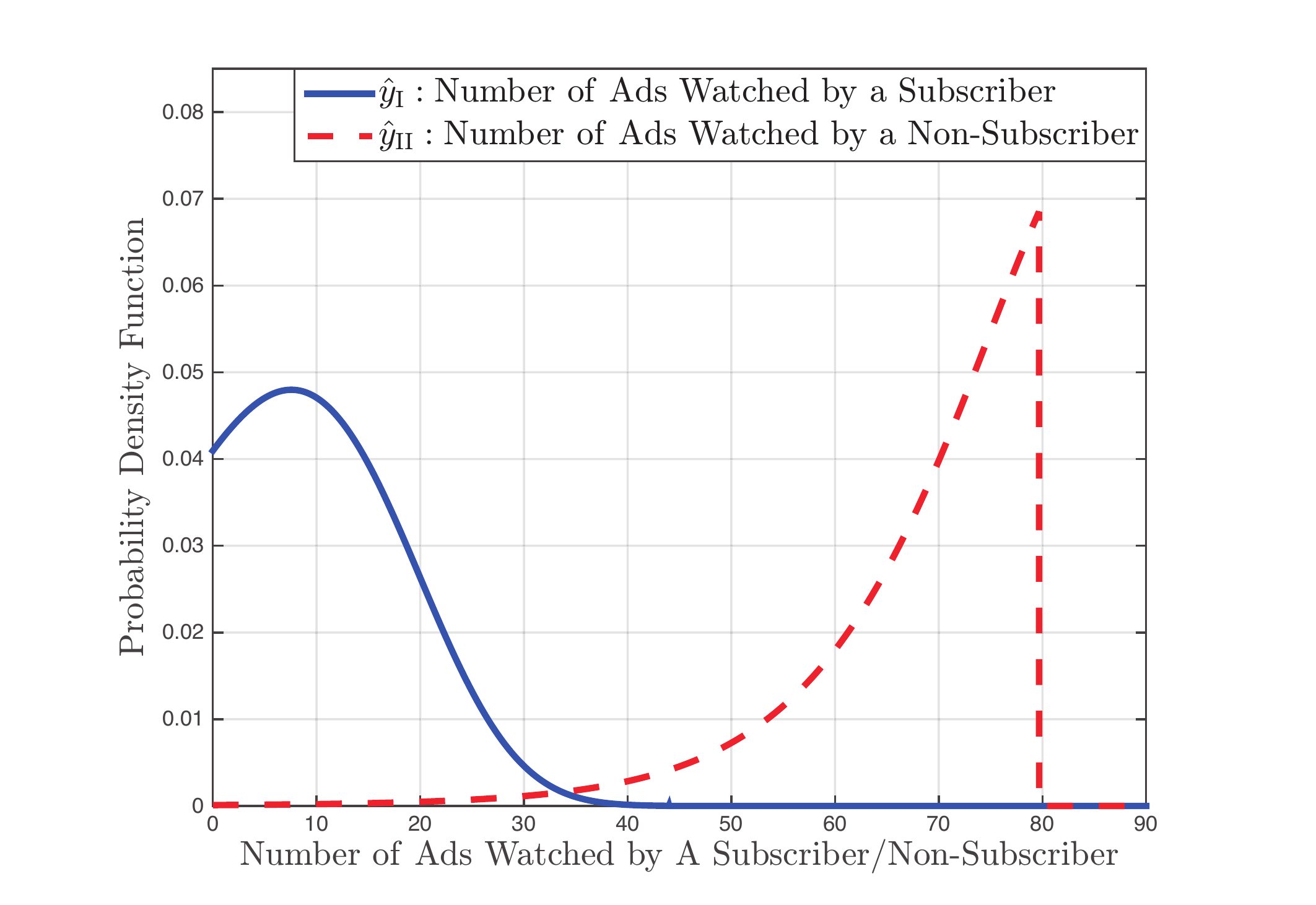}\label{fig:differentiation:tru}}
  \caption{{{Probability Distribution Function of ${\hat y}_{\rm I}$ and ${\hat y}_{\rm II}$.}}}
  \vspace{-0.3cm}
\end{figure*}

First, we compare $\Pi^{\rm SAR}$ and $\Pi^{\rm SUR}$. When $C$ is close to $D\left(0\right)$, {$\Pi^{\rm SAR}$ and $\Pi^{\rm SUR}$ are equal}. In this situation, the operator can only choose a very small unit reward $\omega$. As shown in Case $B$ in Proposition \ref{proposition:stageII:user} and Case $\hat B$ in Proposition \ref{proposition:unaware:user}, the users' optimal decisions under the two schemes are the same, which leads to the same operator's revenue. When $C$ is from $0.84\times 10^7$ to $1.54\times 10^7$, {$\Pi^{\rm SAR}$ is smaller than $\Pi^{\rm SUR}$}. This is because the SUR scheme can motivate two segments of users to watch ads (by setting $\omega\in\left(\frac{\Phi u\left(Q\right)}{F u'\left(0\right)},\frac{\Phi Q}{F}\right)$, as shown in Case $\hat C$ in Proposition \ref{proposition:unaware:user}), which generates a higher ad revenue than the SAR scheme. When $C$ is greater than $1.54\times 10^7$, {$\Pi^{\rm SAR}$ is greater than $\Pi^{\rm SUR}$}. The operator will fully utilize the large network capacity under the SAR scheme, and set a large $\omega$ to motivate more users to both subscribe and watch ads. This is consistent with Theorem \ref{theorem:comparison} (i.e., if $C\rightarrow \infty$, then $\Pi^{\rm SAR}>\Pi^{\rm SUR}$). We summarize the results in Observation \ref{observation:log} (the comparison between $\Pi^{\rm SAR}$ and $\Pi^{\rm SURD}$ is similar to the comparison between $\Pi^{\rm SAR}$ and $\Pi^{\rm SUR}$).
\begin{observation}\label{observation:log}
When $u\left(z\right)=\ln\left(1+z\right)$, if $C$ is small, the SUR scheme achieves a higher operator's revenue; otherwise, the SAR scheme achieves a higher operator's revenue.
\end{observation}
Second, we compare $\Pi^{\rm SUR}$ and $\Pi^{\rm SURD}$. {When $C=1.24\times10^7$, Fig. \ref{fig:insimu:a} shows that the ad slots' differentiation can improve the operator's revenue under the SUR scheme by $9.4\%$. This is because the value of ${\hat \omega}^*$ under the SUR scheme satisfies Case $\hat C$ in Proposition \ref{proposition:unaware:user}, which implies that both subscribers and non-subscribers watch ads. Moreover, the subscribers and non-subscribers have quite different ad watching behaviors. In Fig. \ref{fig:differentiation:uni}, we illustrate the distributions of ${\hat y}_{\rm I}$ (i.e., the number of ads watched by a subscriber) and ${\hat y}_{\rm II}$ (i.e., the number of ads watched by a non-subscriber) when $C=1.24\times10^7$ and the operator uses the SUR scheme. We can see that both ${\hat y}_{\rm I}$ and ${\hat y}_{\rm II}$ follow uniform distributions, but their mean values are significantly different.}

\subsubsection{{Generalized $\alpha$-Fair Utility Function}}
{We assume that $u\left(z\right)=\frac{\left(z+\mu\right)^{1-\alpha}}{1-\alpha}-\frac{\mu^{1-\alpha}}{1-\alpha}$. We choose $\alpha=0.8$ and $\mu=0.8$, and the other settings are the same as those in Fig. \ref{fig:insimu:a}. In Fig. \ref{fig:newsimu:b}, we plot $\Pi^{\rm SAR}$, $\Pi^{\rm SUR}$, and $\Pi^{\rm SURD}$ against $C$. We can see that the comparison among the operator's optimal revenues under different schemes is similar to that in Fig. \ref{fig:insimu:a}. We summarize the key results about the comparison between $\Pi^{\rm SAR}$ and $\Pi^{\rm SUR}$ in the following observation.
\begin{observation}\label{observation:alpha}
When $u\left(z\right)=\frac{\left(z+\mu\right)^{1-\alpha}}{1-\alpha}-\frac{\mu^{1-\alpha}}{1-\alpha}$, if $C$ is small, the SUR scheme achieves a higher operator's revenue; otherwise, the SAR scheme achieves a higher operator's revenue.
\end{observation}
}

\subsubsection{Exponential Utility Function}

{We assume that $u\left(z\right)=1-e^{-\gamma z}$, and choose $\gamma=0.7$, $N=10^7$, $F=45$, $Q=2$, $\theta\sim{\cal U}\left[0,250\right]$, $\Phi=0.3$, $K=23$, and $B=5$.} In Fig. \ref{fig:newsimu:c} and Fig. \ref{fig:newsimu:d}, we show the comparison between $\Pi^{\rm SAR}$, $\Pi^{\rm SUR}$, and $\Pi^{\rm SURD}$ under different degrees of the wear-out effect.

\begin{figure*}[t]
  \centering
  \subfigure[Logarithmic Utility.]{
    \includegraphics[scale=0.25]{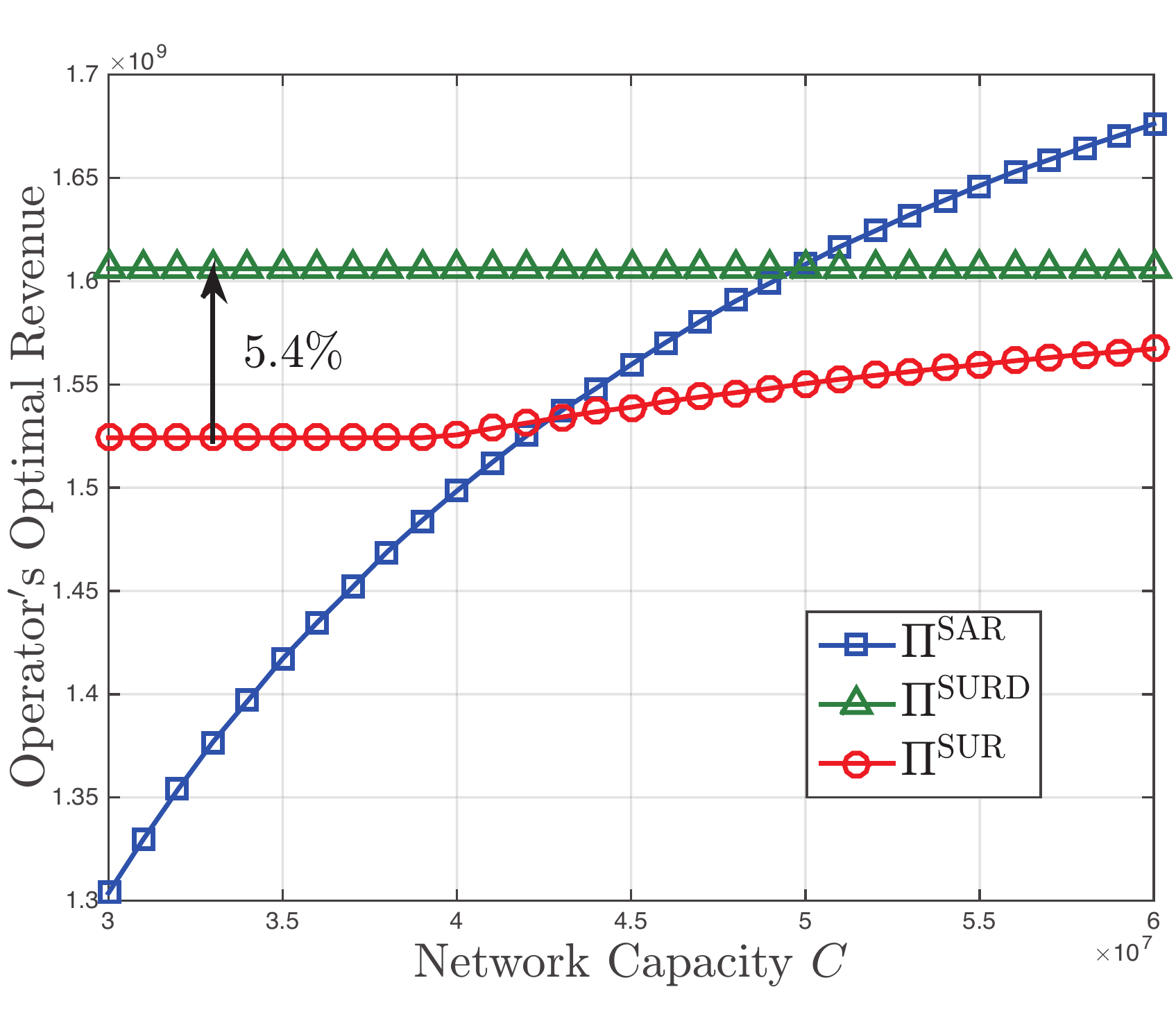}\label{fig:trun:log1}}
  \subfigure[Generalized $\alpha$-Fair Utility.]{
    \includegraphics[scale=0.25]{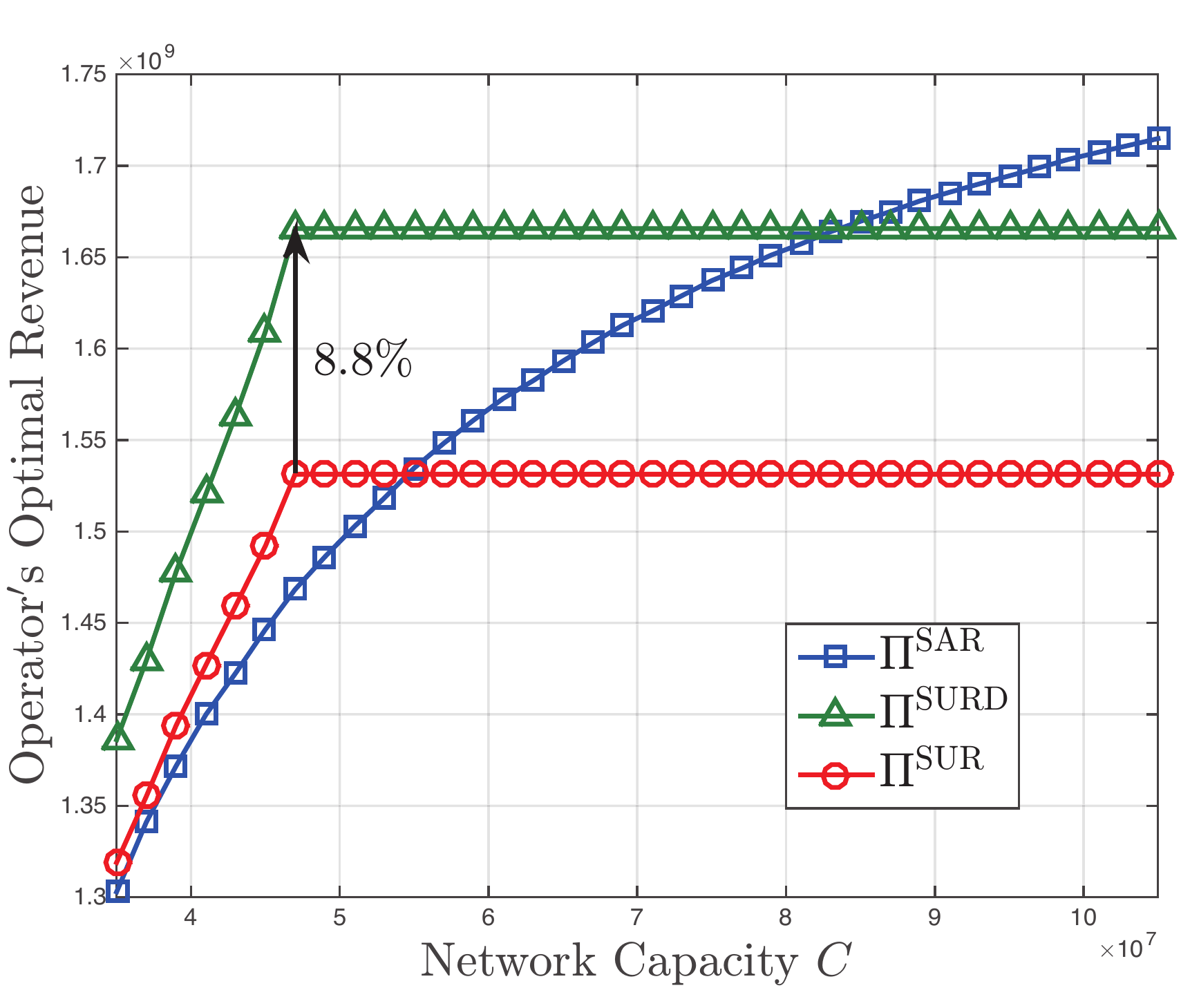}\label{fig:trun:alpha2}}
    \subfigure[Exponential Utility (Large $A$).]{
    \includegraphics[scale=0.25]{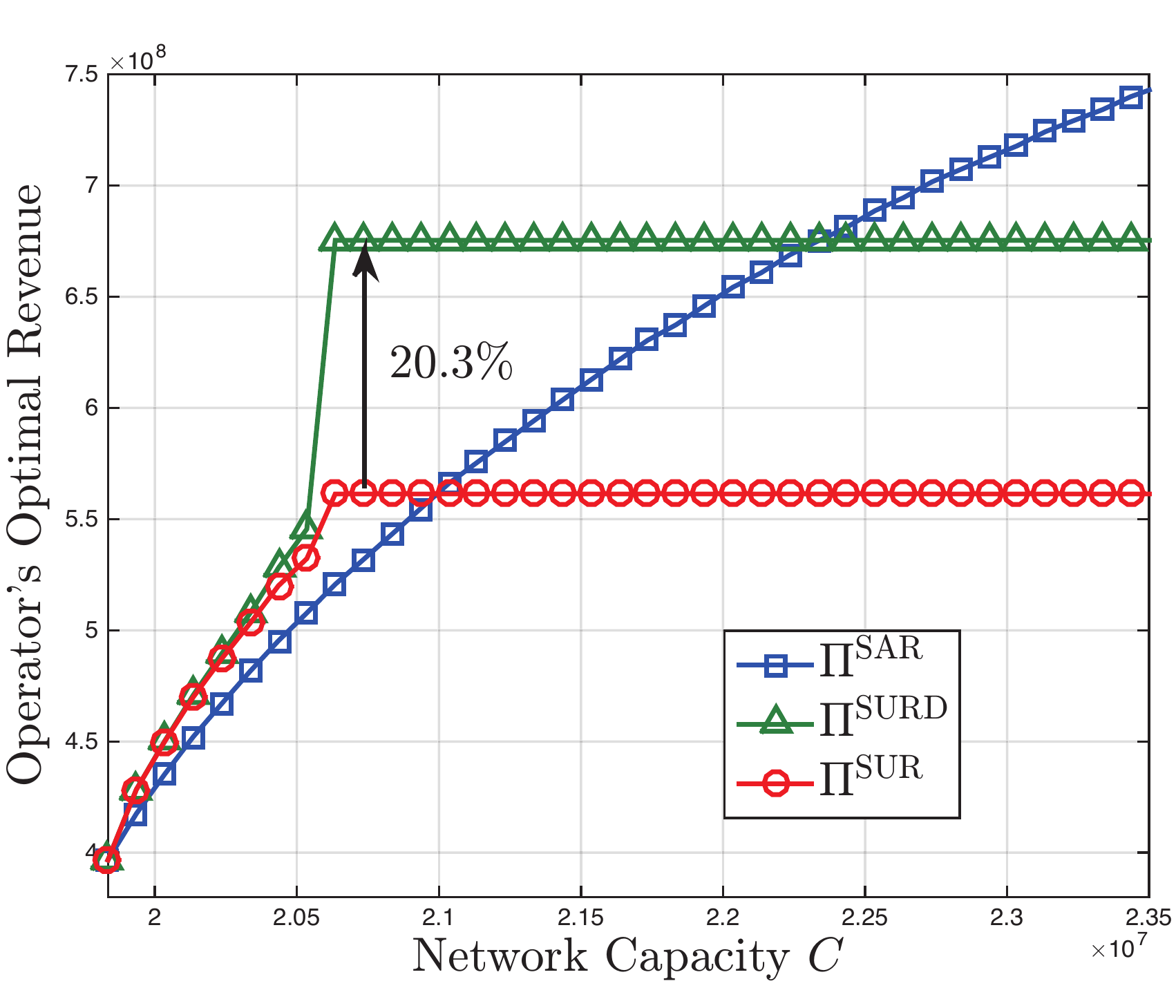}\label{fig:trun:exp3}}
  \subfigure[Exponential Utility (Small $A$).]{
    \includegraphics[scale=0.249]{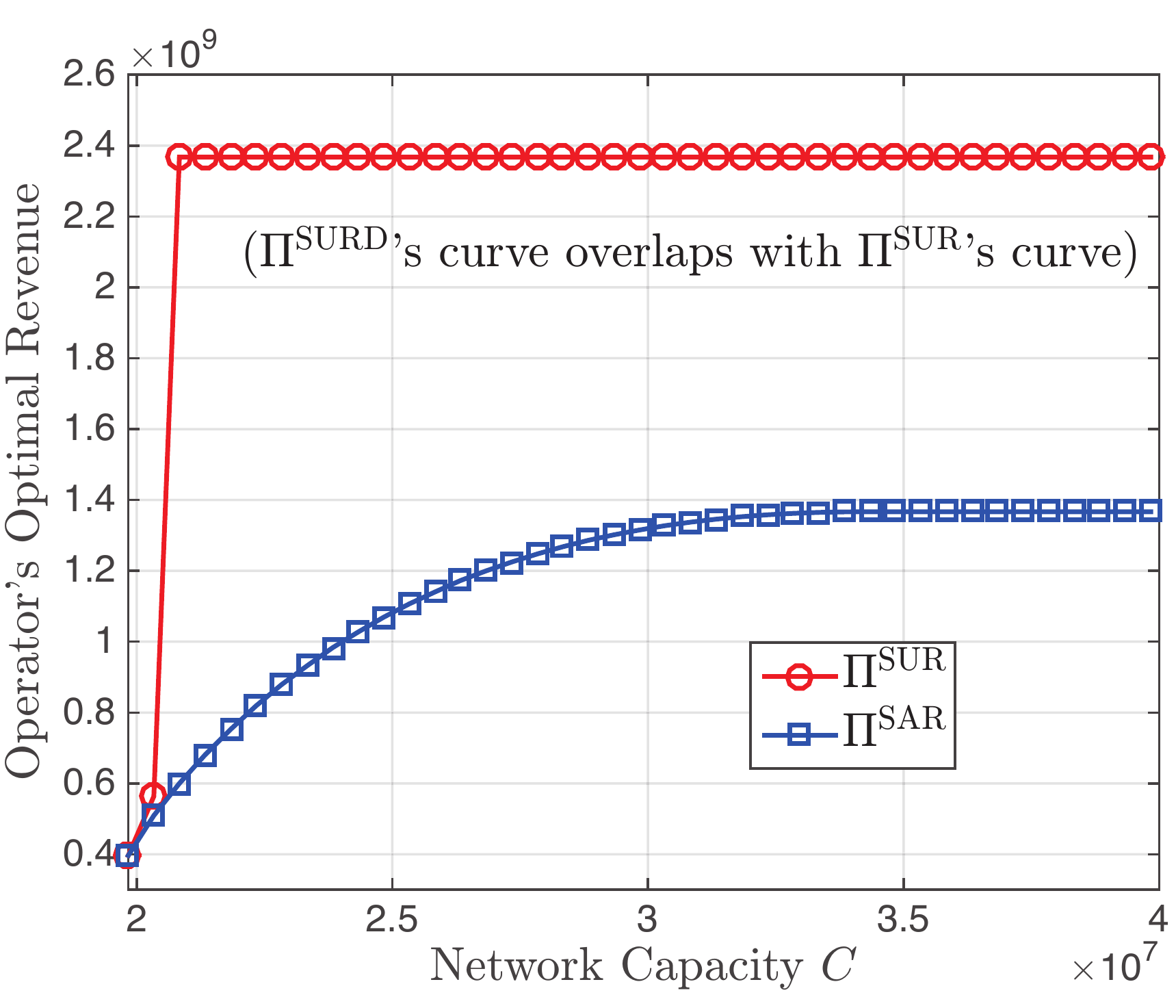}\label{fig:trun:exp4}}
  \caption{$\Pi^{\rm SAR}$, $\Pi^{\rm SUR}$, and $\Pi^{\rm SURD}$ Under Different Network Capacity (Truncated Normally Distributed $\theta$).}
  \label{fig:insimu:tru}
\end{figure*}

In Fig. \ref{fig:newsimu:c}, we consider a large wear-out effect ($A=0.9$). The comparison between $\Pi^{\rm SAR}$ and $\Pi^{\rm SUR}$ (or $\Pi^{\rm SURD}$) is similar to those in Fig. \ref{fig:insimu:a} and Fig. \ref{fig:newsimu:b}. The SAR scheme achieves a higher revenue than the SUR scheme when $C$ is large. Comparing $\Pi^{\rm SUR}$ and $\Pi^{\rm SURD}$ in Fig. \ref{fig:newsimu:c}, we observe that differentiation improves the operator's revenue under the SUR scheme by at most $9.9\%$.  


In Fig. \ref{fig:newsimu:d}, we consider a small wear-out effect ($A=0.2$), and have three observations. First, $\Pi^{\rm SAR}$ may not change with $C$, which is different from the logarithmic utility situation shown in Fig. \ref{fig:insimu:a}. When each user has an exponential utility, the operator may not benefit from the increase of $C$, since it may not use up the capacity for the rewards (as discussed in Section \ref{subsec:aware:operator}). 
Second, $\Pi^{\rm SAR}$ is always no greater than $\Pi^{\rm SUR}$ (even under a large $C$), which is different from the logarithmic utility situation {and the generalized $\alpha$-fair utility situation}. Under the SAR scheme, each user has to pay the subscription fee $F>0$ before receiving the data rewards. The exponential utility function is upper bounded (i.e., $u\left(z\right)=1-e^{-\gamma z}\le1$), and hence the users with $\theta<F$ will never subscribe and watch ads under the SAR scheme, regardless of the unit data reward $\omega$. When $A$ is small, the advertisers are willing to buy more slots, and having more users watching ads significantly increases the operator's revenue. Therefore, the SUR scheme, which can motivate the users with $\theta<F$ to watch ads, achieves a higher revenue than the SAR scheme. 
Third, the $\Pi^{\rm SURD}$ curve overlaps the $\Pi^{\rm SUR}$ curve, because the operator chooses a large $\omega$ to incentivize the users to watch ads under a small $A$. In this situation, all the ad slots are generated by non-subscribers under the SUR scheme (see Case $\hat D$ of Proposition \ref{proposition:unaware:user}), and the differentiation cannot improve the operator's revenue. 


We summarize the key observations below.
\begin{observation}\label{observation:exp}
When $u\left(z\right)=1-e^{-\gamma z}$, (i) under a large $A$, the SUR scheme achieves a higher operator revenue than the SAR scheme if and only if $C$ is below a finite threshold; (ii) under a small $A$, the SUR scheme always achieves a higher operator revenue than the SAR scheme.
\end{observation}


\subsection{{Truncated Normally Distributed User Types}}\label{subsec:trun}
{We next assume that each user's type $\theta$ follows a truncated normal distribution. We show that most observations under the uniformly distributed user types still hold.}

\subsubsection{{Logarithmic Utility Function}}
{We assume that $u\left(z\right)=\ln\left(1+z\right)$, and obtain the distribution of $\theta$ by truncating the normal distribution ${\cal N}\left(75,40\right)$ to interval $\left[0,150\right]$. We choose $N=10^7$, $F=40$, $Q=2$, $\Phi=0.03$, $K=8$, $A=0.5$, and $B=10$. In Fig. \ref{fig:trun:log1}, we plot $\Pi^{\rm SAR}$, $\Pi^{\rm SUR}$, and $\Pi^{\rm SURD}$ against $C$. We can see that the SUR scheme outperforms the SAR scheme if and only if $C$ is below a threshold. This is consistent with Observation \ref{observation:log}.}

\subsubsection{{Generalized $\alpha$-Fair Utility Function}}
{We next assume that $u\left(z\right)=\frac{\left(z+\mu\right)^{1-\alpha}}{1-\alpha}-\frac{\mu^{1-\alpha}}{1-\alpha}$, where $\alpha=0.8$ and $\mu=0.8$. The other settings are the same as those in Fig. \ref{fig:trun:log1}. We plot the operator's optimal revenues under different schemes in Fig. \ref{fig:trun:alpha2}. The influence of $C$ on the comparison is consistent with Observation \ref{observation:alpha}.}

\subsubsection{{Exponential Utility Function}}
{We next assume that $u\left(z\right)=1-e^{-\gamma z}$, and obtain the distribution of $\theta$ by truncating the normal distribution ${\cal N}\left(125,30\right)$ to interval $\left[0,250\right]$. We choose $\gamma=0.7$, $N=10^7$, $F=40$, $Q=2$, $\Phi=0.5$, $K=16$, and $B=5$. Fig. \ref{fig:trun:exp3} and Fig. \ref{fig:trun:exp4} show the comparison between $\Pi^{\rm SAR}$, $\Pi^{\rm SUR}$, and $\Pi^{\rm SURD}$ under $A=0.9$ and $A=0.2$, respectively.

Fig. \ref{fig:trun:exp3} shows that if the wear-out effect is large, the SUR scheme outperforms the SAR scheme under a small $C$. Fig. \ref{fig:trun:exp4} shows that if the wear-out effect is small, the SUR scheme always outperforms the SAR scheme. These results are consistent with Observation \ref{observation:exp}.

In Fig. \ref{fig:trun:exp3}, when $C=2.07\times 10^7$, the differentiation of the ad slots improves the operator's revenue under the SUR scheme by $20.3\%$. To explain this large improvement, we illustrate the distributions of ${\hat y}_{\rm I}$ and ${\hat y}_{\rm II}$ under $C=2.07\times 10^7$ and the SUR scheme in Fig. \ref{fig:differentiation:tru}. We can observe that the difference between the two distributions is greater than that in Fig. \ref{fig:differentiation:uni} (where each user has a logarithmic utility function and a uniformly distributed type). For example, the value of $\frac{{\mathbb E}\left[{\hat y}_{\rm II}\right]}{{\mathbb E}\left[{\hat y}_{\rm I}\right]}$ in Fig. \ref{fig:differentiation:tru} is around $5.7$, and the value of $\frac{{\mathbb E}\left[{\hat y}_{\rm I}\right]}{{\mathbb E}\left[{\hat y}_{\rm II}\right]}$ in Fig. \ref{fig:differentiation:uni} is around $2.9$.{\footnote{{{Note that ${\mathbb E}\left[{\hat y}_{\rm I}\right]$ can be either larger or smaller than ${\mathbb E}\left[{\hat y}_{\rm II}\right]$, which depends on the parameter settings and the assumptions on the utility function and user type distribution.}}}} Intuitively, when the difference between the subscribers' and non-subscribers' ad watching behaviors is larger, the benefit of differentiation is more obvious. Therefore, the improvement of $\Pi^{\rm SURD}$ over $\Pi^{\rm SUR}$ in Fig. \ref{fig:trun:exp3} is greater than the improvement in Fig. \ref{fig:insimu:a} (which is $9.4\%$).
}


\section{Conclusion}\label{sec:conclusion}
Mobile data rewarding is an emerging approach to monetize mobile services. 
We modeled the data rewarding ecosystem and analyzed an operator's rewarding scheme. 
Our results reveal that: (i) increasing the unit data reward may decrease the number of ads watched by the users, and the operator may not use up its network capacity to reward the users; (ii) under the SUR scheme, the operator can improve its revenue by differentiating the ad slots generated by the subscribers and non-subscribers; (iii) the operator's optimal choice between the SAR and SUR schemes is sensitive to the user utility function, network capacity, and advertising's wear-out effect. 

In future work, we plan to first study the operator's joint optimization of the data plan and the data rewards. Under the SAR scheme, the operator can reduce the subscription fee to motivate more users to subscribe and watch ads. Under the SUR scheme, the operator may increase the subscription fee, which (i) extracts more revenue from the users with high $\theta$ and (ii) pushes more users with low $\theta$ to become non-subscribers and watch ads. Second, we are interested in relaxing the assumptions of a monopolistic operator and homogeneous advertisers. For example, when there are multiple operators, they will compete for users as well as advertisers, which may increase the unit data rewards and reduce the ad prices. 
{Third, we can study a general data rewarding scheme where the operator can set different unit data rewards for the subscribers and non-subscribers. The SAR and SUR schemes can be treated as two special cases of this general scheme.}





\bibliographystyle{IEEEtran}
\bibliography{jsacref}

\begin{thebibliography}{10}
\providecommand{\url}[1]{#1}
\csname url@samestyle\endcsname
\providecommand{\newblock}{\relax}
\providecommand{\bibinfo}[2]{#2}
\providecommand{\BIBentrySTDinterwordspacing}{\spaceskip=0pt\relax}
\providecommand{\BIBentryALTinterwordstretchfactor}{4}
\providecommand{\BIBentryALTinterwordspacing}{\spaceskip=\fontdimen2\font plus
\BIBentryALTinterwordstretchfactor\fontdimen3\font minus
  \fontdimen4\font\relax}
\providecommand{\BIBforeignlanguage}[2]{{%
\expandafter\ifx\csname l@#1\endcsname\relax
\typeout{** WARNING: IEEEtran.bst: No hyphenation pattern has been}%
\typeout{** loaded for the language `#1'. Using the pattern for}%
\typeout{** the default language instead.}%
\else
\language=\csname l@#1\endcsname
\fi
#2}}
\providecommand{\BIBdecl}{\relax}
\BIBdecl

\bibitem{yu2019business}
H.~Yu, E.~Wei, and R.~A. Berry, ``A business model analysis of mobile data
  rewards,'' in \emph{Proc. of IEEE INFOCOM}, Paris, France, 2019, pp.
  2098--2106.

\bibitem{Analytics2018}
S.~Analytics, ``Worldwide cellular user forecasts 2018-2023,'' Tech. Rep.,
  2018.

\bibitem{AnalyticsAD}
------, ``Can operator collaboration on sponsored data lead to success?'' Tech.
  Rep., 2018.

\bibitem{emarketer}
eMarketer,
  {https://www.emarketer.com/Article/Want-App-Users-Interact-with-Your-Ads-Reward-Them/1010966},
  accessed on Nov. 27, 2019.

\bibitem{Aquto}
Aquto, {http://www.aquto.com/}, accessed on Nov. 27, 2019.

\bibitem{Unlockdjsac}
N.~Summers,
  {https://www.engadget.com/2016/06/09/tesco-mobile-unlockd-lock-screen-ads/},
  accessed on Nov. 27, 2019.

\bibitem{DOCOMO}
DOCOMO,
  \url{https://www.nttdocomo.co.jp/english/info/media_center/pr/2017/1225_00.html},
  accessed on Nov. 27, 2019.

\bibitem{Optus}
R.~Johnston,
  https://www.gizmodo.com.au/2016/11/optus-wants-you-to-watch-ads-in-return-for-data/,
  accessed on Nov. 27, 2019.

\bibitem{ATTadnew}
AT\&T, \url{https://about.att.com/story/2018/att_appnexus.html}, accessed on
  Nov. 27, 2019.

\bibitem{qqensure}
S.~Dewing, {https://tech.co/news/adwallet-future-advertising-app-2017-08},
  accessed on Nov. 27, 2019.

\bibitem{unlocktimes}
J.~Naftulin,
  https://www.businessinsider.com/dscout-research-people-touch-cell-phones-2617-times-a-day-2016-7,
  accessed on Nov. 27, 2019.

\bibitem{CPMReport}
AdStage, ``Q3 2018 paid search and paid social benchmark report,'' Tech. Rep.,
  2018.

\bibitem{ultramobile}
Ultra{~}Mobile, {https://www.ultramobile.com/monthly-plans/}, accessed on Nov.
  27, 2019.

\bibitem{unlimited}
T.~Haselton,
  {https://www.cnbc.com/2018/07/13/unlimited-data-plan-caps-verizon-att-tmobile-sprint.html},
  accessed on Nov. 27, 2019.

\bibitem{pechmann1988advertising}
C.~Pechmann and D.~W. Stewart, ``Advertising repetition: A critical review of
  wearin and wearout,'' \emph{Current issues and research in advertising},
  vol.~11, no. 1-2, pp. 285--329, 1988.

\bibitem{kirmani1997advertising}
A.~Kirmani, ``Advertising repetition as a signal of quality: If it's advertised
  so much, something must be wrong,'' \emph{Journal of advertising}, vol.~26,
  no.~3, pp. 77--86, 1997.

\bibitem{joe2018sponsoring}
C.~Joe-Wong, S.~Sen, and S.~Ha, ``Sponsoring mobile data: Analyzing the impact
  on internet stakeholders,'' \emph{IEEE/ACM Transactions on Networking},
  vol.~26, no.~3, pp. 1179--1192, 2018.

\bibitem{riggins2002market}
F.~J. Riggins, ``Market segmentation and information development costs in a
  two-tiered fee-based and sponsorship-based web site,'' \emph{Journal of
  Management Information Systems}, vol.~19, no.~3, pp. 69--86, 2002.

\bibitem{yu2017public}
H.~Yu, M.~H. Cheung, L.~Gao, and J.~Huang, ``Public {Wi-Fi} monetization via
  advertising,'' \emph{IEEE/ACM Transactions on Networking}, vol.~25, no.~4,
  pp. 2110--2121, 2017.

\bibitem{guo2017economic}
H.~Guo, X.~Zhao, L.~Hao, and D.~Liu, ``Economic analysis of reward
  advertising,'' \emph{Working Paper}, 2017.

\bibitem{andrews2013economic}
M.~Andrews, U.~{\"O}zen, M.~I. Reiman, and Q.~Wang, ``Economic models of
  sponsored content in wireless networks with uncertain demand,'' in
  \emph{Proc. of IEEE INFOCOM Workshops}, Turin, Italy, 2013, pp. 345--350.

\bibitem{lotfi2017economics}
M.~H. Lotfi, K.~Sundaresan, S.~Sarkar, and M.~A. Khojastepour, ``Economics of
  quality sponsored data in non-neutral networks,'' \emph{IEEE/ACM Transactions
  on Networking}, vol.~25, no.~4, pp. 2068--2081, 2017.

\bibitem{zhang2015sponsored}
L.~Zhang, W.~Wu, and D.~Wang, ``Sponsored data plan: A two-class service model
  in wireless data networks,'' in \emph{Proc. of ACM SIGMETRICS}, Portland, OR,
  USA, 2015.

\bibitem{bangera2017advertising}
P.~Bangera, S.~Hasan, and S.~Gorinsky, ``An advertising revenue model for
  access {ISPs},'' in \emph{Proc. of IEEE ISCC}, Heraklion, Greece, 2017.

\bibitem{sen2017incentive}
S.~Sen, G.~Burtch, A.~Gupta, and R.~Rill, ``Incentive design for ad-sponsored
  content: Results from a randomized trial,'' in \emph{Proc. of IEEE INFOCOM
  Workshops}, Atlanta, GA, USA, 2017, pp. 826--831.

\bibitem{harishankaraccept}
M.~Harishankar, N.~Srinivasan, C.~Joe-Wong, and P.~Tague, ``To accept or not to
  accept: The question of supplemental discount offers in mobile data plans,''
  in \emph{Proc. of IEEE INFOCOM}, Honolulu, HI, USA, 2018.

\bibitem{choi2017online}
H.~Choi, C.~Mela, S.~Balseiro, and A.~Leary, ``Online display advertising
  markets: A literature review and future directions,'' \emph{Working Paper},
  2017.

\bibitem{bergemann2011targeting}
D.~Bergemann and A.~Bonatti, ``Targeting in advertising markets: implications
  for offline versus online media,'' \emph{The RAND Journal of Economics},
  vol.~42, no.~3, pp. 417--443, 2011.

\bibitem{duan2015pricing}
L.~Duan, J.~Huang, and B.~Shou, ``Pricing for local and global {Wi-Fi}
  markets,'' \emph{IEEE Transactions on Mobile Computing}, vol.~14, no.~5, pp.
  1056--1070, 2015.

\bibitem{schmidt2009minimum}
D.~A. Schmidt, C.~Shi, R.~A. Berry, M.~L. Honig, and W.~Utschick, ``Minimum
  mean squared error interference alignment,'' in \emph{Proc. of IEEE ACSSC},
  Pacific Grove, CA, USA, 2009, pp. 1106--1110.

\bibitem{zhou2005utility}
C.~Zhou, M.~L. Honig, and S.~Jordan, ``Utility-based power control for a
  two-cell {CDMA} data network,'' \emph{IEEE Transactions on Wireless
  Communications}, vol.~4, no.~6, pp. 2764--2776, 2005.

\bibitem{anderson2015advertising}
S.~P. Anderson and B.~Jullien, ``The advertising-financed business model in
  two-sided media markets,'' in \emph{Handbook of media economics}, 2015,
  vol.~1, pp. 41--90.

\bibitem{anand2011advertising}
B.~N. Anand and R.~Shachar, ``Advertising, the matchmaker,'' \emph{The RAND
  Journal of Economics}, vol.~42, no.~2, pp. 205--245, 2011.

\bibitem{campbell2003brand}
M.~C. Campbell and K.~L. Keller, ``Brand familiarity and advertising repetition
  effects,'' \emph{Journal of consumer research}, vol.~30, no.~2, pp. 292--304,
  2003.

\bibitem{dewenter2012file}
R.~Dewenter, J.~Haucap, and T.~Wenzel, ``On file sharing with indirect network
  effects between concert ticket sales and music recordings,'' \emph{Journal of
  Media Economics}, vol.~25, no.~3, pp. 168--178, 2012.

\bibitem{rasch2013piracy}
A.~Rasch and T.~Wenzel, ``Piracy in a two-sided software market,''
  \emph{Journal of Economic Behavior \& Organization}, vol.~88, pp. 78--89,
  2013.

\bibitem{yu2019pricing}
H.~Yu, G.~Iosifidis, B.~Shou, and J.~Huang, ``Pricing for collaboration between
  online apps and offline venues,'' \emph{IEEE Transactions on Mobile
  Computing}, 2019.

\bibitem{OptusFill}
Optus, {https://www.optus.com.au/shop/mobile/apps/optusxtra}, accessed on Nov.
  27, 2019.

\bibitem{bertsekas1999nonlinear}
D.~P. Bertsekas, \emph{Nonlinear programming}.\hskip 1em plus 0.5em minus
  0.4em\relax Belmont, MA, USA: Athena Scientific, 1999.

\end{thebibliography}

\begin{IEEEbiography}
[{\includegraphics[width=1in,height=1.25in,clip,keepaspectratio]{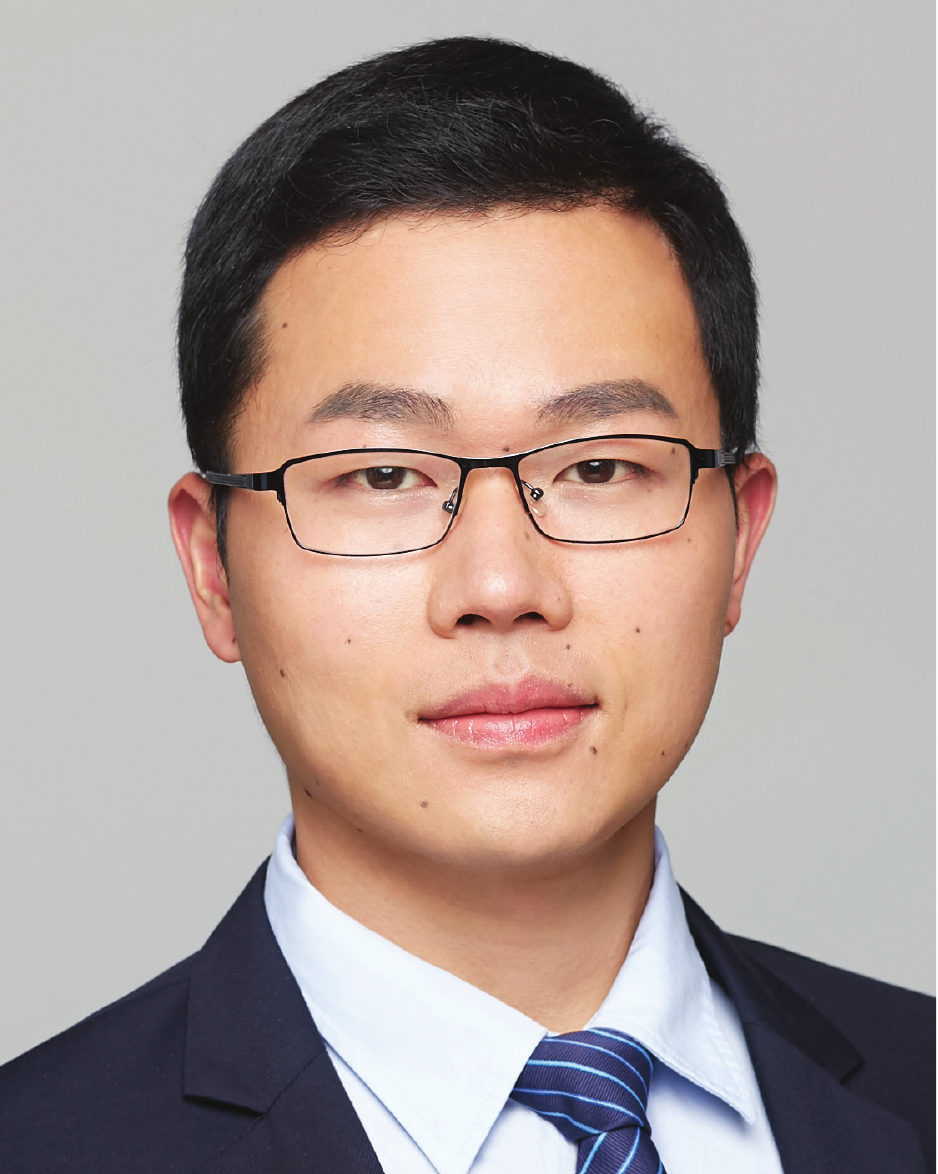}}]
{Haoran Yu} (S'14-M'17) received the Ph.D. degree from the Chinese University of Hong Kong in 2016. During 2015-2016, he was a Visiting Student in the Yale Institute for Network Science and the Department of Electrical Engineering at Yale University. During 2018-2019, he was a Post-Doctoral Fellow in the Department of Electrical and Computer Engineering at Northwestern University. His recent research interests lie in the field of mechanism design for networks. His paper in {\it IEEE INFOCOM 2016} was selected as a Best Paper Award finalist and one of top 5 papers from over 1600 submissions.
\end{IEEEbiography}

\begin{IEEEbiography}[{\includegraphics[width=1in,height=1.25in,clip,keepaspectratio]{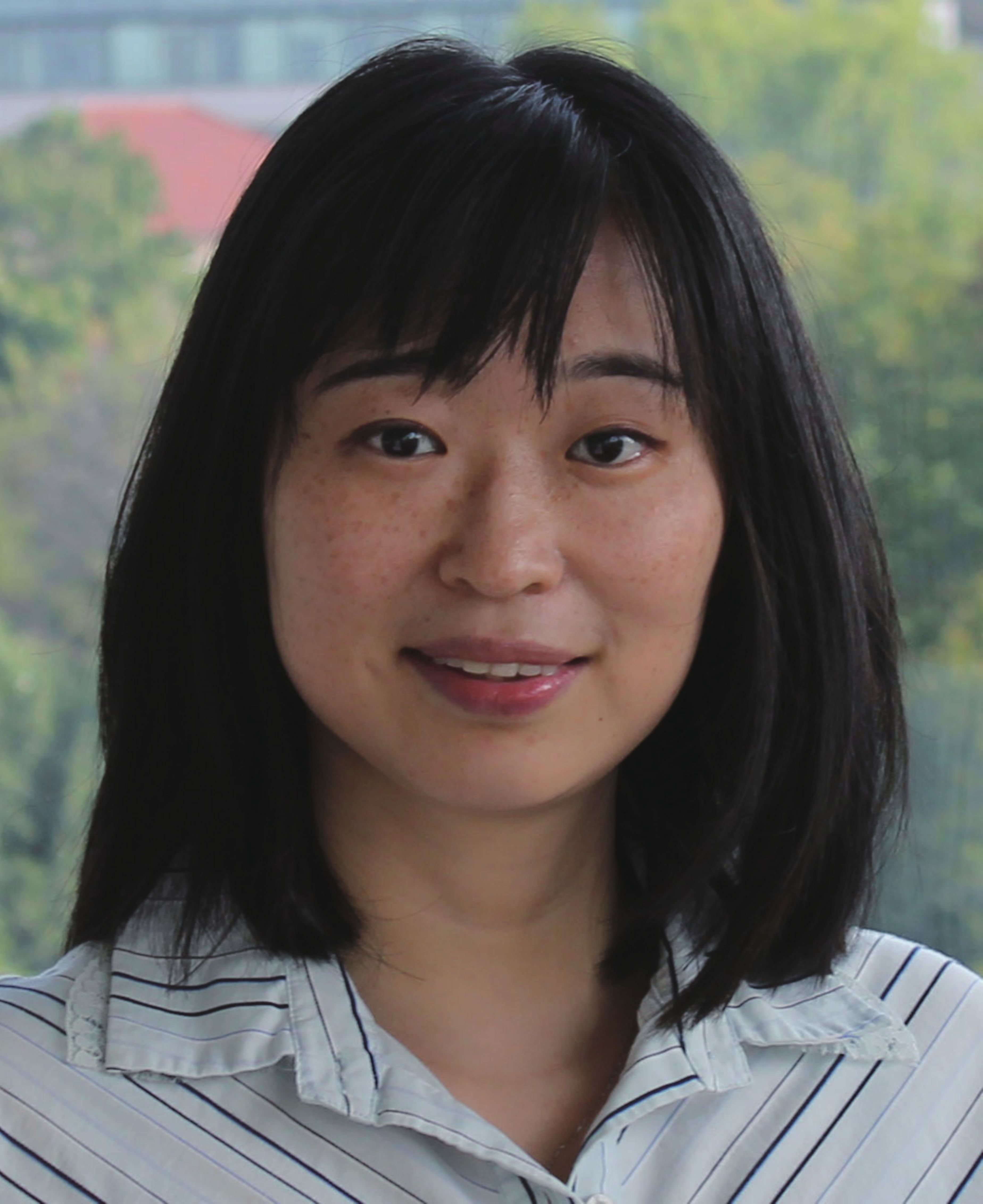}}]{Ermin Wei} is currently an Assistant Professor at the ECE Dept. of Northwestern University. She completed her PhD studies in Electrical Engineering and Computer Science at MIT in 2014, advised by Professor Asu Ozdaglar, where she also obtained her M.S.. She received her undergraduate triple degree in Computer Engineering, Finance and Mathematics with a minor in German, from University of Maryland, College Park. Wei has received many awards, including the Graduate Women of Excellence Award, second place prize in Ernst A. Guillemen Thesis Award and Alpha Lambda Delta National Academic Honor Society Betty Jo Budson Fellowship. Wei's research interests include distributed optimization methods, convex optimization and analysis, smart grid, communication systems and energy networks and market economic analysis.
\end{IEEEbiography}

\begin{IEEEbiography}[{\includegraphics[width=1.1in,height=1.375in,clip,keepaspectratio]{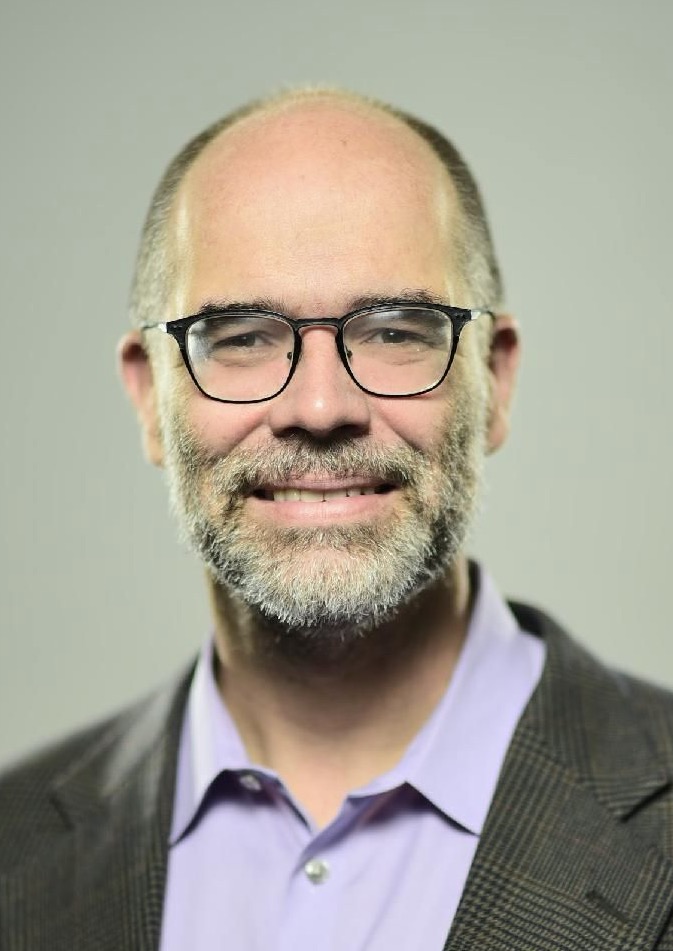}}]
{Randall A. Berry} (S'93-M'00-SM'12-F'14) received the Ph.D. degree from the Massachusetts Institute of Technology in 2000 and joined Northwestern University, where he is currently 
the John A. Dever Professor and Chair of Electrical and Computer Engineering. Dr. Berry's research interests span topics in network economics, wireless communication, computer networking, and information theory. 
He was a recipient of the 2003 CAREER Award from the National Science Foundation.  He served as an Editor for the IEEE TRANSACTIONS ON WIRELESS COMMUNICATIONS from 2006 to 2009 and the IEEE TRANSACTIONS ON INFORMATION THEORY from 2008 to 2012, as well as a Guest Editor for special issues of the IEEE JOURNAL ON SELECTED AREAS IN COMMUNICATIONS in 2017, the IEEE JOURNAL ON SELECTED TOPICS IN SIGNAL PROCESSING in 2008, and the IEEE TRANSACTIONS ON INFORMATION THEORY in 2007. He has also served on the program and organizing committees of numerous conferences, including serving as the Co-Chair for the 2012 IEEE Communication Theory Workshop and 2010 IEEE ICC Wireless Networking Symposium and the TPC Co-Chair for the 2018 ACM MobiHoc conference.
\end{IEEEbiography}



\newpage


\vspace{0.2cm}

\appendices

{{

\begin{centering}
{{\large{\bf{Outline}}}}\\
\end{centering}

\vspace{0.2cm}

\noindent
{\bf Appendix \ref{appendix:notationtable}: Notation Table}.\\
{\bf Appendix \ref{appendix:unique:theta2}: Proof of Lemma \ref{lemma:theta0}}.\\
{\bf Appendix \ref{appendix:proposition31}: Proof of Proposition \ref{proposition:stageII:user}}.\\
{\bf Appendix \ref{appendix:monotonicity:theta2}: $\theta_2$'s Monotonicity with Respect to $\omega$}.\\
{\bf Appendix \ref{appendix:proposition32}: Proof of Proposition \ref{proposition:advertiser}}.\\
{\bf Appendix \ref{appendix:details:SAR}: Example of Computing $m^*\left(\omega,p\right)$}.\\
{\bf Appendix \ref{appendix:theorem1}: Proof of Theorem \ref{theorem:SAR:price}}.\\
{\bf Appendix \ref{appendix:monotonicityD}: Proof of Proposition \ref{proposition:monotonicity}}.\\
{\bf Appendix \ref{appendix:proofoftheorem1}: Proof of Theorem \ref{theorem:subscriptionaware}}.\\
{\bf Appendix \ref{appendix:sec:proposition4}: Proof of Proposition \ref{proposition:uniform:useup}}.\\
{\bf Appendix \ref{appendix:numericalexample}: \!Example \!Where \!${\mathbb E}\!\left[y\right]\! N^{\rm ad}\!\left(\omega\right)\!$ \!Decreases \!with $\omega$}.\\
{\bf Appendix \ref{appendix:uniquetheta4}: Proof of Lemma \ref{lemma:theta3}}.\\
{\bf Appendix \ref{appendix:proposition41}: Proof of Proposition \ref{proposition:unaware:user}}.\\
{\bf Appendix \ref{appendix:lemma3}: Proof of Lemma \ref{lemma:theta4}}.\\
{\bf Appendix \ref{appendix:sufficientcondition}: Proof of Theorem \ref{theorem:notuse}}.\\
{\bf Appendix \ref{appendix:theorem4}: Proof of Theorem \ref{theorem:differentiation}}.\\
{\bf Appendix \ref{appendix:theorem5}: Proof of Theorem \ref{theorem:comparison}}.\\
{\bf Appendix \ref{appendix:morepara}: Numerical Results Under Different Settings}.\\

\section{Notation Table}\label{appendix:notationtable}

We summarize the key notations in Table \ref{table:notation}.

\begin{table}
\caption{{{Key Notations.}}}\label{table:notation}
\begin{tabular}{|p{2.3cm}|p{5.7cm}|}
\hline
\multicolumn{2}{|c|}{{\bf Parameters}}\\
\hline
{$F>0$} & {Subscription fee of data plan}\\
{$Q>0$} & {Amount of data included in data plan}\\
{$N>0$} & {Mass of users}\\
{$\Phi>0$} & {A user's disutility of watching one ad}\\
{$K>0$} & {Number of advertisers}\\
{$A\ge0$, $B>0$} & {Coefficients in (\ref{equ:wearout}) capturing advertising's effectiveness. $A$ is the degree of wear-out effect.}\\
{$C\ge D\left(0\right)$} & {Network capacity}\\
\hline
\hline
\multicolumn{2}{|c|}{{\bf Decision Variables}}\\
\hline
{$\omega\in\left[0,\infty\right)$} & {Operator's unit data reward}\\
{$p\in\left(0,\infty\right)$} & {Operator's ad price}\\
{$r\in\left\{0,1\right\}$} & {A user's subscription decision}\\
{$x\in\left[0,\infty\right)$} & {Number of ads that a user watches}\\
{$m\in\left[0,\infty\right)$} & {Number of ad slots that an advertiser purchases}\\
\hline
\hline
\multicolumn{2}{|c|}{{\bf Random Variables}}\\
\hline
{$\theta\in\left[0,\theta_{\max}\right]$} & {A user's valuation for mobile service}\\
{$y$} & {Number of ads watched by a user who chooses to watch ads (i.e., $x^*\!\left(\theta,\omega\right)\!>\!0$)}\\
\hline
\hline
\multicolumn{2}{|c|}{{\bf Functions}}\\
\hline
{$g\left(\theta\right)$} & {Probability density function of $\theta$}\\
{$u\left(\cdot\right)$} & {A user's utility function}\\
{$\Pi^{\rm user}\left(\theta,r,x,\omega\right)$} & {A type-$\theta$ user's payoff function}\\
{$N^{\rm ad}\left(\omega\right)$} & {Mass of users who watch ads (i.e., with $x^*\left(\theta,\omega\right)>0$)}\\
{$\Pi^{\rm ad}\left(m,\omega,p\right)$} & {An advertiser's payoff function}\\
{$R^{\rm data}\left(\omega\right)$} & {Operator's revenue from data market}\\
{$R^{\rm ad}\left(\omega,p\right)$} & {Operator's ad revenue}\\
{$R^{\rm total}\left(\omega,p\right)$} & {Operator's total revenue}\\
{$D\left(\omega\right)$} & {Total data demand}\\
\hline
\hline
\multicolumn{2}{|c|}{{\bf Other Notations}}\\
\hline
{$\Pi^{\rm SAR}$} & {Operator's optimal total revenue under SAR scheme}\\
{$\Pi^{\rm SUR}$, $\Pi^{\rm SURD}$} & {Operator's optimal total revenues under SUR scheme. $\Pi^{\rm SUR}$: no ad slot differentiation; $\Pi^{\rm SURD}$: with ad slot differentiation.}\\
\hline
\end{tabular}
\end{table}

\section{Proof of Lemma \ref{lemma:theta0}}\label{appendix:unique:theta2}
\begin{proof}
We define function $h\left(\theta\right)$ as
\begin{align}
h\left(\theta\right)\triangleq \theta u\left(\left(u'\right)^{-1}\left(\frac{\Phi}{\omega \theta}\right)\right)-F-\frac{\Phi}{\omega}\left(\left(u'\right)^{-1}\left(\frac{\Phi}{\omega\theta}\right)-Q\right).
\end{align}
First, we compute its derivative. Note that we have $u'\left(\left(u'\right)^{-1}\left(\frac{\Phi}{\omega \theta}\right)\right)=\frac{\Phi}{\omega \theta}$. Next, we apply the chain rule to compute $\frac{d h\left(\theta\right)}{d \theta}$. After the cancellation of the same terms, we get the following result:
\begin{align}
\frac{d h\left(\theta\right)}{d \theta} = u\left(\left(u'\right)^{-1}\left(\frac{\Phi}{\omega \theta}\right)\right).
\end{align}
Note that function $u\left(\cdot\right)$ is strictly concave and $\lim_{z\rightarrow \infty}  u'\left(z\right)=0$. This implies that $u'\left(\cdot\right)$ is a strictly decreasing function. As a result, $\left(u'\right)^{-1}\left(\cdot\right)$, which is the inverse function of $u'\left(\cdot\right)$, is also a strictly decreasing function. Based on this result and the fact that $u\left(\cdot\right)$ is a strictly increasing function, we can see that $\frac{d h\left(\theta\right)}{d \theta}$ is strictly increasing in $\theta$. Furthermore, when $\theta=\theta_1=\frac{\Phi}{\omega u'\left(Q\right)}$, the value of $\frac{d h\left(\theta\right)}{d \theta}$ is given by $\frac{d h\left(\theta\right)}{d \theta}=u\left(\left(u'\right)^{-1}\left({ { u'\left(Q\right)}}\right)\right)=u\left(Q\right)$. Since $u\left(0\right)=0$, $Q>0$, and $u\left(\cdot\right)$ is strictly increasing, we can conclude that $\frac{d h\left(\theta\right)}{d \theta}|_{\theta=\theta_1}>0$. Hence, $\frac{d h\left(\theta\right)}{d \theta}>0$ for $\theta\ge \theta_1=\frac{\Phi}{\omega u'\left(Q\right)}$.

Second, we compute $h\left(\theta_1\right)$. By substituting $\theta_1=\frac{\Phi}{\omega u'\left(Q\right)}$ into $h\left(\theta\right)$, we have 
\begin{align}
h\left(\theta_1\right)= \frac{\Phi}{\omega u'\left(Q\right)} u\left(Q\right)-F.
\end{align}
When $\omega>\frac{\Phi u\left(Q\right)}{F u'\left(Q\right)}$, we can see that $h\left(\theta_1\right)<0$.

Third, we compute $h\left(\theta_0\right)$. By substituting $\theta_0=\frac{F}{u\left(Q\right)}$ into $h\left(\theta\right)$, we have 
\begin{align}
\nonumber
h\left(\theta_0\right)=& \frac{F}{u\left(Q\right)} u\left(\left(u'\right)^{-1}\left(\frac{\Phi {u\left(Q\right)}}{\omega {F}}\right)\right)-F\\
& -\frac{\Phi}{\omega}\left(\left(u'\right)^{-1}\left(\frac{\Phi {u\left(Q\right)}}{\omega {F}}\right)-Q\right).
\end{align}
Next, we compare $h\left(\theta_0\right)$ with $0$. We define a new function $\Delta\left(F\right)
$ as follows:
\begin{align}
\nonumber
\Delta\left(F\right)=& \frac{F}{u\left(Q\right)} u\left(\left(u'\right)^{-1}\left(\frac{\Phi {u\left(Q\right)}}{\omega {F}}\right)\right)-F\\
& -\frac{\Phi}{\omega}\left(\left(u'\right)^{-1}\left(\frac{\Phi {u\left(Q\right)}}{\omega {F}}\right)-Q\right).
\end{align}
We can compute $\Delta\left(\frac{\Phi u\left(Q\right)}{\omega u'\left(Q\right)}\right)=\Delta\left(F\right)|_{F=\frac{\Phi u\left(Q\right)}{\omega u'\left(Q\right)}}$ as
\begin{align}
\nonumber
\Delta\left(\frac{\Phi u\left(Q\right)}{\omega u'\left(Q\right)}\right)=& \frac{\Phi }{\omega u'\left(Q\right) } u\left(\left(u'\right)^{-1}\left({{ u'\left(Q\right)}}\right)\right)\\
\nonumber
& -\frac{\Phi u\left(Q\right)}{\omega u'\left(Q\right)} -\frac{\Phi}{\omega}\left(\left(u'\right)^{-1}\left({{ u'\left(Q\right)}}\right)-Q\right)\\
=& \frac{\Phi u\left(Q\right)}{\omega u'\left(Q\right) }-\frac{\Phi u\left(Q\right)}{\omega u'\left(Q\right)} =0.\label{SM:Delta}
\end{align}
Then, we analyze $\frac{d \Delta\left(F\right)}{d F}$. We compute $\frac{d \Delta\left(F\right)}{d F}$ as follows:
\begin{align}
\frac{d \Delta\left(F\right)}{d F}=\frac{1}{u\left(Q\right)} u\left(\left(u'\right)^{-1}\left(\frac{\Phi {u\left(Q\right)}}{\omega {F}}\right)\right)-1.
\end{align}
Recall that $u\left(\cdot\right)$ is strictly increasing and $\left(u'\right)^{-1}\left(\cdot\right)$ is strictly decreasing. Hence, $\frac{d \Delta\left(F\right)}{d F}$ is strictly increasing in $F$. Moreover, we can compute $\frac{d \Delta\left(F\right)}{d F}|_{F=\frac{\Phi u\left(Q\right)}{\omega u'\left(Q\right)}}$ as $\frac{d \Delta\left(F\right)}{d F}|_{F=\frac{\Phi u\left(Q\right)}{\omega u'\left(Q\right)}}=\frac{1}{u\left(Q\right)} u\left(\left(u'\right)^{-1}\left({ { u'\left(Q\right)}}\right)\right)-1=0$. Therefore, we can see that $\frac{d \Delta\left(F\right)}{d F}$ is zero when $F=\frac{\Phi u\left(Q\right)}{\omega u'\left(Q\right)}$ and is positive when $F>\frac{\Phi u\left(Q\right)}{\omega u'\left(Q\right)}$. Combining this result and $\Delta\left(F\right)|_{F=\frac{\Phi u\left(Q\right)}{\omega u'\left(Q\right)}}=0$ in (\ref{SM:Delta}), we can conclude that $\Delta\left(F\right)>0$ for $F>\frac{\Phi u\left(Q\right)}{\omega u'\left(Q\right)}$. Because $\Delta\left(F\right)$ equals $h\left(\theta_0\right)$, we have $h\left(\theta_0\right)>0$ when $\omega>\frac{\Phi u\left(Q\right)}{F u'\left(Q\right)}$. 

When $\omega>\frac{\Phi u\left(Q\right)}{F u'\left(Q\right)}$, we have proved that $\frac{d h\left(\theta\right)}{d \theta}>0$ for $\theta\ge \theta_1$, $h\left(\theta_1\right)<0$, and $h\left(\theta_0\right)>0$. Hence, there exists a unique $\theta_2\in\left(\theta_1,\theta_0\right)$ satisfying $h\left(\theta_2\right)=0$. 
\end{proof}

\section{Proof of Proposition \ref{proposition:stageII:user}}\label{appendix:proposition31}
\begin{proof}

{\bf Step 1:} We analyze Case A, i.e., $\omega \in\left[0,\frac{\Phi}{u'\left(Q\right)\theta_{\max}}\right]$. 

Suppose a type-$\theta$ user subscribes to the data plan, i.e., $r=1$. Its payoff is $\Pi^{\rm user}\left(\theta,1,x,\omega\right)=\theta u\left(Q+\omega x\right)-F-\Phi x$. We can see that
\begin{align}
\nonumber
 \frac{\partial \Pi^{\rm user}\left(\theta,1,x,\omega\right)}{\partial x}&= \theta \omega u'\left(Q+\omega x\right)-\Phi \\
& \le \frac{  u'\left(Q+\omega x\right)\theta}{u'\left(Q\right) \theta_{\max}}\Phi-\Phi.
\end{align}
Since $u'\left(\cdot\right)$ is a strictly decreasing function, we have $u'\left(Q+\omega x\right)<u'\left(Q\right)$ for $x>0$. Hence, $\frac{\partial \Pi^{\rm user}\left(\theta,1,x,\omega\right)}{\partial x}<0$ for $x>0$. This implies that if a user subscribes to the data plan, it will not watch any ad for rewards. In this case, the user's payoff is $\theta u\left(Q\right)-F$. 

Suppose a type-$\theta$ user does not subscribe, i.e., $r=0$. Under the SAR scheme, the user cannot watch ads for the rewards. Hence, its payoff is $0$.

Comparing $\theta u\left(Q\right)-F$ and $0$, we can see that a user subscribes if and only if $\theta\ge \theta_0 =\frac{F}{u\left(Q\right)}$. Moreover, a user with any $\theta\in\left[0,\theta_{\max}\right]$ will not watch ads. That is to say, we have
\begin{align}
r^*\left(\theta,\omega\right)={\mathbbm 1}_{\left\{\theta\ge \theta_0\right\}},{~~}{~~}x^*\left(\theta,\omega\right)=0, {~~}\theta\in\left[0,\theta_{\max}\right].
\end{align}

{\bf Step 2:} We analyze Case B, i.e., $\omega\in\left(\frac{\Phi}{u'\left(Q\right)\theta_{\max}},\frac{\Phi u\left(Q\right)}{Fu'\left(Q\right)}\right]$.

Suppose a type-$\theta$ user subscribes to the data plan, i.e., $r=1$. Its payoff is $\Pi^{\rm user}\left(\theta,1,x,\omega\right)=\theta u\left(Q+\omega x\right)-F-\Phi x$. By checking $\frac{\partial \Pi^{\rm user}\left(\theta,1,x,\omega\right)}{\partial x}$, we can see the following result:

(i) If $\theta\in\left[0,\frac{\Phi}{\omega u'\left(Q\right)}\right)$, the user will not watch ads (i.e., $x=0$), and its payoff will be $\theta u\left(Q\right)-F$;

(ii) If $\theta\in\left[\frac{\Phi}{\omega u'\left(Q\right)},\theta_{\max}\right]$, the user will choose $x=\frac{1}{\omega}\left(\left(u'\right)^{-1}\left(\frac{\Phi}{\omega\theta}\right)-Q\right)$. Since $\theta\ge \frac{\Phi}{\omega u'\left(Q\right)}$, the value of $\frac{1}{\omega}\left(\left(u'\right)^{-1}\left(\frac{\Phi}{\omega\theta}\right)-Q\right)$ is non-negative. Then, we can see that the user's payoff satisfies 
\begin{align}
\nonumber
\Pi^{\rm user}\left(\theta,1,\frac{1}{\omega}\left(\left(u'\right)^{-1}\left(\frac{\Phi}{\omega\theta}\right)-Q\right),\omega\right)\ge & \Pi^{\rm user}\left(\theta,1,0,\omega\right)\\
= & \theta u\left(Q\right)-F.
\end{align}

Since $\omega\le \frac{\Phi u\left(Q\right)}{F u'\left(Q\right)}$, we have $\frac{\Phi}{\omega u'\left(Q\right)}\ge \frac{F}{u\left(Q\right)}$. Hence, for a user with $\theta\ge \frac{\Phi}{\omega u'\left(Q\right)}$, its payoff under $r=1$ and $x=\frac{1}{\omega}\left(\left(u'\right)^{-1}\left(\frac{\Phi}{\omega\theta}\right)-Q\right)$ is non-negative. 

Suppose a type-$\theta$ user does not subscribe, i.e., $r=0$. Under the SAR scheme, the user cannot watch ads for the rewards and its payoff is $0$.

Next, we can compare the choices of $r=1$ and $r=0$. Recall that $\frac{\Phi}{\omega u'\left(Q\right)}\ge \frac{F}{u\left(Q\right)}$. We discuss the choices of users with $\theta\in\left[0,\frac{F}{u\left(Q\right)}\right)$, $\theta\in\left[\frac{F}{u\left(Q\right)},\frac{\Phi}{\omega u'\left(Q\right)}\right)$, and $\theta\in\left[\frac{\Phi}{\omega u'\left(Q\right)},\theta_{\max}\right)$, separately. If $\theta\in\left[0,\frac{F}{u\left(Q\right)}\right)$, the user has a higher payoff under $r=0$. As discussed above, it cannot watch ads. If $\theta\in\left[\frac{F}{u\left(Q\right)},\frac{\Phi}{\omega u'\left(Q\right)}\right)$, the user has a higher payoff under $r=1$, and it will not watch ads. If $\theta\in\left[\frac{\Phi}{\omega u'\left(Q\right)},\theta_{\max}\right)$, the user has a higher payoff under $r=1$, and it will choose $x=\frac{1}{\omega}\left(\left(u'\right)^{-1}\left(\frac{\Phi}{\omega\theta}\right)-Q\right)$. Recall that we define $\theta_0= \frac{F}{u\left(Q\right)}$ and $\theta_1= \frac{\Phi}{\omega u'\left(Q\right)}$. We can conclude that the following result holds for $\theta\in\left[0,\theta_{\max}\right]$:
\begin{align}
r^*\left(\theta,\omega\right)\!=\!{\mathbbm 1}_{\left\{\theta\ge \theta_0\right\}},x^*\!\left(\theta,\omega\right)\!=\!\frac{1}{\omega}\!\left(\!\left(u'\right)^{-1}\!\left(\frac{\Phi}{\omega\theta}\right)\!-\!Q\right) \!{\mathbbm 1}_{\left\{\theta\ge \theta_1\right\}}.
\end{align}

{\bf Step 3:} We analyze Case C, i.e., $\omega\in \left(\frac{\Phi u\left(Q\right)}{F u'\left(Q\right)},\infty \right)$.

Suppose a type-$\theta$ user subscribes to the data plan, i.e., $r=1$. Its payoff is $\Pi^{\rm user}\left(\theta,1,x,\omega\right)=\theta u\left(Q+\omega x\right)-F-\Phi x$. By checking $\frac{\partial \Pi^{\rm user}\left(\theta,1,x,\omega\right)}{\partial x}$, we can see the following result:

(i) If $\theta\in\left[0,\frac{\Phi}{\omega u'\left(Q\right)}\right)$, the user will not watch ads (i.e., $x=0$), and its payoff will be $\theta u\left(Q\right)-F$;

(ii) If $\theta\in\left[\frac{\Phi}{\omega u'\left(Q\right)},\theta_{\max}\right]$, the user will choose $x=\frac{1}{\omega}\left(\left(u'\right)^{-1}\left(\frac{\Phi}{\omega\theta}\right)-Q\right)$. Since $\theta\ge \frac{\Phi}{\omega u'\left(Q\right)}$, the value of $\frac{1}{\omega}\left(\left(u'\right)^{-1}\left(\frac{\Phi}{\omega\theta}\right)-Q\right)$ is non-negative. The user's corresponding payoff is given by
\begin{align}
\nonumber
\Pi^{\rm user}\left(\theta,1,x,\omega\right)=& \theta u\left(\left(u'\right)^{-1}\left(\frac{\Phi}{\omega\theta}\right)\right)-F \\
& - \frac{\Phi}{\omega}\left(\left(u'\right)^{-1}\left(\frac{\Phi}{\omega\theta}\right)-Q\right)\label{SM:pro1:a}
\end{align}

Suppose a type-$\theta$ user does not subscribe, i.e., $r=0$. Under the SAR scheme, the user cannot watch ads for the rewards and its payoff is $0$.

Next, we can compare the choices of $r=1$ and $r=0$. Since $\omega> \frac{\Phi u\left(Q\right)}{F u'\left(Q\right)}$, we have $\frac{\Phi}{\omega u'\left(Q\right)} < \frac{F}{u\left(Q\right)}$. If $\theta\in\left[0,\frac{\Phi}{\omega u'\left(Q\right)}\right)$, the user has a higher payoff under $r=0$, and it cannot watch ads. If $\theta\in\left[\frac{\Phi}{\omega u'\left(Q\right)},\theta_2\right)$, we can see the following relation based on our proof in Appendix \ref{appendix:unique:theta2}:
\begin{align}
\theta u\left(\left(u'\right)^{-1}\left(\frac{\Phi}{\omega \theta}\right)\right)-F-\frac{\Phi}{\omega}\left(\left(u'\right)^{-1}\left(\frac{\Phi}{\omega\theta}\right)-Q\right)<0.
\end{align}
The left side is the value of $\Pi^{\rm user}\left(\theta,1,x,\omega\right)$ in (\ref{SM:pro1:a}). The inequality implies that the user has a higher payoff under $r=0$. In this case, it cannot watch ads. 

If $\theta\in\left[\theta_2,\theta_{\max}\right]$, we can see the following relation based on our proof in Appendix \ref{appendix:unique:theta2}:
\begin{align}
\theta u\left(\left(u'\right)^{-1}\left(\frac{\Phi}{\omega \theta}\right)\right)-F-\frac{\Phi}{\omega}\left(\left(u'\right)^{-1}\left(\frac{\Phi}{\omega\theta}\right)-Q\right)\ge0.
\end{align}
This implies that the user has a higher payoff under $r=1$. In this case, it chooses $x=\frac{1}{\omega}\left(\left(u'\right)^{-1}\left(\frac{\Phi}{\omega\theta}\right)-Q\right)$. We can conclude that the following result holds for $\theta\in\left[0,\theta_{\max}\right]$:
\begin{align}
r^*\!\left(\theta,\omega\right)\!=\!{\mathbbm 1}_{\left\{\theta\ge \theta_2\right\}},x^*\!\left(\theta,\omega\right)\!=\!\frac{1}{\omega}\!\left(\!\left(u'\right)^{-1}\!\left(\frac{\Phi}{\omega\theta}\right)\!-\!Q\right) \!{\mathbbm 1}_{\left\{\theta\ge \theta_2\right\}}.
\end{align}

Hence, we have proved $r^*\left(\theta,\omega\right)$ and $x^*\left(\theta,\omega\right)$ for Case A, Case B, and Case C. 
\end{proof}

\begin{figure*}
\begin{align}
\frac{d\theta_2}{d\omega} u\left(\left(u'\right)^{-1}\left(\frac{\Phi}{\omega \theta_2}\right)\right) \!+\! \theta_2 u'\left(\left(u'\right)^{-1}\left(\frac{\Phi}{\omega \theta_2}\right)\right)\frac{d \left(\left(u'\right)^{-1}\left(\frac{\Phi}{\omega \theta_2}\right)\right)}{d\omega} +\frac{\Phi}{\omega^2} \left(\left(u'\right)^{-1}\left(\frac{\Phi}{\omega\theta_2}\right)-Q\right) \!-\!\frac{\Phi}{\omega} \frac{d \left(\left(u'\right)^{-1}\left(\frac{\Phi}{\omega \theta_2}\right)\right)}{d\omega}=0.\label{SM:equ:long}
\end{align}
\hrule
\end{figure*}

\section{$\theta_2$'s Monotonicity with Respect to $\omega$}\label{appendix:monotonicity:theta2}
We prove that in Case C, $\theta_2$ decreases as $\omega$ increases.
\begin{proof}
Based on $\theta_2$'s definition, we have
\begin{align}
\theta_2 u\left(\left(u'\right)^{-1}\left(\frac{\Phi}{\omega \theta_2}\right)\right)-F-\frac{\Phi}{\omega}\left(\left(u'\right)^{-1}\left(\frac{\Phi}{\omega\theta_2}\right)-Q\right)=0.
\end{align}
For both sides of the equation, we take their derivatives with respect to $\omega$, and get equation (\ref{SM:equ:long}). After rearrangement, we have the following equation:
\begin{align}
\frac{d\theta_2}{d\omega} u\left(\left(u'\right)^{-1}\left(\frac{\Phi}{\omega \theta_2}\right)\right) +\frac{\Phi}{\omega^2} \left(\left(u'\right)^{-1}\left(\frac{\Phi}{\omega\theta_2}\right)-Q\right) =0.
\end{align}
Hence, we can get the expression of $\frac{d\theta_2}{d\omega}$ as follows:
\begin{align}
\frac{d\theta_2}{d\omega} =\frac{\frac{\Phi}{\omega^2} \left(Q-\left(u'\right)^{-1}\left(\frac{\Phi}{\omega\theta_2}\right)\right) }{u\left(\left(u'\right)^{-1}\left(\frac{\Phi}{\omega \theta_2}\right)\right)}.
\end{align}
Based on $\theta_2$'s definition, we have $\theta_2>\theta_1=\frac{\Phi}{\omega u'\left(Q\right)}$. Recall that $\left(u'\right)^{-1}\left(\cdot\right)$ is strictly decreasing. We can see that $\left(u'\right)^{-1}\left(\frac{\Phi}{\omega \theta_2}\right)>\left(u'\right)^{-1}\left(\frac{\Phi}{\omega \theta_1}\right)=Q$. As a result, the value of $\frac{\Phi}{\omega^2} \left(Q-\left(u'\right)^{-1}\left(\frac{\Phi}{\omega\theta_2}\right)\right)$ is negative, and the value of $u\left(\left(u'\right)^{-1}\left(\frac{\Phi}{\omega \theta_2}\right)\right)$ is positive. Therefore, we can conclude that $\frac{d \theta_2}{d \omega}<0$. 
\end{proof}

\section{Proof of Proposition \ref{proposition:advertiser}}\label{appendix:proposition32}
\begin{proof}
First, we consider the case where $N^{\rm ad}\left(\omega\right)=0$, i.e., the mass of users watching ads is zero. Based on the advertiser payoff's definition, $\Pi^{\rm ad}\left(m,\omega,p\right)= -mp$. Since $p>0$, none of the advertisers will purchase the ad slots, i.e., $m^*\left(\omega,p\right)=0$.

Second, we consider the case where $N^{\rm ad}\left(\omega\right)>0$. An advertiser's payoff is
\begin{multline}
\Pi^{\rm ad}\left(m,\omega,p\right)=\\{\mathbb E}_y\left[B\frac{my}{{\mathbb E}\left[y\right]N^{\rm ad}\left(\omega\right)}-A\left(\frac{my}{{\mathbb E}\left[y\right]N^{\rm ad}\left(\omega\right)}\right)^2\right]N^{\rm ad}\left(\omega\right)-mp.
\end{multline}
After rearrangement, we have
\begin{align}
\Pi^{\rm ad}\left(m,\omega,p\right)=-\frac{A}{N^{\rm ad}\left(\omega\right)} \frac{ {\mathbb E} \left[y^2\right]}{\left({\mathbb E}\left[y\right]\right)^2 } m^2+\left(B-p\right)m,
\end{align}
which is a quadratic function of $m\ge0$. We can easily see that
\begin{align}
m^*\left(\omega,p\right)=\max\left\{\frac{\left(B-p\right)N^{\rm ad}\left(\omega\right) \left({\mathbb E}\left[y\right]\right)^2}{2A{\mathbb E} \left[y^2\right]  },0\right\}.
\end{align}

Combining the analysis for $N^{\rm ad}\left(\omega\right)=0$ and $N^{\rm ad}\left(\omega\right)>0$, we can see the following result: If $N^{\rm ad}\left(\omega\right)=0$ or $p\ge B$, then $m^*\left(\omega,p\right)=0$; otherwise, $m^*\left(\omega,p\right)=\frac{B-p}{2A} \frac{\left({\mathbb E}\left[y\right]\right)^2}{{\mathbb E}\left[y^2\right]} N^{\rm ad}\left(\omega\right)$.
\end{proof}


\section{Example of Computing $m^*\left(\omega,p\right)$}\label{appendix:details:SAR}
\vspace{-0.1cm}
In this section, we assume that each user has a logarithmic utility function, i.e., $u\left(z\right)=\ln\left(1+z\right)$, and a uniformly distributed type, i.e., $\theta\sim{\cal U}\left[0,\theta_{\max}\right]$. We compute the value of $N^{\rm ad}\left(\omega\right)$, the distribution of $y$, and the expression of $m^*\left(\omega,p\right)$ when $\omega$ satisfies Case A, Case B, and Case C.

{\bf Step 1:} We consider Case A. Since $u\left(z\right)=\ln\left(1+z\right)$, the condition of $\omega$ in Case A becomes $\omega \in\left[0,\frac{\left(1+Q\right)\Phi}{\theta_{\max}}\right]$. In this case, there is no user watching ads. Hence, $N^{\rm ad}\left(\omega\right)=0$. From Proposition \ref{proposition:advertiser}, we have $m^*\left(\omega,p\right)=0$.

{\bf Step 2:} We consider Case B. Since $u\left(z\right)=\ln\left(1+z\right)$, the condition of $\omega$ in Case B becomes $\omega\in \left(\frac{\left(1+Q\right)\Phi}{\theta_{\max}},\frac{\Phi}{F}\left(1+Q\right)\ln\left(1+Q\right) \right]$. Based on Proposition \ref{proposition:stageII:user}, the users' ad watching decisions in Case B are characterized by the following equation:
\begin{align}
\nonumber
x^*\left(\theta,\omega\right)=&\frac{1}{\omega}\left(\left(u'\right)^{-1}\left(\frac{\Phi}{\omega\theta}\right)-Q\right) {\mathbbm 1}_{\left\{\theta\ge \theta_1\right\}}\\
\nonumber
\overset{(a)}{=}& \frac{1}{\omega} \left(\frac{\theta \omega}{\Phi}-1-Q\right) {\mathbbm 1}_{\left\{\theta\ge \theta_1\right\}}\\
\overset{(b)}{=}&  \frac{\theta-\theta_1}{\Phi} {\mathbbm 1}_{\left\{\theta\ge \theta_1\right\}}.\label{SM:equ:x:caseB}
\end{align}
Here, equality (a) is due to $u\left(z\right)=\ln\left(1+z\right)$, and equality (b) is due to $\theta_1=\frac{\Phi\left(1+Q\right)}{\omega}$ in Case B. Hence, only the users with $\theta\ge\theta_1$ watch ads. Because $\theta\sim{\cal U}\left[0,\theta_{\max}\right]$, we can compute $N^{\rm ad}\left(\omega\right)$ as follows:
\begin{align}
N^{\rm ad}\left(\omega\right)=\frac{\theta_{\max}-\theta_1}{\theta_{\max}} N.
\end{align}
Moreover, according to (\ref{SM:equ:x:caseB}) and the fact that $\theta\sim{\cal U}\left[0,\theta_{\max}\right]$, we can see that the number of ads watched by one of the $N^{\rm ad}\left(\omega\right)$ users is uniformly distributed in $\left[0,\frac{\theta_{\max}-\theta_1}{\Phi}\right]$. This implies that $y$ is uniformly distributed in $\left[0,\frac{\theta_{\max}-\theta_1}{\Phi}\right]$. Then, we can compute ${\mathbb E}\left[y\right]$ and ${\mathbb E}\left[y^2\right]$ as follows:
\begin{align}
{\mathbb E}\left[y\right]=\frac{1}{2}\frac{\theta_{\max}-\theta_1}{\Phi},{\mathbb E}\left[y^2\right]=\frac{1}{3} \left(\frac{\theta_{\max}-\theta_1}{\Phi}\right)^2.
\end{align}
Based on Proposition \ref{proposition:advertiser}, we can derive $m^*\left(\omega,p\right)$'s expression as
\begin{align}
m^*\left(\omega,p\right)=\frac{3}{8}\frac{\max\left\{B-p,0\right\}}{A}  \frac{\theta_{\max}-\theta_1}{\theta_{\max}} N.
\end{align}

{\bf Step 3:} We consider Case C. Since $u\left(z\right)=\ln\left(1+z\right)$, the condition of $\omega$ in Case C becomes $\omega\in \left(\frac{\Phi}{F}\left(1+Q\right)\ln\left(1+Q\right),\infty \right)$. Based on Proposition \ref{proposition:stageII:user}, the users' ad watching decisions in Case C are characterized by the following equation:
\begin{align}
\nonumber
x^*\left(\theta,\omega\right)=&\frac{1}{\omega}\left(\left(u'\right)^{-1}\left(\frac{\Phi}{\omega\theta}\right)-Q\right) {\mathbbm 1}_{\left\{\theta\ge \theta_2\right\}}\\
=&  \frac{\theta-\theta_1}{\Phi} {\mathbbm 1}_{\left\{\theta\ge \theta_2\right\}}.\label{SM:equ:x:caseC}
\end{align}
Hence, only the users with $\theta\ge\theta_2$ watch ads. Because $\theta\sim{\cal U}\left[0,\theta_{\max}\right]$, we can compute $N^{\rm ad}\left(\omega\right)$ as follows:
\begin{align}
N^{\rm ad}\left(\omega\right)=\frac{\theta_{\max}-\theta_2}{\theta_{\max}} N.
\end{align}
Moreover, according to (\ref{SM:equ:x:caseC}) and the fact that $\theta\sim{\cal U}\left[0,\theta_{\max}\right]$, we can see that the number of ads watched by one of the $N^{\rm ad}\left(\omega\right)$ users is uniformly distributed in $\left[\frac{\theta_2-\theta_1}{\Phi},\frac{\theta_{\max}-\theta_1}{\Phi}\right]$. This means that $y$ is uniformly distributed in $\left[\frac{\theta_2-\theta_1}{\Phi},\frac{\theta_{\max}-\theta_1}{\Phi}\right]$. Then, we can compute ${\mathbb E}\left[y\right]$ and ${\mathbb E}\left[y^2\right]$ as follows:
\begin{align}
& {\mathbb E}\left[y\right]=\frac{\theta_2-\theta_1+\theta_{\max}-\theta_1}{2\Phi},\\
& {\mathbb E}\left[y^2\right]=\left(\frac{\theta_2-\theta_1+\theta_{\max}-\theta_1}{2\Phi}\right)^2+\frac{\left(\frac{\theta_{\max}-\theta_1-\left(\theta_2-\theta_1\right)}{\Phi}\right)^2}{12}.
\end{align}
To simplify the presentation, we let $\lambda_a\triangleq \theta_2-\theta_1$ and $\lambda_b\triangleq \theta_{\max}-\theta_1$. We can further have the following result:
\begin{align}
\nonumber
\frac{\left({\mathbb E}\left[y\right]\right)^2}{{\mathbb E}\left[y^2\right]}&=\frac{\left(\frac{\lambda_a+\lambda_b}{2\Phi}\right)^2}{\left(\frac{\lambda_a+\lambda_b}{2\Phi}\right)^2+\frac{1}{12}\left(\frac{\lambda_b-\lambda_a}{\Phi}\right)^2}\\
&=\frac{3\left({\lambda_a+\lambda_b}\right)^2}{3\left({\lambda_a+\lambda_b}\right)^2+\left({\lambda_b-\lambda_a}\right)^2}.\label{SM:equ:sim:a}
\end{align}
Based on Proposition \ref{proposition:advertiser}, we can derive $m^*\left(\omega,p\right)$'s expression as
\begin{align}
\nonumber
& m^*\left(\omega,p\right)\\
\nonumber
& =\frac{\max\left\{B-p,0\right\}}{2A}\frac{3\left({\lambda_a+\lambda_b}\right)^2}{3\left({\lambda_a+\lambda_b}\right)^2+\left({\lambda_b-\lambda_a}\right)^2}\frac{\theta_{\max}-\theta_2}{\theta_{\max}} N \\
\nonumber
& =\frac{\max\left\{B-p,0\right\}}{2A}\frac{3\left({\lambda_a+\lambda_b}\right)^2}{4\lambda_a^2+4\lambda_b^2+4\lambda_a\lambda_b}\frac{\lambda_b-\lambda_a}{\theta_{\max}} N \\
\nonumber
& =\frac{3\max\left\{B-p,0\right\}}{8A}\frac{N}{\theta_{\max}}\frac{\left({\lambda_a+\lambda_b}\right)^2}{\lambda_a^2+\lambda_b^2+\lambda_a\lambda_b}\frac{\left(\lambda_b-\lambda_a\right)^2}{\lambda_b-\lambda_a} \\
\nonumber
& =\frac{3\max\left\{B-p,0\right\}}{8A}\frac{N}{\theta_{\max}}\frac{\left({\lambda_b^2-\lambda_a^2}\right)^2}{\lambda_b^3-\lambda_a^3} \\
& =\frac{3}{8}\frac{\max\left\{B-p,0\right\}}{A}  \frac{N}{\theta_{\max}} \frac{\left(\left(\theta_{\max}-\theta_1\right)^2-\left(\theta_2-\theta_1\right)^2\right)^2}{\left(\theta_{\max}-\theta_1\right)^3-\left(\theta_2-\theta_1\right)^3}.\label{SM:equ:sim:b}
\end{align}
This completes our analysis of $m^*\left(\omega,p\right)$ under three cases when $u\left(z\right)=\ln\left(1+z\right)$ and $\theta\sim{\cal U}\left[0,\theta_{\max}\right]$.

\section{Proof of Theorem \ref{theorem:SAR:price}}\label{appendix:theorem1}
\begin{proof}
First, we consider the case where $\omega\in\left[0,\frac{\Phi}{u'\left(Q\right) \theta_{\max}}\right]$. According to Proposition \ref{proposition:stageII:user}, no user watches ads, i.e., $N^{\rm ad}\left(\omega\right)=0$. From Proposition \ref{proposition:advertiser}, we can see that $m^*\left(\omega,p\right)=0$ for any $p>0$. This means that the operator's ad revenue $R^{\rm ad}\left(\omega,p\right)$ is zero, regardless of the ad price. Hence, all positive prices lead to the same ad revenue, and any positive price is optimal.  

Second, we consider the case where $\omega\in\left(\frac{\Phi}{u'\left(Q\right) \theta_{\max}},\infty\right)$. From Proposition \ref{proposition:stageII:user}, the number of users watching ads is positive, i.e., $N^{\rm ad}\left(\omega\right)>0$. If $p\ge B$, the value of $m^*\left(\omega,p\right)$ is zero based on Proposition \ref{proposition:advertiser}. In this situation, the value of $R^{\rm ad}\left(\omega,p\right)$ is also zero. If $p<B$, we have $m^*\left(\omega,p\right)=\frac{B-p}{2A} \frac{\left({\mathbb E}\left[y\right]\right)^2}{{\mathbb E}\left[y^2\right]} N^{\rm ad}\left(\omega\right)$. Then, when $\omega$ is given, the operator's problem of deciding $p$ can be rewritten as follows:

\begin{align}
& \max_{p>0} K \frac{B-p}{2A} \frac{\left({\mathbb E}\left[y\right]\right)^2}{{\mathbb E}\left[y^2\right]} N^{\rm ad}\left(\omega\right) p\\
& {~~}{~~}{\rm s.t.}{~~} K \frac{B-p}{2A} \frac{\left({\mathbb E}\left[y\right]\right)^2}{{\mathbb E}\left[y^2\right]} N^{\rm ad}\left(\omega\right)\le {\mathbb E}\left[y\right] N^{\rm ad}\left(\omega\right).
\end{align}

After rearrangement, the problem becomes:
\begin{align}
& \max_{p>0} K \frac{B-p}{2A} \frac{\left({\mathbb E}\left[y\right]\right)^2}{{\mathbb E}\left[y^2\right]} N^{\rm ad}\left(\omega\right) p\\
& {~~}{~~}{\rm s.t.}{~~}  p  \ge B-\frac{2A {\mathbb E}\left[y^2\right]}{K{\mathbb E}\left[y\right]}.
\end{align}
The objective function is quadratic in $p$ and achieves the maximum value at $p=\frac{B}{2}$. Hence, we can see that the optimal price under a given $\omega$ is given by
\begin{align}
p^*\left(\omega\right)=\max\left\{\frac{B}{2},B-\frac{2A {\mathbb E}\left[y^2\right]}{K {\mathbb E}\left[y\right]}\right\}.
\end{align}
\end{proof}

\section{Proof of Proposition \ref{proposition:monotonicity}}\label{appendix:monotonicityD}
\begin{proof}
{\bf Step 1:} We analyze the monotonicity of $D\left(\omega\right)$ in three cases. 

First, when $\omega\in\left[0,\frac{\Phi}{u'\left(Q\right)\theta_{\max}}\right]$, the value of $D\left(\omega\right)$ is given by
\begin{align}
D\left(\omega\right)=NQ \int_{\theta_0}^{\theta_{\max}} g\left(\theta\right) d\theta = NQ \int_{\frac{F}{u\left(Q\right)}}^{\theta_{\max}} g\left(\theta\right) d\theta,\label{SM:equ:D:a}
\end{align}
which is independent of $\omega$. 

Second, when $\omega\in\left(\frac{\Phi}{u'\left(Q\right)\theta_{\max}},\frac{\Phi u\left(Q\right)}{Fu'\left(Q\right)}\right]$, the expression of $D\left(\omega\right)$ is given by
\begin{align}
\nonumber
D\left(\omega\right)=& NQ \int_{\frac{F}{u\left(Q\right)}}^{\theta_{\max}} g\left(\theta\right) d\theta \\
& + N  \int_{\frac{\Phi}{\omega u'\left(Q\right)}}^{\theta_{\max}} \left(\left(u'\right)^{-1}\left(\frac{\Phi}{\omega\theta}\right)-Q\right) g\left(\theta\right)d\theta.\label{SM:equ:D:b}
\end{align}
Based on Leibniz's rule, we can further compute $\frac{d D\left(\omega\right)}{d \omega}$ as follows:
\begin{align}
\nonumber
& \frac{d D\left(\omega\right)}{d \omega} =N  \int_{\frac{\Phi}{\omega u'\left(Q\right)}}^{\theta_{\max}} \frac{d\left(\left(u'\right)^{-1}\left(\frac{\Phi}{\omega\theta}\right)\right)}{d\omega} g\left(\theta\right)d\theta\\
\nonumber
& -N \frac{d \left(\frac{\Phi}{\omega u'\left(Q\right)}\right)}{d\omega}  \left(\left(u'\right)^{-1}\left(\frac{\Phi}{\omega \frac{\Phi}{\omega u'\left(Q\right)}}\right)-Q\right) g\left( \frac{\Phi}{\omega u'\left(Q\right)}\right) \\
& =N  \int_{\frac{\Phi}{\omega u'\left(Q\right)}}^{\theta_{\max}} \frac{d\left(\left(u'\right)^{-1}\left(\frac{\Phi}{\omega\theta}\right)\right)}{d\omega} g\left(\theta\right)d\theta.
\end{align}
From our proof in Lemma \ref{lemma:theta0}, function $\left(u'\right)^{-1}\left(\cdot\right)$ is strictly decreasing. Hence, $\frac{d\left(\left(u'\right)^{-1}\left(\frac{\Phi}{\omega\theta}\right)\right)}{d\omega}$ is positive. This implies that $\frac{d D\left(\omega\right)}{d \omega}$ is positive. 


Third, when $\omega\in\left(\frac{\Phi u\left(Q\right)}{Fu'\left(Q\right)},\infty\right)$, the expression of $D\left(\omega\right)$ is given by
\begin{align}
\nonumber
D\left(\omega\right)=& NQ \int_{\theta_2}^{\theta_{\max}} g\left(\theta\right) d\theta \\
& + N  \int_{\theta_2}^{\theta_{\max}} \left(\left(u'\right)^{-1}\left(\frac{\Phi}{\omega\theta}\right)-Q\right) g\left(\theta\right)d\theta.\label{SM:equ:D:c}
\end{align}
We can further compute $\frac{d D\left(\omega\right)}{d \omega}$ as follows:
\begin{align}
\nonumber
\frac{d D\left(\omega\right)}{d \omega} = &-NQ g\left(\theta_2\right) \frac{d\theta_2}{d\omega}\!+\! N\!  \int_{\theta_2}^{\theta_{\max}} \!\frac{d \left(\left(u'\right)^{-1}\left(\frac{\Phi}{\omega\theta}\right)\right)}{d\omega} g\left(\theta\right)\!d\theta\\
&-N \left(\left(u'\right)^{-1}\left(\frac{\Phi}{\omega\theta_2}\right)-Q\right) g\left(\theta_2\right) \frac{d\theta_2}{d\omega}.\label{SM:equ:dDfunction}
\end{align}
From our proof in Appendix \ref{appendix:monotonicity:theta2}, we have $\frac{d\theta_2}{d\omega}<0$. Hence, the first term in the right side of (\ref{SM:equ:dDfunction}) is positive. Since function $\left(u'\right)^{-1}\left(\cdot\right)$ is strictly decreasing, we have $\frac{d \left(\left(u'\right)^{-1}\left(\frac{\Phi}{\omega\theta}\right)\right)}{d\omega}>0$. Hence, the second term in the right side of (\ref{SM:equ:dDfunction}) is positive. Furthermore, from the definition of $\theta_2$, we have $\theta_2>\theta_1=\frac{\Phi}{\omega u'\left(Q\right)}$. As a result, the value of $\left(u'\right)^{-1}\left(\frac{\Phi}{\omega\theta_2}\right)-Q$ is positive. Considering that $\frac{d\theta_2}{d\omega}<0$, the third term in the right side of (\ref{SM:equ:dDfunction}) is positive. Based on the above analysis, we can see that $\frac{d D\left(\omega\right)}{d \omega} >0$. 

{\bf Step 2:} We analyze the continuity of $D\left(\omega\right)$. Since $u\left(\cdot\right)$ is twice differentiable, we can see that $D\left(\omega\right)$ is continuous for $\omega\in\left[0,\frac{\Phi}{u'\left(Q\right)\theta_{\max}}\right)$, $\omega\in\left(\frac{\Phi}{u'\left(Q\right)\theta_{\max}},\frac{\Phi u\left(Q\right)}{Fu'\left(Q\right)}\right)$, and $\omega\in\left(\frac{\Phi u\left(Q\right)}{Fu'\left(Q\right)},\infty\right)$, based on (\ref{SM:equ:D:a}), (\ref{SM:equ:D:b}), and (\ref{SM:equ:D:c}). Next, we analyze the continuity of $D\left(\omega\right)$ at $\omega=\frac{\Phi}{u'\left(Q\right)\theta_{\max}}$ and $\omega=\frac{\Phi u\left(Q\right)}{Fu'\left(Q\right)}$.

When $\omega=\frac{\Phi}{u'\left(Q\right)\theta_{\max}}$, the value of $\frac{\Phi}{\omega u'\left(Q\right)}$ equals $\theta_{\max}$. Based on (\ref{SM:equ:D:b}), we can compute $\lim_{\omega\searrow \frac{\Phi}{u'\left(Q\right)\theta_{\max}}} D\left(\omega\right)$ as 
\begin{align}
\nonumber
\lim_{\omega\searrow \frac{\Phi}{u'\left(Q\right)\theta_{\max}}} D\left(\omega\right) = NQ \int_{\frac{F}{u\left(Q\right)}}^{\theta_{\max}} g\left(\theta\right) d\theta.
\end{align}
From (\ref{SM:equ:D:a}), we can see that this equals $D\left(\omega\right)|_{\omega=\frac{\Phi}{u'\left(Q\right)\theta_{\max}}}$. Based on (\ref{SM:equ:D:a}), we can also see that $\lim_{\omega\nearrow \frac{\Phi}{u'\left(Q\right)\theta_{\max}}} D\left(\omega\right)=D\left(\omega\right)|_{\omega=\frac{\Phi}{u'\left(Q\right)\theta_{\max}}}$. Hence, $D\left(\omega\right)$ is continuous at $\omega=\frac{\Phi}{u'\left(Q\right)\theta_{\max}}$. 

Then, we analyze the continuity of $D\left(\omega\right)$ at $\omega=\frac{\Phi u\left(Q\right)}{Fu'\left(Q\right)}$. Recall that $\theta_0=\frac{F}{u\left(Q\right)}$ and $\theta_1=\frac{\Phi}{\omega u'\left(Q\right)}$. We can see that $\lim_{\omega \searrow\frac{\Phi u\left(Q\right)}{Fu'\left(Q\right)}} \theta_1=\theta_0$. According to the definition of $\theta_2$, we have $\theta_2\in\left(\theta_1,\theta_0\right)$. Hence, we have the relation that $\lim_{\omega \searrow\frac{\Phi u\left(Q\right)}{Fu'\left(Q\right)}} \theta_2=
\lim_{\omega \searrow\frac{\Phi u\left(Q\right)}{Fu'\left(Q\right)}} \theta_1=\theta_0$. Based on this relation, (\ref{SM:equ:D:b}), and (\ref{SM:equ:D:c}), we can derive the following equation:
\begin{align}
\lim_{\omega \searrow\frac{\Phi u\left(Q\right)}{Fu'\left(Q\right)}} D\left(\omega\right)=D\left(\omega\right)|_{\omega=\frac{\Phi u\left(Q\right)}{Fu'\left(Q\right)}}.
\end{align}
From (\ref{SM:equ:D:b}), we can also see that $\lim_{\omega \nearrow\frac{\Phi u\left(Q\right)}{Fu'\left(Q\right)}} D\left(\omega\right)=D\left(\omega\right)|_{\omega=\frac{\Phi u\left(Q\right)}{Fu'\left(Q\right)}}$. Therefore,  $D\left(\omega\right)$ is continuous at $\omega=\frac{\Phi u\left(Q\right)}{Fu'\left(Q\right)}$.

The above analysis shows that $D\left(\omega\right)$ is continuous for $\omega\ge0$.

{\bf Step 3:} We prove that $\lim_{\omega\rightarrow \infty} D\left(\omega\right)=\infty$. 

First, we analyze $\lim_{\omega\rightarrow \infty} \left(u'\right)^{-1} \left(\frac{\Phi}{\omega \theta}\right)$. Recall that $u\left(\cdot\right)$ is strictly concave and $\lim_{z\rightarrow\infty} u'\left(z\right)=0$. We can see that as $z$ increases from $0$ to $\infty$, the value of $u'\left(z\right)$ strictly decreases from $u'\left(0\right)$ to $0$. Hence, as ${\hat z}$ increases from $0$ to $u'\left(0\right)$, the value of $\left(u'\right)^{-1}\left({\hat z}\right)$ strictly decreases from $\infty$ to $0$. Hence, we have $\lim_{{\hat z}\rightarrow 0} \left(u'\right)^{-1}\left({\hat z}\right)=\infty$. This implies the following relation:
\begin{align}
\lim_{\omega\rightarrow \infty} \left(u'\right)^{-1} \left(\frac{\Phi}{\omega \theta}\right)=\infty.\label{SM:equ:D:inf}
\end{align}

When $\omega>\frac{\Phi u\left(Q\right)}{Fu'\left(Q\right)}$, we can derive the following inequality from (\ref{SM:equ:D:c}):
\begin{align}
\nonumber
D\left(\omega\right)>N  \int_{\theta_0}^{\theta_{\max}} \left(\left(u'\right)^{-1}\left(\frac{\Phi}{\omega\theta}\right)-Q\right) g\left(\theta\right)d\theta.\label{SM:inequ:D}
\end{align}
Recall that $\theta_0=\frac{F}{u\left(Q\right)}$, which is independent of $\omega$. From (\ref{SM:equ:D:inf}), we can see that when $\omega\rightarrow\infty$, the right side of (\ref{SM:inequ:D}) approaches infinity. Therefore, we have $\lim_{\omega\rightarrow \infty} D\left(\omega\right)=\infty$.

So far, we have shown that (i) $D\left(\omega\right)$ is continuous in $\omega\in\left[0,\infty\right)$; (ii) $D\left(\omega\right)$ does not change with $\omega$ for $\omega\in\left[0,\frac{\Phi}{u'\left(Q\right)\theta_{\max}}\right]$, and strictly increases with $\omega$ for $\omega\in\left(\frac{\Phi}{u'\left(Q\right)\theta_{\max}},\infty\right)$; (iii) when $\omega$ approaches infinity, $D\left(\omega\right)$ approaches infinity. Therefore, we can conclude that given $C\ge D\left(0\right)$, there is a unique $\omega\in\left[\frac{\Phi}{u'\left(Q\right)\theta_{\max}},\infty\right)$ such that $D\left(\omega\right)=C$. Furthermore, this $\omega$ strictly increases with $C$.
\end{proof}

\section{Proof of Theorem \ref{theorem:subscriptionaware}}\label{appendix:proofoftheorem1}
\begin{proof}
{\bf Step 1:} We analyze the case where $C=D\left(0\right)$. From constraint $D\left(\omega\right)\le C$, we can see that $\omega$ needs to satisfy $0\le\omega\le\frac{\Phi}{u'\left(Q\right)\theta_{\max}}$. Based on Proposition \ref{proposition:stageII:user}, no user watches ads in this case. As a result, the value of $N^{\rm ad}\left(\omega\right)$ is zero. From Proposition \ref{proposition:advertiser}, the value of $m^*\left(\omega,p\right)$ is also zero. Based on (\ref{equ:obj})-(\ref{equ:capacity:ad}), we can easily see that the operator's problem of deciding $\omega$ becomes:
\begin{align}
\max_{\omega\in\left[0,\frac{\Phi}{u'\left(Q\right)\theta_{\max}}\right]} R^{\rm data}\left(\omega\right),
\end{align}
where $R^{\rm data}\left(\omega\right)$ equals $NF\int_{\frac{F}{u\left(Q\right)}}^{\theta_{\max}} g\left(\theta\right)d\theta$ and is independent of $\omega$. Therefore, when $C=D\left(0\right)$, any $\omega$ from set $\left[0,\frac{\Phi}{u'\left(Q\right)\theta_{\max}}\right]$ is optimal. 

Next, we show that when $C=D\left(0\right)$, the value of $D^{-1}\left(C\right)$ is in the set $\left[0,\frac{\Phi}{u'\left(Q\right)\theta_{\max}}\right]$. According to the definition of $D^{-1}\left(C\right)$ in Proposition \ref{proposition:monotonicity}, $D^{-1}\left(C\right)$ takes the value from set $\left[\frac{\Phi}{u'\left(Q\right)\theta_{\max}},\infty\right)$. Since $D\left(0\right)=D\left(\frac{\Phi}{u'\left(Q\right)\theta_{\max}}\right)$, when $C=D\left(0\right)$, the value of $D^{-1}\left(C\right)$ is $\frac{\Phi}{u'\left(Q\right)\theta_{\max}}$. Therefore, the value of $D^{-1}\left(C\right)$ is in the set $\left[0,\frac{\Phi}{u'\left(Q\right)\theta_{\max}}\right]$, and $\omega=D^{-1}\left(C\right)$ is one optimal solution to the operator's problem. 

{\bf Step 2:} We analyze the case where $C>D\left(0\right)$. First, we prove that $\omega^*$ should lie in the interval $\left(\frac{\Phi}{u'\left(Q\right)\theta_{\max}},D^{-1}\left(C\right)\right]$. If $\omega\in\left[0,\frac{\Phi}{u'\left(Q\right)\theta_{\max}}\right]$, no user watches ads (from our analysis in {\bf Step 1}). The operator's revenue only consists of the revenue from the data market, and the value is $NF\int_{\frac{F}{u\left(Q\right)}}^{\theta_{\max}} g\left(\theta\right)d\theta$. If $\omega\in\left(\frac{\Phi}{u'\left(Q\right)\theta_{\max}},D^{-1}\left(C\right)\right]$, the operator can choose $p^*\left(\omega\right)=\max\left\{\frac{B}{2},B-\frac{2A {\mathbb E}\left[y^2\right]}{K {\mathbb E}\left[y\right]}\right\}$ according to Theorem \ref{theorem:SAR:price}. Then, we can easily verify that the operator's revenue from the data market is no less than $NF\int_{\frac{F}{u\left(Q\right)}}^{\theta_{\max}} g\left(\theta\right)d\theta$ and its ad revenue is positive. As a result, when $C>D\left(0\right)$, the value of $\omega^*$ should be in the interval $\left(\frac{\Phi}{u'\left(Q\right)\theta_{\max}},D^{-1}\left(C\right)\right]$.

Since $\omega^*\in\left(\frac{\Phi}{u'\left(Q\right)\theta_{\max}},D^{-1}\left(C\right)\right]$, we can utilize Proposition \ref{proposition:advertiser} to simplify the operator's optimization problem as follows:

\begin{align}
& \max R^{\rm data}\left(\omega\right)+K p\frac{B-p}{2A} \frac{\left({\mathbb E}\left[y\right]\right)^2}{{\mathbb E}\left[y^2\right]} N^{\rm ad}\left(\omega\right)\\
& {~~}{~~}{\rm s.t.}{~~} K \frac{B-p}{2A} \frac{\left({\mathbb E}\left[y\right]\right)^2}{{\mathbb E}\left[y^2\right]} N^{\rm ad}\left(\omega\right) \le {\mathbb E}\left[y\right] N^{\rm ad}\left(\omega\right) \label{SM:equ:opt:con}\\
& {~~}{~~}{\rm var.}{~~} \omega\in\left(\frac{\Phi}{u'\left(Q\right)\theta_{\max}},D^{-1}\left(C\right)\right],p>0.
\end{align}

We prove $\omega^*=D^{-1}\left(C\right)$ by contradiction. Suppose that $\omega^*<D^{-1}\left(C\right)$ and the corresponding ad price is $p^*\left(\omega^*\right)$. Next, we prove that we can find $\left(\omega,p\right)$ that is feasible and generates a higher total revenue than $\left(\omega^*,p^*\left(\omega^*\right)\right)$. 

Since $\omega^*<D^{-1}\left(C\right)$, we let $\tilde \omega$ be a value in $\left(\omega^*,D^{-1}\left(C\right)\right]$. When the unit data reward is $\tilde \omega$, the number of users watching ads is $N^{\rm ad}\left({\tilde \omega}\right)$. We use a random variable $\tilde y$ to denote the number of ads watched by one of these $N^{\rm ad}\left({\tilde \omega}\right)$ users under the unit data reward $\tilde \omega$. Then, we can choose $\tilde p$ to satisfy the following equation:
\begin{align}
\frac{B-{\tilde p}}{2A} \frac{\left({\mathbb E}\left[{\tilde y}\right]\right)^2}{{\mathbb E}\left[{\tilde y}^2\right]} N^{\rm ad}\left({\tilde \omega}\right)=\frac{B-p^*\left(\omega^*\right)}{2A} \frac{\left({\mathbb E}\left[y^*\right]\right)^2}{{\mathbb E}\left[\left(y^*\right)^2\right]} N^{\rm ad}\left(\omega^*\right).\label{SM:equ:fe3}
\end{align}
Here, we use a random variable $y^*$ to denote the number of ads watched by one of the users choosing to watch ads under the unit data reward $\omega^*$. 

Next, we prove that $\left({\tilde \omega},{\tilde p}\right)$ satisfies constraint (\ref{SM:equ:opt:con}). Based on the feasibility of solution $\left(\omega^*,p^*\left(\omega^*\right)\right)$, the following inequality holds:
\begin{align}
K \frac{B-p^*\left(\omega^*\right)}{2A} \frac{\left({\mathbb E}\left[y^*\right]\right)^2}{{\mathbb E}\left[\left(y^*\right)^2\right]} N^{\rm ad}\left(\omega^*\right) \le {\mathbb E}\left[y^*\right]N^{\rm ad}\left(\omega^*\right).\label{SM:inequ:fe1}
\end{align}
One condition of Theorem \ref{theorem:subscriptionaware} is that ${\mathbb E}\left[y\right]N^{\rm ad}\left(\omega\right)$ increases with $\omega$ for $\omega>\frac{\Phi}{u'\left(Q\right)\theta_{\max}}$. From ${\tilde \omega}>{\omega^*}$, we have the following relation:
\begin{align}
{\mathbb E}\left[\tilde y\right]N^{\rm ad}\left(\tilde \omega\right)> {\mathbb E}\left[y^*\right]N^{\rm ad}\left(\omega^*\right).\label{SM:inequ:fe2}
\end{align}
Based on (\ref{SM:inequ:fe1}) and (\ref{SM:inequ:fe2}), we have the following result:
\begin{align}
K \frac{B-p^*\left(\omega^*\right)}{2A} \frac{\left({\mathbb E}\left[y^*\right]\right)^2}{{\mathbb E}\left[\left(y^*\right)^2\right]} N^{\rm ad}\left(\omega^*\right) < {\mathbb E}\left[\tilde y\right]N^{\rm ad}\left(\tilde \omega\right).
\end{align}
From the above inequality and (\ref{SM:equ:fe3}), we can derive the following relation:
\begin{align}
K \frac{B-{\tilde p}}{2A} \frac{\left({\mathbb E}\left[{\tilde y}\right]\right)^2}{{\mathbb E}\left[{\tilde y}^2\right]} N^{\rm ad}\left({\tilde \omega}\right)< {\mathbb E}\left[\tilde y\right]N^{\rm ad}\left(\tilde \omega\right).
\end{align}
This implies that $\left({\tilde \omega},{\tilde p}\right)$ satisfies constraint (\ref{SM:equ:opt:con}).

Then, we prove that $\left({\tilde \omega},{\tilde p}\right)$ generates a larger objective value than $\left(\omega^*,{p^*}\left(\omega^*\right)\right)$. One condition of Theorem \ref{theorem:subscriptionaware} is that $\frac{\left({\mathbb E}\left[y\right]\right)^2}{{\mathbb E}\left[y^2\right]} N^{\rm ad}\left(\omega\right)$ increases with $\omega$ for $\omega>\frac{\Phi}{u'\left(Q\right)\theta_{\max}}$. From ${\tilde \omega}>{\omega^*}$, we have the following relation:
\begin{align}
\frac{\left({\mathbb E}\left[\tilde y\right]\right)^2}{{\mathbb E}\left[{\tilde y}^2\right]} N^{\rm ad}\left(\tilde \omega\right)> \frac{\left({\mathbb E}\left[y^*\right]\right)^2}{{\mathbb E}\left[\left(y^*\right)^2\right]} N^{\rm ad}\left(\omega^*\right).
\end{align}
Based on this inequality and (\ref{SM:equ:fe3}), we can see that
\begin{align}
{\tilde p}> p^*\left(\omega^*\right).
\end{align}
From ${\tilde p}> p^*\left(\omega^*\right)$ and (\ref{SM:equ:fe3}), we can see that the ad revenue under $\left({\tilde \omega},{\tilde p}\right)$ is greater than that under $\left(\omega^*,{p^*}\left(\omega^*\right)\right)$:
\begin{align}
\nonumber
& K {\tilde p}\frac{B-{\tilde p}}{2A} \frac{\left({\mathbb E}\left[\tilde y\right]\right)^2}{{\mathbb E}\left[{\tilde y}^2\right]} N^{\rm ad}\left(\tilde \omega\right)>\\
& K {p^*}\left(\omega^*\right)\frac{B-{p^*}\left(\omega^*\right)}{2A} \frac{\left({\mathbb E}\left[y^*\right]\right)^2}{{\mathbb E}\left[\left(y^*\right)^2\right]} N^{\rm ad}\left(\omega^*\right).\label{SM:equ:fe4}
\end{align}
Moreover, from Proposition \ref{proposition:stageII:user} and ${\tilde \omega}>{\omega^*}$, the number of subscribers under unit reward ${\tilde \omega}$ is no less than the number of subscribers under unit reward $\omega^*$. This implies that $R^{\rm data}\left(\tilde \omega\right) \ge R^{\rm data}\left(\omega^*\right)$. Combining (\ref{SM:equ:fe4}) and $R^{\rm data}\left(\tilde \omega\right) \ge R^{\rm data}\left(\omega^*\right)$, we can conclude that $\left({\tilde \omega},{\tilde p}\right)$ generates a larger objective value than $\left(\omega^*,{p^*}\left(\omega^*\right)\right)$.

Therefore, we have proved that (i) $\left({\tilde \omega},{\tilde p}\right)$ is feasible and (ii) $\left({\tilde \omega},{\tilde p}\right)$ generates a larger objective value than $\left(\omega^*,{p^*}\left(\omega^*\right)\right)$. This contradicts with the assumption that $\left(\omega^*,{p^*}\left(\omega^*\right)\right)$ is the optimal solution. As a result, when the two conditions in Theorem \ref{theorem:subscriptionaware} hold, $\omega^*$ should equal $D^{-1}\left(C\right)$. 

According to our results in {\bf Step 1} and {\bf Step 2}, when the two conditions in Theorem \ref{theorem:subscriptionaware} hold, the optimal unit data reward is given by $\omega^*=D^{-1}\left(C\right)$.
\end{proof}


\section{Proof of Proposition \ref{proposition:uniform:useup}}\label{appendix:sec:proposition4}
\begin{proof}
We prove that when $u\left(z\right)=\ln\left(1+z\right)$ and $\theta\sim{\mathcal U}\left[0,\theta_{\max}\right]$, both $\frac{\left({\mathbb E}\left[y\right]\right)^2}{{\mathbb E}\left[y^2\right]} N^{\rm ad}\left(\omega\right)$ and ${\mathbb E}\left[y\right] N^{\rm ad}\left(\omega\right)$ increase with $\omega$ for $\omega\in\left(\frac{\Phi}{u'\left(Q\right)\theta_{\max}},\infty\right)$.

{\bf Step 1:} We consider the case where $\omega\in\left(\frac{\Phi}{u'\left(Q\right)\theta_{\max}},\frac{\Phi u\left(Q\right)}{F u'\left(Q\right)}\right]$. Based on our analysis in Appendix \ref{appendix:details:SAR}, when $u\left(z\right)=\ln\left(1+z\right)$ and $\theta\sim{\mathcal U}\left[0,\theta_{\max}\right]$, we have the following results:
\begin{align}
& \theta_1=\frac{\Phi\left(1+Q\right)}{\omega},N^{\rm ad}\left(\omega\right)=\frac{\theta_{\max}-\theta_1}{\theta_{\max}} N,\\
& {\mathbb E}\left[y\right]=\frac{1}{2}\frac{\theta_{\max}-\theta_1}{\Phi},{\mathbb E}\left[y^2\right]=\frac{1}{3} \left(\frac{\theta_{\max}-\theta_1}{\Phi}\right)^2.
\end{align}
Hence, we can compute $\frac{\left({\mathbb E}\left[y\right]\right)^2}{{\mathbb E}\left[y^2\right]} N^{\rm ad}\left(\omega\right)$ as follows:
\begin{align}
\frac{\left({\mathbb E}\left[y\right]\right)^2}{{\mathbb E}\left[y^2\right]} N^{\rm ad}\left(\omega\right)=\frac{3}{4}\frac{\theta_{\max}-\theta_1}{\theta_{\max}} N,
\end{align}
which increases with $\omega$. Then, we can compute ${\mathbb E}\left[y\right] N^{\rm ad}\left(\omega\right)$ as follows:
\begin{align}
{\mathbb E}\left[y\right] N^{\rm ad}\left(\omega\right)=\frac{1}{2}\frac{\left(\theta_{\max}-\theta_1\right)^2}{\Phi \theta_{\max}} N,
\end{align}
which also increases with $\omega$.

{\bf Step 2:} We consider the case where $\omega\in\left(\frac{\Phi u\left(Q\right)}{F u'\left(Q\right)},\infty\right)$. Based on our analysis in Appendix \ref{appendix:details:SAR}, when $u\left(z\right)=\ln\left(1+z\right)$ and $\theta\sim{\mathcal U}\left[0,\theta_{\max}\right]$, we have the following results:
\begin{align}
& N^{\rm ad}\left(\omega\right)=\frac{\theta_{\max}-\theta_2}{\theta_{\max}} N,{\mathbb E}\left[y\right]=\frac{\theta_2-\theta_1+\theta_{\max}-\theta_1}{2\Phi},\\
& {\mathbb E}\left[y^2\right]=\left(\frac{\theta_2-\theta_1+\theta_{\max}-\theta_1}{2\Phi}\right)^2+\frac{\left(\frac{\theta_{\max}-\theta_1-\left(\theta_2-\theta_1\right)}{\Phi}\right)^2}{12}.
\end{align}
To simplify the presentation, we let $y_h\triangleq \frac{\theta_{\max}-\theta_1}{\Phi}$ and $y_l \triangleq \frac{\theta_2-\theta_1}{\Phi}$. Based on the relation $y_h^3-y_l^3=\left(y_l^2+y_h^2+y_l y_h\right)\left(y_h-y_l\right)$, we can derive the following equation:
\begin{align}
\frac{\left({\mathbb E}\left[y\right]\right)^2}{{\mathbb E}\left[y^2\right]} N^{\rm ad}\left(\omega\right)=\frac{3}{4}\frac{N\Phi}{\theta_{\max}}\frac{\left({y_h^2-y_l^2}\right)^2}{y_h^3-y_l^3}.
\end{align}

We can compute the derivative of $\frac{\left(y_h^2-y_l^2\right)^2}{y_h^3-y_l^3}$ with respect to $\omega$ as follows:
\begin{align}
\frac{d \left(\frac{\left(y_h^2-y_l^2\right)^2}{y_h^3-y_l^3}\right)}{d \omega}=\frac{-\frac{d \theta_2}{d\omega}\frac{1}{\Phi}\left(y_h y_l^2 + 4 y_h^2 y_l + y_l^3\right)+\frac{\theta_1}{\omega\Phi}\left(y_h-y_l\right)^3}{\left(y_h^2+y_l^2+y_hy_l\right)^2}.
\end{align}
From our analysis in Appendix \ref{appendix:monotonicity:theta2}, we have $\frac{d \theta_2}{d\omega}<0$. Moreover, we have $y_h>y_l$. We can see that the above derivative is positive. Therefore, $\frac{\left({\mathbb E}\left[y\right]\right)^2}{{\mathbb E}\left[y^2\right]} N^{\rm ad}\left(\omega\right)$ increases with $\omega$.

Then, we can compute ${\mathbb E}\left[y\right] N^{\rm ad}\left(\omega\right)$ as follows:
\begin{align}
{\mathbb E}\left[y\right] N^{\rm ad}\left(\omega\right)=\frac{\theta_2-\theta_1+\theta_{\max}-\theta_1}{2\Phi} \frac{\theta_{\max}-\theta_2}{\theta_{\max}} N.
\end{align}
The derivative with respect to $\omega$ is computed as
\begin{align}
\nonumber
&\frac{d \left({\mathbb E}\left[y\right] N^{\rm ad}\left(\omega\right)\right)}{d \omega} =\\
\nonumber
&\frac{N}{2\Phi \theta_{\max}}\left(\frac{d \theta_2}{d\omega}\theta_{\max} -2\frac{d\theta_1}{d\omega}\theta_{\max}-\frac{d \theta_2}{d\omega}\theta_2 +2\frac{d\theta_1}{d\omega}\theta_2  \right)\\
\nonumber
&+\frac{N}{2\Phi \theta_{\max}}\left( -\frac{d\theta_2}{d\omega}\theta_2  -\frac{d\theta_2}{d\omega}\theta_{\max} +2\frac{d\theta_2}{d\omega}\theta_1 \right) \\
& =\frac{N}{2\Phi \theta_{\max}}\left( 2\left(\theta_2-\theta_{\max}\right)\frac{d\theta_1}{d\omega}+2\left(\theta_1-\theta_2\right)\frac{d\theta_2}{d\omega} \right).
\end{align}
Since $\theta_{\max}>\theta_2>\theta_1$, $\frac{d\theta_1}{d\omega}<0$, and $\frac{d\theta_2}{d\omega}<0$, we can see that ${\mathbb E}\left[y\right] N^{\rm ad}\left(\omega\right)$ increases with $\omega$.

Combing {\bf Step 1} and {\bf Step 2}, we can see that when $u\left(z\right)=\ln\left(1+z\right)$ and $\theta\sim{\mathcal U}\left[0,\theta_{\max}\right]$, both $\frac{\left({\mathbb E}\left[y\right]\right)^2}{{\mathbb E}\left[y^2\right]} N^{\rm ad}\left(\omega\right)$ and ${\mathbb E}\left[y\right] N^{\rm ad}\left(\omega\right)$ increase with $\omega$ for $\omega\in\left(\frac{\Phi}{u'\left(Q\right)\theta_{\max}},\infty\right)$. According to Theorem \ref{theorem:subscriptionaware}, the optimal unit data reward is given by $\omega^*=D^{-1}\left(C\right)$.
\end{proof}

\section{Example where ${\mathbb E}\left[y\right] N^{\rm ad}\left(\omega\right)$ decreases with $\omega$}\label{appendix:numericalexample}

{
We use a numerical example to show that when each user has an exponential utility function, it is possible that ${\mathbb E}\left[y\right] N^{\rm ad}\left(\omega\right)$ (i.e., the number of available ad slots) decreases with $\omega$ and $\omega^*<D^{-1}\left(C\right)$.

We assume that $u\left(z\right)=1-e^{-\gamma z}$, and obtain the distribution of $\theta$ by truncating the normal distribution ${\cal N}\left(30,60\right)$ to interval $\left[0,320\right]$. We choose $\gamma=0.95$, $N=10^7$, $F=40$, $Q=2$, $\Phi=0.5$, $K=16$, $A=0.9$, $B=5$, and $C=2.15\times10^7$. 

In Fig. \ref{appendix:fig:example}, we plot the value of ${\mathbb E}\left[y\right] N^{\rm ad}\left(\omega\right)$ against $\omega$ (under the SAR scheme). We can see that ${\mathbb E}\left[y\right] N^{\rm ad}\left(\omega\right)$ is decreasing in $\omega\in\left[0.117,0.217\right]$. Moreover, we can numerically compute that $\omega^*=0.137$ and $D\left(\omega^*\right)=1.846\times 10^7$. Since $D\left(\omega^*\right)$ is smaller than $C$ and $D\left(\omega\right)$ increases with $\omega$ under the SAR scheme, we can see that $\omega^*<D^{-1}\left(C\right)$.
}

\begin{figure}[t]
  \centering
  \includegraphics[scale=0.4]{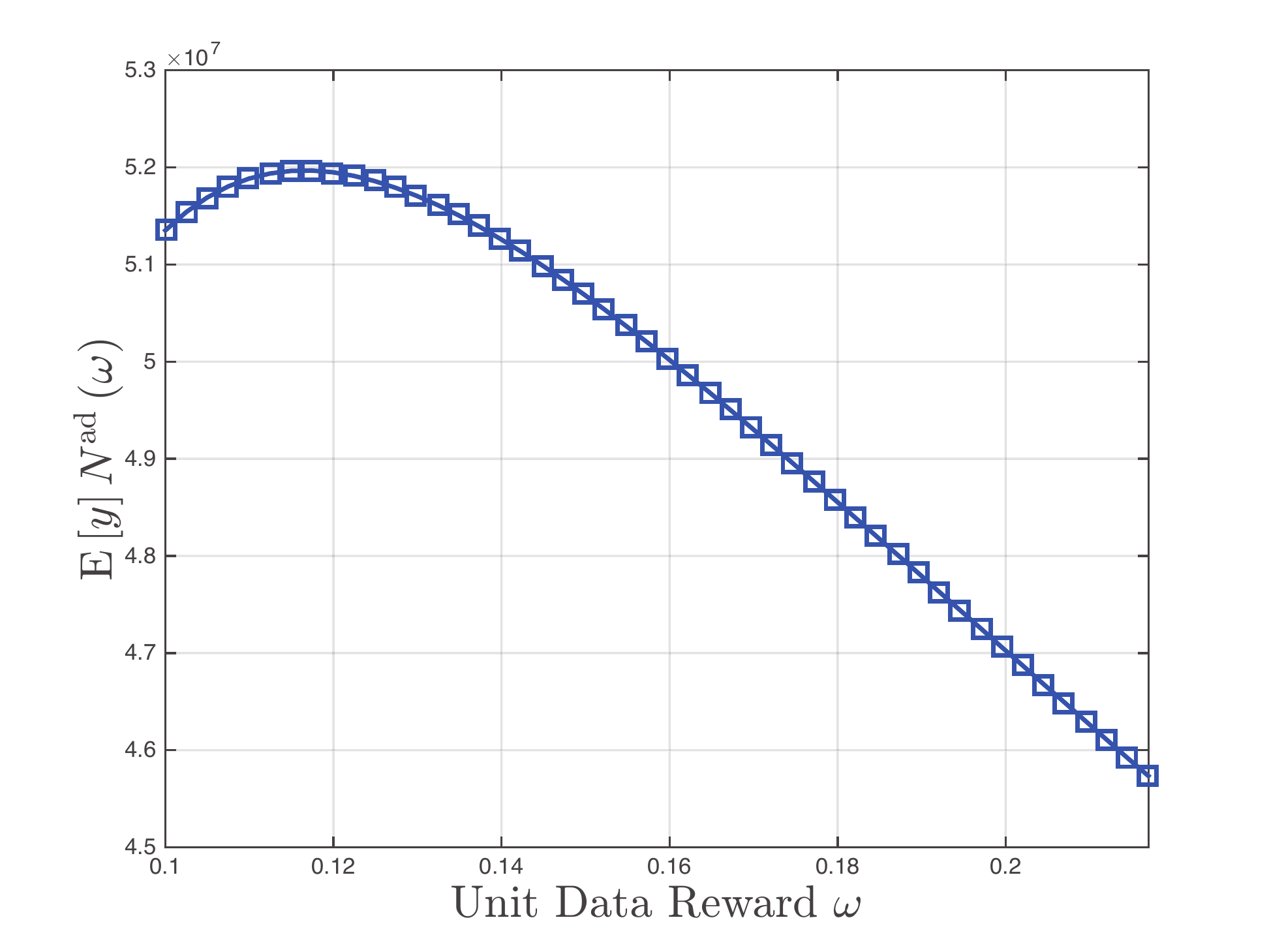}\\
  \caption{Impact of $\omega$ on ${\mathbb E}\left[y\right] N^{\rm ad}\left(\omega\right)$.}
  \label{appendix:fig:example}
\end{figure}

\section{Proof of Lemma \ref{lemma:theta3}}\label{appendix:uniquetheta4}


\begin{proof}

Let $v\left(\theta\right)=\theta u\left( \left(u'\right)^{-1}\left(\frac{\Phi}{\omega\theta}\right) \right)- \frac{\Phi}{\omega} \left(u'\right)^{-1}\left(\frac{\Phi}{\omega\theta}\right)-\theta u\left(Q\right)+F$. First, since $u'\left(\left(u'\right)^{-1}\left(\frac{\Phi}{\omega \theta}\right)\right)=\frac{\Phi}{\omega \theta}$, we can compute $\frac{d v\left(\theta\right)}{d\theta}$ as
\begin{align}
\frac{d v\left(\theta\right)}{d\theta}=u\left( \left(u'\right)^{-1}\left(\frac{\Phi}{\omega\theta}\right) \right)-u\left(Q\right).
\end{align}
Recall that $u\left(\cdot\right)$ is increasing and $\left(u'\right)^{-1}\left(\cdot\right)$ is decreasing. When $\theta=\theta_1=\frac{\Phi}{\omega u'\left(Q\right)}$, we have $\frac{d v\left(\theta\right)}{d\theta}=0$; when $\theta<\theta_1$, we can see that $\frac{d v\left(\theta\right)}{d\theta}<\frac{d v\left(\theta\right)}{d\theta}|_{\theta=\theta_1}=0$. 

Second, since $\theta_3=\frac{\Phi}{\omega u'\left(0\right) }$, we can compute $v\left(\theta_3\right)$ as follows:
\begin{align}
v\left(\theta_3\right)=-\frac{\Phi}{\omega u'\left(0\right) } u\left(Q\right)+F.
\end{align}
When $\omega>\frac{\Phi u\left(Q\right)}{F u'\left(0\right)}$, we can see that $v\left(\theta_3\right)>0$. 

Third, we compute $v\left(\theta_1\right)$ as follows:
\begin{align}
v\left(\theta_1\right)=- \frac{\Phi}{\omega} Q+F.
\end{align}
When $\omega<\frac{\Phi Q}{F}$, we can see that $v\left(\theta_1\right)<0$. 

Hence, when $\omega\in\left(\frac{\Phi u\left(Q\right)}{F u'\left(0\right)},\frac{\Phi Q}{F}\right)$, we have $\frac{d v\left(\theta\right)}{d\theta}<0$ for $\theta<\theta_1$, $v\left(\theta_3\right)>0$, and $v\left(\theta_1\right)<0$. Moreover, we have $\frac{d v\left(\theta\right)}{d\theta}|_{\theta=\theta_1}=0$. We can conclude that there exists a unique $\theta_4\in\left(\theta_3,\theta_1\right)$ satisfying $v\left(\theta_4\right)=0$.
\end{proof}

\section{Proof of Proposition \ref{proposition:unaware:user}}\label{appendix:proposition41}
\begin{proof}
{\bf Step 1:} We analyze Case $\hat A$, i.e., $\omega\in\left[0,\frac{\Phi}{u'\left(Q\right)\theta_{\max}}\right]$. 
Suppose a type-$\theta$ user subscribes to the data plan, i.e., $r=1$. Its payoff is $\Pi^{\rm user}\left(\theta,1,x,\omega\right)=\theta u\left(Q+\omega x\right)-F-\Phi x$. We can see that
\begin{align}
\nonumber
 \frac{\partial \Pi^{\rm user}\left(\theta,1,x,\omega\right)}{\partial x}&= \theta \omega u'\left(Q+\omega x\right)-\Phi \\
& \le \frac{  u'\left(Q+\omega x\right)\theta}{u'\left(Q\right) \theta_{\max}}\Phi-\Phi.
\end{align}
Since $u'\left(\cdot\right)$ is a strictly decreasing function, we have $u'\left(Q+\omega x\right)<u'\left(Q\right)$ for $x>0$. Hence, $\frac{\partial \Pi^{\rm user}\left(\theta,1,x,\omega\right)}{\partial x}<0$ for $x>0$. This implies that if a user subscribes to the data plan, it will not watch any ad for rewards. In this case, the user's payoff is $\theta u\left(Q\right)-F$. 

Suppose a type-$\theta$ user does not subscribe, i.e., $r=0$. Its payoff is $\Pi^{\rm user}\left(\theta,0,x,\omega\right)=\theta u\left(\omega x\right)-\Phi x$. We can see that 
\begin{align}
\nonumber
\frac{\partial \Pi^{\rm user}\left(\theta,0,x,\omega\right)}{\partial x}=\omega\theta u'\left(\omega x\right)-\Phi.
\end{align}
Hence, if $\theta\in\left[\frac{\Phi}{\omega u'\left(0\right)},\theta_{\max}\right]$, the user will watch ads with $x=\frac{1}{\omega}\left(u'\right)^{-1}\left(\frac{\Phi}{\omega\theta}\right)$, and its payoff will be $\theta u\left(\left(u'\right)^{-1}\left(\frac{\Phi}{\omega\theta}\right)\right)- \frac{\Phi}{\omega}\left(u'\right)^{-1}\left(\frac{\Phi}{\omega\theta}\right)$; otherwise, the user will not watch ads, and its payoff will be zero. 

Next, we compare the choices of $r=1$ and $r=0$. Recall that we assume $\theta_{\max}>\frac{u'\left(0\right)F}{u'\left(Q\right)u\left(Q\right)}$ in Section \ref{subsec:gamestructure}. Hence, when $\omega\le \frac{\Phi}{u'\left(Q\right)\theta_{\max}}$, we have
\begin{align}
\frac{\Phi}{\omega u'\left(0\right)}\ge \frac{u'\left(Q\right)\theta_{\max}}{u'\left(0\right)}>\frac{F}{u\left(Q\right)}.
\end{align}
Consider a user with $\theta\in\left[0,\frac{F}{u\left(Q\right)}\right)$. As we discussed above, if it subscribes, it will not watch ads and its payoff will be $\theta u\left(Q\right)-F$. Since $\theta<\frac{F}{u\left(Q\right)}$, this payoff is negative. If it does not subscribe, since $\theta<\frac{F}{u\left(Q\right)}<\frac{\Phi}{\omega u'\left(0\right)}$, it will not watch ads and its payoff will be zero. Comparing $r=0$ and $r=1$, this user will not subscribe or watch ads, i.e., ${\hat r}^*\left(\theta,\omega\right)=0$ and ${\hat x}^*\left(\theta,\omega\right)=0$. 

Consider a user with $\theta\in\left[\frac{F}{u\left(Q\right)},\frac{\Phi}{\omega u'\left(0\right)}\right)$. As we discussed above, if it subscribes, it will not watch ads and its payoff will be $\theta u\left(Q\right)-F$. Since $\theta\ge\frac{F}{u\left(Q\right)}$, this payoff is non-negative. If it does not subscribe, since $\theta<\frac{\Phi}{\omega u'\left(0\right)}$, it will not watch ads and its payoff will be zero. Comparing $r=0$ and $r=1$, this user will subscribe without watching any ad, i.e., ${\hat r}^*\left(\theta,\omega\right)=1$ and ${\hat x}^*\left(\theta,\omega\right)=0$. 

Consider a user with $\theta\in\left[\frac{\Phi}{\omega u'\left(0\right)},\theta_{\max}\right]$. As we discussed above, if it subscribes, it will not watch ads and its payoff will be $\theta u\left(Q\right)-F$. If it does not subscribe, it will watch ads and its payoff will be $\theta u\left(\left(u'\right)^{-1}\left(\frac{\Phi}{\omega\theta}\right)\right)- \frac{\Phi}{\omega}\left(u'\right)^{-1}\left(\frac{\Phi}{\omega\theta}\right)$. Next, we compare these two payoffs. We define 
\begin{align}
s\left(\theta\right)\triangleq \theta u\left(Q\right)-F-\theta u\left(\left(u'\right)^{-1}\left(\frac{\Phi}{\omega\theta}\right)\right)+ \frac{\Phi}{\omega}\left(u'\right)^{-1}\left(\frac{\Phi}{\omega\theta}\right).\label{SM:equ:pro5:caseB}
\end{align}
From $u'\left(\left(u'\right)^{-1}\left(\frac{\Phi}{\omega \theta}\right)\right)=\frac{\Phi}{\omega \theta}$, we can compute the derivative of $s\left(\theta\right)$ as follows:
\begin{align}
\frac{d s\left(\theta\right)}{d\theta}= u\left(Q\right) -  u\left(\left(u'\right)^{-1}\left(\frac{\Phi}{\omega\theta}\right)\right).
\end{align}
Since $\left(u'\right)^{-1}\left(\cdot\right)$ is decreasing and $\omega\le\frac{\Phi}{u'\left(Q\right)\theta_{\max}}\le \frac{\Phi}{u'\left(Q\right)\theta}$, we can see that $\frac{d s\left(\theta\right)}{d\theta}\ge0$. Moreover, we can compute $s\left(\frac{\Phi}{\omega u'\left(0\right)}\right)$ as follows:
\begin{align}
s\left(\frac{\Phi}{\omega u'\left(0\right)}\right)=\frac{\Phi}{\omega u'\left(0\right)}u\left(Q\right)-F.
\end{align}
When $\omega\le \frac{\Phi}{u'\left(Q\right)\theta_{\max}}$, the value of $s\left(\frac{\Phi}{\omega u'\left(0\right)}\right)$ is no less than $\frac{ u'\left(Q\right)\theta_{\max}}{ u'\left(0\right)}u\left(Q\right)-F$. Based on $\theta_{\max}>\frac{u'\left(0\right)F}{u'\left(Q\right)u\left(Q\right)}$, we further have the relation that $s\left(\frac{\Phi}{\omega u'\left(0\right)}\right)>\frac{ u'\left(Q\right)}{ u'\left(0\right)}u\left(Q\right)\frac{u'\left(0\right)F}{u'\left(Q\right)u\left(Q\right)}-F=0$. From $\frac{d s\left(\theta\right)}{d\theta}\ge0$ and $s\left(\frac{\Phi}{\omega u'\left(0\right)}\right)>0$, we can see that when $\theta\ge \frac{\Phi}{\omega u'\left(0\right)}$, the value of $s\left(\theta\right)$ is positive. This implies that the user should subscribe as it achieves a higher payoff. Hence, we have ${\hat r}^*\left(\theta,\omega\right)=1$ and ${\hat x}^*\left(\theta,\omega\right)=0$. 

Combining the above three regions of $\theta$, we can conclude that (note that $\theta_0= \frac{F}{u\left(Q\right)}$)
\begin{align}
{\hat r}^*\left(\theta,\omega\right)={\mathbbm 1}_{\left\{\theta\ge \theta_0\right\}},{~~}{~~}{\hat x}^*\left(\theta,\omega\right)=0,{~~}\theta\in\left[0,\theta_{\max}\right].
\end{align}

{\bf Step 2:} We analyze Case $\hat B$, i.e., $\omega\in\left(\frac{\Phi}{u'\left(Q\right)\theta_{\max}},\frac{\Phi u\left(Q\right)}{F u'\left(0\right)}\right]$. 
Suppose a type-$\theta$ user subscribes to the data plan, i.e., $r=1$. Its payoff is $\Pi^{\rm user}\left(\theta,1,x,\omega\right)=\theta u\left(Q+\omega x\right)-F-\Phi x$. We can see that
\begin{align}
\frac{\partial \Pi^{\rm user}\left(\theta,1,x,\omega\right)}{\partial x}= \theta \omega u'\left(Q+\omega x\right)-\Phi.
\end{align}

(i) If $\theta\in\left[0,\frac{\Phi}{\omega u'\left(Q\right)}\right)$, the user will not watch ads (i.e., $x=0$), and its payoff will be $\theta u\left(Q\right)-F$;

(ii) If $\theta\in\left[\frac{\Phi}{\omega u'\left(Q\right)},\theta_{\max}\right]$, the user will choose $x=\frac{1}{\omega}\left(u'\right)^{-1}\left(\frac{\Phi}{\omega\theta}\right)-\frac{Q}{\omega}\ge0$. Then, we can see that the user's payoff is
\begin{align}
\nonumber
& \Pi^{\rm user}\left(\theta,1,\frac{1}{\omega}\left(u'\right)^{-1}\left(\frac{\Phi}{\omega\theta}\right)-\frac{Q}{\omega},\omega\right) =\\
& \theta u\left(\left(u'\right)^{-1}\left(\frac{\Phi}{\omega\theta}\right)\right)-F-\Phi \left(\frac{1}{\omega}\left(u'\right)^{-1}\left(\frac{\Phi}{\omega\theta}\right)-\frac{Q}{\omega}\right).
\end{align}

Suppose a type-$\theta$ user does not subscribe, i.e., $r=0$. Its payoff is $\Pi^{\rm user}\left(\theta,0,x,\omega\right)=\theta u\left(\omega x\right)-\Phi x$. We can see that 
\begin{align}
\nonumber
\frac{\partial \Pi^{\rm user}\left(\theta,0,x,\omega\right)}{\partial x}=\omega\theta u'\left(\omega x\right)-\Phi.
\end{align}
Hence, if $\theta\in\left[\frac{\Phi}{\omega u'\left(0\right)},\theta_{\max}\right]$, the user will watch ads with $x=\frac{1}{\omega}\left(u'\right)^{-1}\left(\frac{\Phi}{\omega\theta}\right)$, and its payoff will be $\theta u\left(\left(u'\right)^{-1}\left(\frac{\Phi}{\omega\theta}\right)\right)- \frac{\Phi}{\omega}\left(u'\right)^{-1}\left(\frac{\Phi}{\omega\theta}\right)$; otherwise, the user will not watch ads, and its payoff will be zero. 

Next, we compare the choices of $r=1$ and $r=0$. From the concavity of $u\left(\cdot\right)$ and $\lim_{z\rightarrow \infty} u'\left(z\right)=0$, we have $u'\left(0\right)>u'\left(Q\right)>0$. Hence, when $\omega\le \frac{\Phi u\left(Q\right)}{F u'\left(0\right)}$, we have $\frac{F}{u\left(Q\right)}\le\frac{\Phi}{\omega u'\left(0\right)}<\frac{\Phi}{\omega u'\left(Q\right)}$.

Consider a user with $\theta\in\left[0,\frac{F}{u\left(Q\right)}\right)$. If this user chooses $r=1$, it will not watch ads and its payoff will be $\theta u\left(Q\right)-F$, which is negative. If this user chooses $r=0$, it will not watch ads and its payoff will be zero. Hence, the user will not subscribe or watch ads, i.e., ${\hat r}^*\left(\theta,\omega\right)=0$ and ${\hat x}^*\left(\theta,\omega\right)=0$. 

Consider a user with $\theta\in\left[\frac{F}{u\left(Q\right)},\frac{\Phi}{\omega u'\left(0\right)}\right)$. If this user chooses $r=1$, it will not watch ads and its payoff will be $\theta u\left(Q\right)-F$, which is non-negative. If this user chooses $r=0$, it will not watch ads and its payoff will be zero. Hence, the user will subscribe without watching ads, i.e., ${\hat r}^*\left(\theta,\omega\right)=1$ and ${\hat x}^*\left(\theta,\omega\right)=0$. 

Consider a user with $\theta\in\left[\frac{\Phi}{\omega u'\left(0\right)},\frac{\Phi}{\omega u'\left(Q\right)}\right)$.  If this user chooses $r=1$, it will not watch ads and its payoff will be $\theta u\left(Q\right)-F$, which is non-negative. If this user chooses $r=0$, it will watch ads and its payoff will be $\theta u\left(\left(u'\right)^{-1}\left(\frac{\Phi}{\omega\theta}\right)\right)- \frac{\Phi}{\omega}\left(u'\right)^{-1}\left(\frac{\Phi}{\omega\theta}\right)$. Next we compare these two payoffs. We can see that the difference between these two payoffs is captured by $s\left(\theta\right)$ defined in (\ref{SM:equ:pro5:caseB}). Based on our analysis in {\bf Step 1}, we can see that when $\theta<\frac{\Phi}{\omega u'\left(Q\right)}$, we have $\frac{d s\left(\theta\right)}{d\theta}>0$ and $s\left(\frac{\Phi}{\omega u'\left(0\right)}\right)=\frac{\Phi}{\omega u'\left(0\right)}u\left(Q\right)-F$. Because $\omega\le \frac{\Phi u\left(Q\right)}{F u'\left(0\right)}$ in Case $\hat B$, we can see that $s\left(\frac{\Phi}{\omega u'\left(0\right)}\right)\ge0$. Therefore, $s\left(\theta\right)$ is non-negative for $\theta\in\left[\frac{\Phi}{\omega u'\left(0\right)},\frac{\Phi}{\omega u'\left(Q\right)}\right)$. This implies that the user should subscribe without watching ads, i.e., ${\hat r}^*\left(\theta,\omega\right)=1$ and ${\hat x}^*\left(\theta,\omega\right)=0$.

Consider a user with $\theta\in\left[\frac{\Phi}{\omega u'\left(Q\right)},\theta_{\max}\right]$. If this user chooses $r=1$, it will watch ads and its payoff will be $\theta u\left(\left(u'\right)^{-1}\left(\frac{\Phi}{\omega\theta}\right)\right)-F-\Phi \left(\frac{1}{\omega}\left(u'\right)^{-1}\left(\frac{\Phi}{\omega\theta}\right)-\frac{Q}{\omega}\right)$. If this user chooses $r=0$, it will watch ads and its payoff will be $\theta u\left(\left(u'\right)^{-1}\left(\frac{\Phi}{\omega\theta}\right)\right)- \frac{\Phi}{\omega}\left(u'\right)^{-1}\left(\frac{\Phi}{\omega\theta}\right)$. Next we compare this two payoffs. We can see that the difference between these two payoffs is $-F+\frac{\Phi Q}{\omega}$. We first analyze the value of $u\left(Q\right)-u'\left(0\right) Q$. We can easily see that the following equation holds:
\begin{align}
u\left(Q\right)-u'\left(0\right) Q=u\left(0\right)+\int_{0}^{Q} u'\left(q\right) dq -\int_{0}^{Q} u'\left(0\right) dq.
\end{align}
From $u\left(0\right)=0$ and the strict concavity of $u\left(\cdot\right)$, we have $u\left(Q\right)-u'\left(0\right) Q<0$. Then, since $\omega\le \frac{\Phi u\left(Q\right)}{F u'\left(0\right)}$ in Case $\hat B$, we have $\omega< \frac{\Phi Q}{F}$. Recall that the difference in the two payoffs is $\frac{\Phi Q}{\omega}-F$. We can see that the user can get a higher payoff by subscription and watching ads, i.e., ${\hat r}^*\left(\theta,\omega\right)=1$ and ${\hat x}^*\left(\theta,\omega\right)=\frac{1}{\omega}\left(u'\right)^{-1}\left(\frac{\Phi}{\omega\theta}\right)-\frac{Q}{\omega}$.

Combining the above four regions of $\theta$, we can conclude that the following result holds for $\theta\in\left[0,\theta_{\max}\right]$ (recall that $\theta_0= \frac{F}{u\left(Q\right)}$ and $\theta_1= \frac{\Phi}{\omega u'\left(Q\right)}$):
\begin{align}
{\hat r}^*\!\left(\theta,\omega\right)\!=\!\!{\mathbbm 1}_{\left\{\theta\ge \theta_0\right\}},\!{\hat x}^*\!\left(\theta,\omega\right)\!\!=\!\!\frac{1}{\omega}\!\left(\!\left(u'\right)^{-1}\!\!\left(\frac{\Phi}{\omega\theta}\right)\!-\!Q\!\right) \!\!{\mathbbm 1}_{\!\left\{\theta\ge \theta_1\!\right\}}.
\end{align}

{\bf Step 3:} We analyze Case $\hat C$, i.e., $\omega\in\left(\frac{\Phi u\left(Q\right)}{F u'\left(0\right)},\frac{\Phi Q}{F}\right)$. 

Suppose a type-$\theta$ user subscribes to the data plan, i.e., $r=1$. Based on our prior analysis in {\bf Step 2}, we have the following results:

(i) If $\theta\in\left[0,\frac{\Phi}{\omega u'\left(Q\right)}\right)$, the user will not watch ads (i.e., $x=0$), and its payoff will be $\theta u\left(Q\right)-F$;

(ii) If $\theta\in\left[\frac{\Phi}{\omega u'\left(Q\right)},\theta_{\max}\right]$, the user will choose $x=\frac{1}{\omega}\left(u'\right)^{-1}\left(\frac{\Phi}{\omega\theta}\right)-\frac{Q}{\omega}$, and its payoff will be $\theta u\left(\left(u'\right)^{-1}\left(\frac{\Phi}{\omega\theta}\right)\right)-F-\Phi \left(\frac{1}{\omega}\left(u'\right)^{-1}\left(\frac{\Phi}{\omega\theta}\right)-\frac{Q}{\omega}\right)$.

Suppose a type-$\theta$ user does not subscribe, i.e., $r=0$. Based on our prior analysis in {\bf Step 2}, we have the following results:

(i) If $\theta\in\left[0,\frac{\Phi}{\omega u'\left(0\right)}\right)$, the user will not watch ads, and its payoff will be zero. 

(ii) If $\theta\in\left[\frac{\Phi}{\omega u'\left(0\right)},\theta_{\max}\right]$, the user will watch ads with $x=\frac{1}{\omega}\left(u'\right)^{-1}\left(\frac{\Phi}{\omega\theta}\right)$, and its payoff will be $\theta u\left(\left(u'\right)^{-1}\left(\frac{\Phi}{\omega\theta}\right)\right)- \frac{\Phi}{\omega}\left(u'\right)^{-1}\left(\frac{\Phi}{\omega\theta}\right)$.

Next, we compare the choices of $r=1$ and $r=0$. Since $\omega>\frac{\Phi u\left(Q\right)}{F u'\left(0\right)}$, we have $\frac{\Phi}{\omega u'\left(0\right)}<\frac{F}{u\left(Q\right)}$. Based on our analysis in Appendix \ref{appendix:uniquetheta4}, when $\omega\in\left(\frac{\Phi u\left(Q\right)}{F u'\left(0\right)},\frac{\Phi Q}{F}\right)$, there is a unique $\theta_4\in\left(\frac{\Phi}{\omega u'\left(0\right)} , \frac{\Phi}{\omega u'\left(Q\right)}\right)$ satisfying $\theta_4 u\left( \left(u'\right)^{-1}\left(\frac{\Phi}{\omega\theta_4}\right) \right)- \frac{\Phi}{\omega} \left(u'\right)^{-1}\left(\frac{\Phi}{\omega\theta_4}\right)-\theta_4 u\left(Q\right)+F=0$.

Consider a user with $\theta\in\left[0,\frac{\Phi}{\omega u'\left(0\right)}\right)$. If this user chooses $r=1$, it will not watch ads and its payoff will be $\theta u\left(Q\right)-F$, which is negative due to $\theta<\frac{\Phi}{\omega u'\left(0\right)}<\frac{F}{u\left(Q\right)}$. If this user chooses $r=0$, it will not watch ads and its payoff will be zero. Hence, the user will not subscribe or watch ads, i.e., ${\hat r}^*\left(\theta,\omega\right)=0$ and ${\hat x}^*\left(\theta,\omega\right)=0$. 

Consider a user with $\theta\in\left[\frac{\Phi}{\omega u'\left(0\right)},\theta_4\right)$. If this user chooses $r=1$, it will not watch ads (due to $\theta_4<\frac{\Phi}{\omega u'\left(Q\right)}$) and its payoff will be $\theta u\left(Q\right)-F$. If this user chooses $r=0$, it will watch ads and its payoff will be $\theta u\left(\left(u'\right)^{-1}\left(\frac{\Phi}{\omega\theta}\right)\right)- \frac{\Phi}{\omega}\left(u'\right)^{-1}\left(\frac{\Phi}{\omega\theta}\right)$. Next, we compare these two payoffs. In Appendix \ref{appendix:uniquetheta4}, we have proved that function $v\left(\theta\right)=\theta u\left( \left(u'\right)^{-1}\left(\frac{\Phi}{\omega\theta}\right) \right)- \frac{\Phi}{\omega} \left(u'\right)^{-1}\left(\frac{\Phi}{\omega\theta}\right)-\theta u\left(Q\right)+F$ is decreasing for $\theta<\frac{\Phi}{\omega u'\left(Q\right)}$ and its value is zero when $\theta=\theta_4$. Therefore, when $\theta<\theta_4$, we have $v\left(\theta\right)>0$. After rearrangement, we have $\theta u\left( \left(u'\right)^{-1}\left(\frac{\Phi}{\omega\theta}\right) \right)- \frac{\Phi}{\omega} \left(u'\right)^{-1}\left(\frac{\Phi}{\omega\theta}\right)>\theta u\left(Q\right)-F$. This implies that the user will not subscribe but it will watch ads, i.e., ${\hat r}^*\left(\theta,\omega\right)=0$ and ${\hat x}^*\left(\theta,\omega\right)=\frac{1}{\omega}\left(u'\right)^{-1}\left(\frac{\Phi}{\omega\theta}\right)$. 

Consider a user with $\theta\in\left[\theta_4,\frac{\Phi}{\omega u'\left(Q\right)}\right)$. Compared with the user with $\theta\in\left[\frac{\Phi}{\omega u'\left(Q\right)},\theta_4\right)$, the only difference here is the comparison between $\theta u\left(\left(u'\right)^{-1}\left(\frac{\Phi}{\omega\theta}\right)\right)- \frac{\Phi}{\omega}\left(u'\right)^{-1}\left(\frac{\Phi}{\omega\theta}\right)$ and $\theta u\left(Q\right)-F$. Based on our analysis of $v\left(\theta\right)$ in Appendix \ref{appendix:uniquetheta4}, we can see that $v\left(\theta\right)\le0$ for $\theta\ge\theta_4$. Hence, we have $\theta u\left( \left(u'\right)^{-1}\left(\frac{\Phi}{\omega\theta}\right) \right)- \frac{\Phi}{\omega} \left(u'\right)^{-1}\left(\frac{\Phi}{\omega\theta}\right)\le\theta u\left(Q\right)-F$. This implies that the user will subscribe without watching ads, i.e., ${\hat r}^*\left(\theta,\omega\right)=1$ and ${\hat x}^*\left(\theta,\omega\right)=0$. 

Consider a user with $\theta\in\left[\frac{\Phi}{\omega u'\left(Q\right)},\theta_{\max}\right]$. If this user chooses $r=1$, it will watch ads and its payoff will be $\theta u\left(\left(u'\right)^{-1}\left(\frac{\Phi}{\omega\theta}\right)\right)-F-\Phi \left(\frac{1}{\omega}\left(u'\right)^{-1}\left(\frac{\Phi}{\omega\theta}\right)-\frac{Q}{\omega}\right)$. If this user chooses $r=0$, it will watch ads and its payoff will be $\theta u\left(\left(u'\right)^{-1}\left(\frac{\Phi}{\omega\theta}\right)\right)- \frac{\Phi}{\omega}\left(u'\right)^{-1}\left(\frac{\Phi}{\omega\theta}\right)$. When $\omega<\frac{\Phi Q}{F}$, the former payoff is greater. That is to say, the user should subscribe and watch ads, i.e., ${\hat r}^*\left(\theta,\omega\right)=1$ and ${\hat x}^*\left(\theta,\omega\right)=\frac{1}{\omega}\left(u'\right)^{-1}\left(\frac{\Phi}{\omega\theta}\right)-\frac{Q}{\omega}$.

Combining the above four regions of $\theta$, we can conclude the following result. For $\theta\in\left[0,\theta_{\max}\right]$, we have (recall that $\theta_3= \frac{\Phi}{\omega u'\left(0\right)}$ and $\theta_1= \frac{\Phi}{\omega u'\left(Q\right)}$)
\begin{align}
& {\hat r}^*\!\left(\theta,\!\omega\right)\!\!=\!\!{\mathbbm 1}_{\left\{\theta\ge \theta_4\right\}},\\
\nonumber
& {\hat x}^*\!\!\left(\theta,\!\omega\right)\!\!=\!\!\frac{1}{\omega}\!\!\left(u'\right)^{-1}\!\!\left(\!\frac{\Phi}{\omega\theta}\!\right) \!\!{\mathbbm 1}_{\left\{\theta_3 \le \theta< \theta_4\right\}}\!\!+\!\!\frac{1}{\omega}\!\!\left(\!\!\left(u'\right)^{-1}\!\!\left(\!\frac{\Phi}{\omega\theta}\!\right)\!\!-\!Q\!\!\right) \!\!{\mathbbm 1}_{\!\left\{\!\theta\ge \theta_1\!\right\}}\!.
\end{align}

{\bf Step 4:} We analyze Case $\hat D$, i.e., $\omega\ge\frac{\Phi Q}{F}$. 

Note that the cost of getting a unit data by subscription is $\frac{F}{Q}$, and the ``cost'' of getting a unit data by watching ads is $\frac{\Phi}{\omega}$ (i.e., the disutility of watching one ad over the corresponding data reward). When $\omega\ge\frac{\Phi Q}{F}$, we have $\frac{F}{Q}\ge\frac{\Phi}{\omega}$. This implies that a user should never consider the subscription, since the user can always watch ads to win the same amount of data but incur a total cost that is no greater than that under the subscription. Hence, we have ${\hat r}^*\left(\theta,\omega\right)=0$ for $\theta\in\left[0,\theta_{\max}\right]$. 

Now we can write a type-$\theta$ user's payoff as $\Pi^{\rm user}\left(\theta,0,x,\omega\right)=\theta u\left(\omega x\right)-\Phi x$. We can see that 
\begin{align}
\nonumber
\frac{\partial \Pi^{\rm user}\left(\theta,0,x,\omega\right)}{\partial x}=\theta \omega u'\left(\omega x\right)-\Phi.
\end{align}

Hence, if $\theta\in\left[0,\frac{\Phi}{\omega u'\left(0\right)}\right)$, the user will not watch ads; otherwise, it will watch ads with $x=\frac{1}{\omega}\left(u'\right)^{-1}\left(\frac{\Phi}{\omega\theta}\right)$. That is to say, we have the following result for $\theta\in\left[0,\theta_{\max}\right]$ (recall that $\theta_3=\frac{\Phi}{\omega u'\left(0\right)}$):
\begin{align}
{\hat r}^*\left(\theta,\omega\right)=0,{~~}{\hat x}^*\left(\theta,\omega\right)=\frac{1}{\omega}\left(u'\right)^{-1}\left(\frac{\Phi}{\omega\theta}\right){\mathbbm 1}_{\left\{\theta \ge \theta_3 \right\}}.
\end{align}

Hence, we have proved ${\hat r}^*\left(\theta,\omega\right)$ and ${\hat x}^*\left(\theta,\omega\right)$ for Case $\hat A$, Case $\hat B$, Case $\hat C$, and Case $\hat D$. 
\end{proof}

\section{Proof of Lemma \ref{lemma:theta4}}\label{appendix:lemma3}
\begin{proof}
{\bf Step 1:} We first prove that when $\omega\in\left(\frac{\Phi u\left(Q\right)}{F u'\left(0\right)},\frac{\Phi Q}{F}\right)$, the relation that $\theta_4>\theta_0$ holds. 

Based on our analysis in Appendix \ref{appendix:uniquetheta4}, if we let $v\left(\theta\right)=\theta u\left( \left(u'\right)^{-1}\left(\frac{\Phi}{\omega\theta}\right) \right)- \frac{\Phi}{\omega} \left(u'\right)^{-1}\left(\frac{\Phi}{\omega\theta}\right)-\theta u\left(Q\right)+F$, then $v\left(\theta\right)$ is decreasing when $\theta<\theta_1=\frac{\Phi}{\omega u'\left(Q\right)}$. Moreover, $\theta_4$ satisfies $\theta_3<\theta_4<\theta_1$ and $v\left(\theta_4\right)=0$.

Now we check the value of $v\left(\theta_0\right)$. We can see that
\begin{align}
v\left(\theta_0\right)\!=\!\frac{F}{u\left(Q\right)} u\left( \left(u'\right)^{-1}\left(\frac{\Phi {u\left(Q\right)}}{\omega F}\right) \right)\!-\! \frac{\Phi}{\omega} \left(u'\right)^{-1}\left(\frac{\Phi {u\left(Q\right)}}{\omega {F}}\right).
\end{align}

Let $\rho\left(\omega\right)\triangleq \frac{F}{u\left(Q\right)} u\left( \left(u'\right)^{-1}\left(\frac{\Phi {u\left(Q\right)}}{\omega F}\right) \right)\!-\! \frac{\Phi}{\omega} \left(u'\right)^{-1}\left(\frac{\Phi {u\left(Q\right)}}{\omega {F}}\right)$. From $u'\left( \left(u'\right)^{-1}\left(\frac{\Phi {u\left(Q\right)}}{\omega F}\right) \right)=\frac{\Phi {u\left(Q\right)}}{\omega F}$, we can compute $\frac{d \rho\left(\omega\right)}{d \omega}$ as
\begin{align}
\frac{d \rho\left(\omega\right)}{d \omega}=\frac{\Phi}{\omega^2} \left(u'\right)^{-1}\left(\frac{\Phi {u\left(Q\right)}}{\omega {F}}\right).
\end{align}
When $\omega=\frac{\Phi u\left(Q\right)}{F u'\left(0\right)}$, the value of $\left(u'\right)^{-1}\left(\frac{\Phi {u\left(Q\right)}}{\omega {F}}\right)$ is zero. When $\omega>\frac{\Phi u\left(Q\right)}{F u'\left(0\right)}$, the value of $\left(u'\right)^{-1}\left(\frac{\Phi {u\left(Q\right)}}{\omega {F}}\right)$ is greater than zero. Therefore, when $\omega=\frac{\Phi u\left(Q\right)}{F u'\left(0\right)}$, we have $\frac{d \rho\left(\omega\right)}{d \omega}=0$; when $\omega>\frac{\Phi u\left(Q\right)}{F u'\left(0\right)}$, we have $\frac{d \rho\left(\omega\right)}{d \omega}>0$. Moreover, by substituting $\omega=\frac{\Phi u\left(Q\right)}{F u'\left(0\right)}$ into the expression of $\rho\left(\omega\right)$, we can see that $\rho\left(\frac{\Phi u\left(Q\right)}{F u'\left(0\right)}\right)=0$. Then, we can conclude that when $\omega>\frac{\Phi u\left(Q\right)}{F u'\left(0\right)}$, we have $\rho\left(\omega\right)>0$. Because $\rho\left(\omega\right)$ has the same expression as $v\left(\theta_0\right)$. When $\omega>\frac{\Phi u\left(Q\right)}{F u'\left(0\right)}$, the value of $v\left(\theta_0\right)$ is positive. 

Next, we compare $\theta_0=\frac{F}{u\left(Q\right)}$ with $\theta_1=\frac{\Phi}{\omega u'\left(Q\right)}$. We can compute $\frac{\theta_0}{\theta_1}$ as $\frac{\theta_0}{\theta_1}=\frac{F \omega u'\left(Q\right)}{u\left(Q\right)\Phi}$. When $\omega<\frac{\Phi Q}{F}$, we have $\frac{\theta_0}{\theta_1}<\frac{{ Q} u'\left(Q\right)}{u\left(Q\right)}$. We can rewrite $\frac{{ Q} u'\left(Q\right)}{u\left(Q\right)}$ as follows:
\begin{align}
\frac{{ Q} u'\left(Q\right)}{u\left(Q\right)}=\frac{\int_{0}^{Q} u'\left(Q\right) dq}{u\left(0\right)+\int_{0}^{Q} u'\left(q\right) dq}.
\end{align}
From $u\left(0\right)=0$ and the strict concavity of $u\left(\cdot\right)$, we have $\frac{{ Q} u'\left(Q\right)}{u\left(Q\right)}<1$. The value of $\frac{\theta_0}{\theta_1}$ is less than $1$, i.e., $\theta_0$ is smaller than $\theta_1$. 

So far, we have shown the following results. First, both $\theta_0$ and $\theta_4$ are smaller than $\theta_1$. Second, $v\left(\theta\right)$ is decreasing for $\theta<\theta_1$. Third, $v\left(\theta_0\right)>0$ and $v\left(\theta_4\right)=0$. It is easy to see that $\theta_0<\theta_4$. 

{\bf Step 2:} We prove that in Case $\hat C$, $\theta_4$ increases as $\omega$ increases.

Based on $\theta_4$'s definition, we have
\begin{align}
\theta_4 u\left( \left(u'\right)^{-1}\left(\frac{\Phi}{\omega\theta_4}\right) \right)- \frac{\Phi}{\omega} \left(u'\right)^{-1}\left(\frac{\Phi}{\omega\theta_4}\right)-\theta_4 u\left(Q\right)+F=0.
\end{align}
For both sides of the equation, we take their derivatives with respect to $\omega$. Since $u'\left( \left(u'\right)^{-1}\left(\frac{\Phi}{\omega\theta_4}\right) \right)=\frac{\Phi}{\omega\theta_4}$, we can get
\begin{align}
\frac{d \theta_4}{d \omega} u\left( \left(u'\right)^{-1}\left(\frac{\Phi}{\omega\theta_4}\right) \right)+\frac{\Phi}{\omega^2}\left(u'\right)^{-1}\left(\frac{\Phi}{\omega\theta_4}\right)-\frac{d \theta_4}{d \omega} u\left(Q\right)=0.
\end{align}
After rearrangement, we can see that
\begin{align}
\frac{d \theta_4}{d \omega}=\frac{\frac{\Phi}{\omega^2}\left(u'\right)^{-1}\left(\frac{\Phi}{\omega\theta_4}\right)}{u\left(Q\right)-u\left( \left(u'\right)^{-1}\left(\frac{\Phi}{\omega\theta_4}\right) \right)}.
\end{align}
Since $\theta_4>\theta_3=\frac{\Phi}{\omega u'\left(0\right)}$, we have $\left(u'\right)^{-1}\left(\frac{\Phi}{\omega\theta_4}\right)>0$. Moreover, since $\theta_4<\theta_1=\frac{\Phi}{\omega u'\left(Q\right)}$, we have $u\left( \left(u'\right)^{-1}\left(\frac{\Phi}{\omega\theta_4}\right) \right)<u\left(Q\right)$. Hence, $\frac{d \theta_4}{d \omega}>0$. 
\end{proof}

\section{Proof of Theorem \ref{theorem:notuse}}\label{appendix:sufficientcondition}
\begin{proof}
{\bf Step 1:} We prove that when $C>N\left(u'\right)^{-1}\left(\frac{F}{\theta_{\max}Q}\right)$, we have ${\hat D}\left(\omega\right)<C$ for all $\omega\in\left[0,\frac{\Phi Q}{F}\right)$.

First, since $\left(u'\right)^{-1}\left(\cdot\right)$ is a decreasing function and $\theta_{\max}>\frac{u'\left(0\right) F}{u'\left(Q\right)u\left(Q\right)}$ (our assumption in Section \ref{subsec:gamestructure}), we can derive the following inequality:
\begin{align}
N\left(u'\right)^{-1}\left(\frac{F}{\theta_{\max}Q}\right)>N\left(u'\right)^{-1}\left(\frac{u'\left(Q\right)u\left(Q\right)}{u'\left(0\right)Q}\right).\label{SM:equ:theorem3}
\end{align}
Then, we analyze the value of $\frac{u\left(Q\right)}{u'\left(0\right)Q}$. The fraction can be rewritten as follows:
\begin{align}
\frac{u\left(Q\right)}{u'\left(0\right)Q}=\frac{u\left(0\right)+\int_{0}^{Q} u'\left(q\right) dq}{\int_{0}^{Q} u'\left(0\right) dq}.
\end{align}
Since $u\left(0\right)=0$ and $u\left(\cdot\right)$ is a strict increasing and concave function, we can see that $\frac{u\left(Q\right)}{u'\left(0\right)Q}<1$. Based on this inequality and (\ref{SM:equ:theorem3}), we have the following result:
\begin{align}
N\left(u'\right)^{-1}\left(\frac{F}{\theta_{\max}Q}\right)>N\left(u'\right)^{-1}\left(u'\left(Q\right)\right)=NQ.\label{SM:equ:theorem3:c}
\end{align}
Therefore, when $C>N\left(u'\right)^{-1}\left(\frac{F}{\theta_{\max}Q}\right)$, we also have $C>NQ$.

Next, we analyze the values of ${\hat D}\left(\omega\right)$ in Case $\hat A$, Case $\hat B$, and Case $\hat C$. When $\omega\in\left[0,\frac{\Phi}{u'\left(Q\right)\theta_{\max}}\right]$, we can compute ${\hat D}\left(\omega\right)$ as follows:
\begin{align}
{\hat D}\left(\omega\right)=NQ\int_{\theta_0}^{\theta_{\max}} g\left(\theta\right) d\theta.
\end{align}
Since $\int_{\theta_0}^{\theta_{\max}} g\left(\theta\right) d\theta<1$ and $C>NQ$, the value of ${\hat D}\left(\omega\right)$ is smaller than $C$ when $\omega\in\left[0,\frac{\Phi}{u'\left(Q\right)\theta_{\max}}\right]$.

When $\omega\in\left(\frac{\Phi}{u'\left(Q\right)\theta_{\max}},\frac{\Phi u\left(Q\right)}{F u'\left(0\right)}\right]$, we can compute ${\hat D}\left(\omega\right)$ as follows:
\begin{align}
\nonumber
& {\hat D}\left(\omega\right)= NQ \int_{\theta_0}^{\theta_{\max}} g\left(\theta\right) d\theta \\
\nonumber
& + N  \int_{\theta_1}^{\theta_{\max}} \left(\left(u'\right)^{-1}\left(\frac{\Phi}{\omega\theta}\right)-Q\right) g\left(\theta\right)d\theta\\
&= NQ \int_{\theta_0}^{\theta_1} g\left(\theta\right) d\theta+N  \int_{\theta_1}^{\theta_{\max}} \left(u'\right)^{-1}\left(\frac{\Phi}{\omega\theta}\right) g\left(\theta\right)d\theta.\label{SM:equ:theorem3:b}
\end{align}
As proved above, we have $\frac{u\left(Q\right)}{u'\left(0\right)Q}<1$. Since $\omega\le\frac{\Phi u\left(Q\right)}{F u'\left(0\right)}$, we can see that $\omega<\frac{\Phi Q}{F}$. Moreover, $\left(u'\right)^{-1}\left(\cdot\right)$ is decreasing and $\theta\le\theta_{\max}$. We can get the relation that $\left(u'\right)^{-1}\left(\frac{\Phi}{\omega\theta}\right)< \left(u'\right)^{-1}\left(\frac{ F}{ Q\theta_{\max}}\right)$. Based on this relation and (\ref{SM:equ:theorem3:b}), we can derive the following inequality:
\begin{align}
\nonumber
{\hat D}\left(\omega\right)\! &<\! NQ \int_{\theta_0}^{\theta_1} g\left(\theta\right) d\theta\!+\!N \left(u'\right)^{-1}\left(\frac{ F}{ Q\theta_{\max}}\right)\! \int_{\theta_1}^{\theta_{\max}}  \!g\left(\theta\right)d\theta \\
\nonumber
& \overset{(a)}{<}\!N \left(u'\right)^{-1}\left(\frac{ F}{ Q\theta_{\max}}\right)\! \left(\int_{\theta_0}^{\theta_1} g\left(\theta\right) d\theta+\int_{\theta_1}^{\theta_{\max}}  \!g\left(\theta\right)d\theta \right)\\
& <N \left(u'\right)^{-1}\left(\frac{ F}{ Q\theta_{\max}}\right)<C.
\end{align}
Here, inequality (a) is due to (\ref{SM:equ:theorem3:c}). Hence, we have shown that the value of ${\hat D}\left(\omega\right)$ is smaller than $C$ when $\omega\in\left(\frac{\Phi}{u'\left(Q\right)\theta_{\max}},\frac{\Phi u\left(Q\right)}{F u'\left(0\right)}\right]$.

When $\omega\in\left(\frac{\Phi u\left(Q\right)}{F u'\left(0\right)},\frac{\Phi Q}{F}\right)$, we can compute ${\hat D}\left(\omega\right)$ as follows:
\begin{align}
\nonumber
{\hat D}\left(\omega\right)=&NQ \int_{\theta_4}^{\theta_{\max}} g\left(\theta\right) d\theta + N  \int_{\theta_3}^{\theta_4} \left(u'\right)^{-1}\left(\frac{\Phi}{\omega\theta}\right) g\left(\theta\right)d\theta\\
\nonumber
& + N  \int_{\theta_1}^{\theta_{\max}} \left(\left(u'\right)^{-1}\left(\frac{\Phi}{\omega\theta}\right)-Q\right) g\left(\theta\right)d\theta \\
\nonumber
=& NQ \int_{\theta_4}^{\theta_1} g\left(\theta\right) d\theta + N  \int_{\theta_3}^{\theta_4} \left(u'\right)^{-1}\left(\frac{\Phi}{\omega\theta}\right) g\left(\theta\right)d\theta\\
&+N  \int_{\theta_1}^{\theta_{\max}} \left(u'\right)^{-1}\left(\frac{\Phi}{\omega\theta}\right) g\left(\theta\right)d\theta.
\end{align}
Based on a similar discussion as that after (\ref{SM:equ:theorem3:b}), we can prove that $\left(u'\right)^{-1}\left(\frac{\Phi}{\omega\theta}\right)< \left(u'\right)^{-1}\left(\frac{ F}{ Q\theta_{\max}}\right)$. Then, we have the following result:
\begin{align}
\nonumber
{\hat D}\left(\omega\right)&<NQ \int_{\theta_4}^{\theta_1} g\left(\theta\right) d\theta \\
&+N \left(u'\right)^{-1}\left(\frac{ F}{ Q\theta_{\max}}\right) \!\left(\int_{\theta_3}^{\theta_4} g\left(\theta\right)d\theta\!+\!\int_{\theta_1}^{\theta_{\max}} g\left(\theta\right)d\theta\right).
\end{align}
From this inequality, (\ref{SM:equ:theorem3:c}), and $C>N\left(u'\right)^{-1}\left(\frac{F}{\theta_{\max}Q}\right)$, we can see that the value of ${\hat D}\left(\omega\right)$ is smaller than $C$.

Combing our analysis for the cases where $\omega\in\left[0,\frac{\Phi}{u'\left(Q\right)\theta_{\max}}\right]$, $\omega\in\left(\frac{\Phi}{u'\left(Q\right)\theta_{\max}},\frac{\Phi u\left(Q\right)}{F u'\left(0\right)}\right]$, and $\omega\in\left(\frac{\Phi u\left(Q\right)}{F u'\left(0\right)},\frac{\Phi Q}{F}\right)$, we can conclude that ${\hat D}\left(\omega\right)<C$ for all $\omega\in\left[0,\frac{\Phi Q}{F}\right)$.

{\bf Step 2:} We prove that when $A>\frac{B^2K}{8F \int_{\theta_0}^{\theta_{\max}}g\left(\theta\right) d\theta}$, the value of ${\hat\omega}^*$ lies in interval $\left[0,\frac{\Phi Q}{F}\right)$.

When $\omega\in\left[\frac{\Phi Q}{F},\infty\right)$, all users watch ads without subscription. In this case, the operator's revenue only consists of the ad revenue. We can compute the operator's revenue as follows:
\begin{align}
\nonumber
{\hat R}^{\rm total}\left(\omega,{\hat p}^*\left(\omega\right)\right)& ={\hat R}^{\rm ad}\left(\omega,{\hat p}^*\left(\omega\right)\right) \\
\nonumber
&= K {\hat m}^*\left(\omega,{\hat p}^*\left(\omega\right)\right) {\hat p}^*\left(\omega\right)\\
&= K  \frac{\left({\mathbb E}\left[\hat y\right]\right)^2}{{\mathbb E}\left[{\hat y}^2\right]} {\hat N}^{\rm ad}\left(\omega\right)\frac{B-{\hat p}^*\left(\omega\right)}{2A} {\hat p}^*\left(\omega\right).
\end{align}
Note that $\frac{\left({\mathbb E}\left[\hat y\right]\right)^2}{{\mathbb E}\left[{\hat y}^2\right]}=\frac{\left({\mathbb E}\left[\hat y\right]\right)^2}{\left({\mathbb E}\left[\hat y\right]\right)^2+{\rm Var}\left[\hat y\right]}\le 1$, where ${\rm Var}\left[\hat y\right]$ is the variance of $\hat y$. Moreover, since $\frac{B-{\hat p}^*\left(\omega\right)}{2A} {\hat p}^*\left(\omega\right)$ is quadratic in ${\hat p}^*\left(\omega\right)$, we have the following relation:
\begin{align}
\frac{B-{\hat p}^*\left(\omega\right)}{2A} {\hat p}^*\left(\omega\right)\le \frac{B^2}{8A}.
\end{align}
Based on the above inequality, $\frac{\left({\mathbb E}\left[\hat y\right]\right)^2}{{\mathbb E}\left[{\hat y}^2\right]}\le1$, and ${\hat N}^{\rm ad}\left(\omega\right)\le N$, we can derive the following inequality:
\begin{align}
{\hat R}^{\rm total}\left(\omega,{\hat p}^*\left(\omega\right)\right)\le\frac{B^2 K}{8A} N.\label{SM:the3:com1}
\end{align}
Hence, when $\omega\in\left[\frac{\Phi Q}{F},\infty\right)$, the value of ${\hat R}^{\rm total}\left(\omega,{\hat p}^*\left(\omega\right)\right)$ is upper bounded by $\frac{B^2 K}{8A} N$.

Next, we analyze the lower bound of ${\hat R}^{\rm total}\left(\omega,{\hat p}^*\left(\omega\right)\right)$ when $\omega\in\left(\frac{\Phi}{u'\left(Q\right)\theta_{\max}},\frac{\Phi u\left(Q\right)}{F u'\left(0\right)}\right]$. In this case, the operator's ad revenue (i.e., ${\hat R}^{\rm ad}\left(\omega,{\hat p}^*\left(\omega\right)\right)$) and the revenue from the data market (i.e., ${\hat R}^{\rm data}\left(\omega\right)$) are positive. We can derive the following relation:
\begin{align}
{\hat R}^{\rm total}\left(\omega,{\hat p}^*\left(\omega\right)\right)>{\hat R}^{\rm data}\left(\omega\right)=NF\int_{\theta_0}^{\theta_{\max}} g\left(\theta\right)d\theta.\label{SM:the3:com2}
\end{align}
If $A>\frac{B^2K}{8F \int_{\theta_0}^{\theta_{\max}}g\left(\theta\right) d\theta}$, the right side of (\ref{SM:the3:com2}) is greater than the right side of (\ref{SM:the3:com1}). This implies that the optimal unit reward ${\hat \omega}^*\notin\left[\frac{\Phi Q}{F},\infty\right)$. In other words, the value of ${\hat \omega}^*$ lies in interval $\left[0,\frac{\Phi Q}{F}\right)$. 

Combining {\bf Step 1} and {\bf Step 2}, we can conclude that ${\hat D}\left({\hat \omega}^*\right)<C$. 
\end{proof}


\section{Proof of Theorem \ref{theorem:differentiation}}\label{appendix:theorem4}
\begin{proof}
Recall that $\Pi^{\rm SURD}$ is the optimal objective value of problem (\ref{equ:diff:obj})-(\ref{equ:diff:adc2}), and $\Pi^{\rm SUR}$ is the optimal objective value of problem (\ref{equ:unaware:obj})-(\ref{equ:unaware:constraint}). Suppose that $\left(\omega,p\right)$ is a feasible solution to (\ref{equ:unaware:obj})-(\ref{equ:unaware:constraint}). To prove that $\Pi^{\rm SURD}\ge \Pi^{\rm SUR}$, we can simply show that for any $\left(\omega,p\right)$, we are able to find a corresponding $\left(\omega,p_{\rm I},p_{\rm II}\right)$ such that (i) $\left(\omega,p_{\rm I},p_{\rm II}\right)$ is feasible to problem (\ref{equ:diff:obj})-(\ref{equ:diff:adc2}) and (ii) the value of objective (\ref{equ:diff:obj}) under $\left(\omega,p_{\rm I},p_{\rm II}\right)$ equals the value of (\ref{equ:unaware:obj}) under $\left(\omega,p\right)$.

Note that when $\omega$ satisfies Case $\hat A$, Case $\hat B$, or Case $\hat D$, the subscribers and non-subscribers will not simultaneously choose to watch ads. In this situation, the operator cannot differentiate ad slots based on the subscription decisions of the users watching ads. As we mentioned in Section \ref{subsec:differentiation}, problem (\ref{equ:unaware:obj})-(\ref{equ:unaware:constraint}) and problem (\ref{equ:diff:obj})-(\ref{equ:diff:adc2}) are the same. For example, when $\omega$ satisfies Case $\hat B$, only the subscribers watch ads. Then, both ${\hat N}_{\rm II}^{\rm ad}\left(\omega\right)$ and ${\hat m}_{\rm II}^*\left(\omega,p_{\rm II}\right)$ are zeros. This allows us to remove the term $K {\hat m}_{\rm II}^*\left(\omega,p_{\rm II}\right)p_{\rm II}$ in objective (\ref{equ:diff:obj}) and also ignore constraint (\ref{equ:diff:adc2}). We can see that problem (\ref{equ:diff:obj})-(\ref{equ:diff:adc2}) reduces to problem (\ref{equ:unaware:obj})-(\ref{equ:unaware:constraint}).

Next, we focus on the situations where $\omega$ satisfies Case $\hat C$. There are two possible situations.

If $\left(\omega,p\right)$ is a feasible solution to problem (\ref{equ:unaware:obj})-(\ref{equ:unaware:constraint}) and $p\ge B$, we can see that ${\hat m}^*\left(\omega,p\right)=0$ and ${\hat R}^{\rm ad}\left(\omega,p\right)=0$, which implies that the operator's total revenue only consists of the revenue from the data market. Then, we can construct a solution $\left(\omega,p_{\rm I},p_{\rm II}\right)$ where $p_{\rm I},p_{\rm II}\ge B$. Under $p_{\rm I}$ and $p_{\rm II}$, we have ${\hat m}_{\rm I}^*\left(\omega,p_{\rm I}\right)={\hat m}_{\rm II}^*\left(\omega,p_{\rm II}\right)=0$. We can easily verify that (i) the solution $\left(\omega,p_{\rm I},p_{\rm II}\right)$ is feasible to problem (\ref{equ:diff:obj})-(\ref{equ:diff:adc2}) and (ii) the value of objective (\ref{equ:diff:obj}) under $\left(\omega,p_{\rm I},p_{\rm II}\right)$ equals the value of (\ref{equ:unaware:obj}) under $\left(\omega,p\right)$ (i.e., both values equal ${\hat R}^{\rm data}\left(\omega\right)$).

If $\left(\omega,p\right)$ is a feasible solution to problem (\ref{equ:unaware:obj})-(\ref{equ:unaware:constraint}) and $p< B$, we can construct a solution $\left(\omega,p_{\rm I},p_{\rm II}\right)$ where $p_{\rm I}$ and $p_{\rm II}$ are given by
\begin{align}
& p_{\rm I}=B-\left(B-p\right) \frac{{\mathbb E}\left[{\hat y}\right]}{{\mathbb E}\left[{\hat y}^2\right]} \frac{{\mathbb E}\left[{\hat y}_{\rm I}^2\right]}{{\mathbb E}\left[{\hat y}_{\rm I}\right]},\label{SM:equ:the4:p1}\\
& p_{\rm II}=B-\left(B-p\right) \frac{{\mathbb E}\left[{\hat y}\right]}{{\mathbb E}\left[{\hat y}^2\right]} \frac{{\mathbb E}\left[{\hat y}_{\rm II}^2\right]}{{\mathbb E}\left[{\hat y}_{\rm II}\right]}.\label{SM:equ:the4:p2}
\end{align}

\begin{figure*}
\begin{align}
\nonumber
&{\hat N}_{\rm I}^{\rm ad}\left(\omega\right) {\mathbb E}\left[{\hat y}_{\rm I}\right] p_{\rm I} +  {\hat N}_{\rm II}^{\rm ad}\left(\omega\right) {\mathbb E}\left[{\hat y}_{\rm II}\right] p_{\rm II}\\
\nonumber
=& {\hat N}_{\rm I}^{\rm ad}\left(\omega\right) {\mathbb E}\left[{\hat y}_{\rm I}\right] \left(B-\left(B-p\right) \frac{{\mathbb E}\left[{\hat y}\right]}{{\mathbb E}\left[{\hat y}^2\right]} \frac{{\mathbb E}\left[{\hat y}_{\rm I}^2\right]}{{\mathbb E}\left[{\hat y}_{\rm I}\right]}\right) + {\hat N}_{\rm II}^{\rm ad}\left(\omega\right) {\mathbb E}\left[{\hat y}_{\rm II}\right] \left(B-\left(B-p\right) \frac{{\mathbb E}\left[{\hat y}\right]}{{\mathbb E}\left[{\hat y}^2\right]} \frac{{\mathbb E}\left[{\hat y}_{\rm II}^2\right]}{{\mathbb E}\left[{\hat y}_{\rm II}\right]}\right) \\
\nonumber
=& B\left({\hat N}_{\rm I}^{\rm ad}\left(\omega\right) {\mathbb E}\left[{\hat y}_{\rm I}\right]+{\hat N}_{\rm II}^{\rm ad}\left(\omega\right) {\mathbb E}\left[{\hat y}_{\rm II}\right]\right) -\left(B-p\right) \frac{{\mathbb E}\left[{\hat y}\right]}{{\mathbb E}\left[{\hat y}^2\right]} \left({\hat N}_{\rm I}^{\rm ad}\left(\omega\right) {\mathbb E}\left[{\hat y}_{\rm I}^2\right] + {\hat N}_{\rm II}^{\rm ad}\left(\omega\right) {\mathbb E}\left[{\hat y}_{\rm II}^2\right]\right)\\
\nonumber
\overset{(a)}{=}& B\!\left(\frac{{\hat N}^{\rm ad}\left(\omega\right) \int_{\theta_1}^{\theta_{\max}} g\left(\theta\right) \!d\theta }{\int_{\theta_1}^{\theta_{\max}} g\left(\theta\right) \!d\theta + \int_{\theta_3}^{\theta_4} g\left(\theta\right) \!d\theta}\! \int_{\theta_1}^{\theta_{\max}} \!{\hat x}^*\left(\theta,\omega\right) \frac{g\left(\theta\right)}{\int_{\theta_1}^{\theta_{\max}} g\left(\theta\right)\!d\theta} \!d\theta \!+\!\frac{{\hat N}^{\rm ad}\left(\omega\right) \int_{\theta_3}^{\theta_4} g\left(\theta\right) \!d\theta }{\int_{\theta_1}^{\theta_{\max}} g\left(\theta\right) \!d\theta + \int_{\theta_3}^{\theta_4} g\left(\theta\right) \!d\theta} \!\int_{\theta_3}^{\theta_4} \!{\hat x}^*\left(\theta,\omega\right) \frac{g\left(\theta\right)}{\int_{\theta_3}^{\theta_4} g\left(\theta\right)\!d\theta} \!d\theta \!\right) \\
\nonumber
 -&\left(\!B\!-\!p\right)\! \frac{{\mathbb E}\left[{\hat y}\right]}{{\mathbb E}\left[{\hat y}^2\right]} \!\!\left(\!\frac{{\hat N}^{\rm ad}\left(\omega\right) \int_{\theta_1}^{\theta_{\max}} g\left(\theta\right) \!d\theta }{\int_{\theta_1}^{\theta_{\max}} g\left(\theta\right) \!d\theta + \int_{\theta_3}^{\theta_4} g\left(\theta\right) \!d\theta}\! \int_{\theta_1}^{\theta_{\max}} \! \frac{\left({\hat x}^*\left(\theta,\omega\right)\right)^2 g\left(\theta\right)}{\int_{\theta_1}^{\theta_{\max}} g\left(\theta\right)\!d\theta} \!d\theta \!+\!\frac{{\hat N}^{\rm ad}\left(\omega\right) \int_{\theta_3}^{\theta_4} g\left(\theta\right) \!d\theta }{\int_{\theta_1}^{\theta_{\max}} g\left(\theta\right) \!d\theta + \int_{\theta_3}^{\theta_4} g\left(\theta\right) \!d\theta} \!\int_{\theta_3}^{\theta_4} \! \frac{\left({\hat x}^*\left(\theta,\omega\right)\right)^2 g\left(\theta\right)}{\int_{\theta_3}^{\theta_4} g\left(\theta\right)\!d\theta} \!d\theta\!\right)\\
 \nonumber
=& B \frac{{\hat N}^{\rm ad}\left(\omega\right)}{\int_{\theta_1}^{\theta_{\max}} g\left(\theta\right) d\theta + \int_{\theta_3}^{\theta_4} g\left(\theta\right) d\theta}\left( \int_{\theta_1}^{\theta_{\max}} {\hat x}^*\left(\theta,\omega\right) g\left(\theta\right) d\theta + \int_{\theta_3}^{\theta_4} {\hat x}^*\left(\theta,\omega\right) {g\left(\theta\right)} d\theta \right) \\
\nonumber
 &-\left(B-p\right) \frac{{\mathbb E}\left[{\hat y}\right]}{{\mathbb E}\left[{\hat y}^2\right]} \frac{{\hat N}^{\rm ad}\left(\omega\right)}{\int_{\theta_1}^{\theta_{\max}} g\left(\theta\right) d\theta + \int_{\theta_3}^{\theta_4} g\left(\theta\right) d\theta}\left( \int_{\theta_1}^{\theta_{\max}}  \left({\hat x}^*\left(\theta,\omega\right)\right)^2 g\left(\theta\right) d\theta + \int_{\theta_3}^{\theta_4}  {\left({\hat x}^*\left(\theta,\omega\right)\right)^2 g\left(\theta\right)} d\theta\right)\\ 
\nonumber
\overset{(b)}{=}& B {\hat N}^{\rm ad}\left(\omega\right) {\mathbb E}\left[{\hat y}\right] -\left(B-p\right) \frac{{\mathbb E}\left[{\hat y}\right]}{{\mathbb E}\left[{\hat y}^2\right]} {\hat N}^{\rm ad}\left(\omega\right) {\mathbb E}\left[{\hat y}^2\right] \\
=& B {\hat N}^{\rm ad}\left(\omega\right) {\mathbb E}\left[{\hat y}\right] -\left(B-p\right) {\hat N}^{\rm ad}\left(\omega\right) {\mathbb E}\left[{\hat y}\right] = p {\hat N}^{\rm ad}\left(\omega\right) {\mathbb E}\left[{\hat y}\right]. \label{SM:equ:the4:long}
\end{align}
\hrule
\end{figure*}

{\bf Step 1:} We can prove that this solution is feasible to problem (\ref{equ:diff:obj})-(\ref{equ:diff:adc2}). Specifically, since $p< B$, we can see that $p_{\rm I},p_{\rm II}<B$. Moreover, since $\omega$ satisfies Case $\hat C$, we have ${\hat N}_{\rm I}^{\rm ad}\left(\omega\right),{\hat N}_{\rm II}^{\rm ad}\left(\omega\right)>0$. Based on our discussions about ${\hat m}_{\rm I}^*\left(\omega,p_{\rm I}\right)$ and ${\hat m}_{\rm II}^*\left(\omega,p_{\rm II}\right)$ in Section \ref{subsec:differentiation}, we have the following relations:
\begin{align}
\nonumber
{\hat m}_{\rm I}^*\left(\omega,p_{\rm I}\right)&=\frac{\max\left\{B-p_{\rm I},0\right\}}{2A} \frac{\left({\mathbb E}\left[{\hat y}_{\rm I}\right]\right)^2}{{\mathbb E}\left[{\hat y}_{\rm I}^2\right]} {\hat N}_{\rm I}^{\rm ad}\left(\omega\right)\\
\nonumber
&=\frac{B-p}{2A} \frac{{\mathbb E}\left[{\hat y}\right]}{{\mathbb E}\left[{\hat y}^2\right]} \frac{{\mathbb E}\left[{\hat y}_{\rm I}^2\right]}{{\mathbb E}\left[{\hat y}_{\rm I}\right]}\frac{\left({\mathbb E}\left[{\hat y}_{\rm I}\right]\right)^2}{{\mathbb E}\left[{\hat y}_{\rm I}^2\right]} {\hat N}_{\rm I}^{\rm ad}\left(\omega\right) \\
&=\frac{B-p}{2A} \frac{{\mathbb E}\left[{\hat y}\right] {\mathbb E}\left[{\hat y}_{\rm I}\right]}{{\mathbb E}\left[{\hat y}^2\right]} {\hat N}_{\rm I}^{\rm ad}\left(\omega\right),\label{SM:equ:the4:a}\\
{\hat m}_{\rm II}^*\left(\omega,p_{\rm II}\right)&=\frac{B-p}{2A} \frac{{\mathbb E}\left[{\hat y}\right] {\mathbb E}\left[{\hat y}_{\rm II}\right]}{{\mathbb E}\left[{\hat y}^2\right]} {\hat N}_{\rm II}^{\rm ad}\left(\omega\right).\label{SM:equ:the4:c}
\end{align}
Then, we can verify the feasibility of constraints (\ref{equ:diff:adc1}) and (\ref{equ:diff:adc2}). 
By substituting the expression of ${\hat m}_{\rm I}^*\left(\omega,p_{\rm I}\right)$ in (\ref{SM:equ:the4:a}), we can compute the ratio between $K {\hat m}_{\rm I}^*\left(\omega,p_{\rm I}\right)$ and ${\mathbb E}\left[{\hat y}_{\rm I}\right] {\hat N}_{\rm I}^{\rm ad}\left(\omega\right)$ as follows:
\begin{align}
\frac{K {\hat m}_{\rm I}^*\left(\omega,p_{\rm I}\right)} {{\mathbb E}\left[{\hat y}_{\rm I}\right] {\hat N}_{\rm I}^{\rm ad}\left(\omega\right)}=K\frac{B-p}{2A} \frac{{\mathbb E}\left[{\hat y}\right]}{{\mathbb E}\left[{\hat y}^2\right]}.\label{SM:equ:the4:b}
\end{align}
Because $\left(\omega,p\right)$ is a feasible solution to problem (\ref{equ:unaware:obj})-(\ref{equ:unaware:constraint}) and $p< B$, we have the relation that $\frac{K {\hat m}^*\left(\omega,p\right)}{{\hat N}^{\rm ad}\left(\omega\right) {\mathbb E}\left[{\hat y}\right]
}\le 1$. Since ${\hat m}^*\left(\omega,p\right)=\frac{B-p}{2A} \frac{\left({\mathbb E}\left[{\hat y}\right]\right)^2}{{\mathbb E}\left[{\hat y}^2\right]} {\hat N}^{\rm ad}\left(\omega\right)$, we can see that $K \frac{B-p}{2A} \frac{{\mathbb E}\left[{\hat y}\right]}{{\mathbb E}\left[{\hat y}^2\right]}\le 1$. This means the left side of (\ref{SM:equ:the4:b}) is no greater than $1$. Therefore, $\frac{K {\hat m}_{\rm I}^*\left(\omega,p_{\rm I}\right)} {{\mathbb E}\left[{\hat y}_{\rm I}\right] {\hat N}_{\rm I}^{\rm ad}\left(\omega\right)}$ is no greater than $1$, which implies the feasibility of constraint (\ref{equ:diff:adc1}). 

Based on a similar approach, we can prove that $\omega$ and $p_{\rm II}$ given in (\ref{SM:equ:the4:p2}) satisfy constraint (\ref{equ:diff:adc2}). Because constraint (\ref{equ:diff:capacity}) (i.e., ${\hat D}\left(\omega\right) \le C$) is the same as the capacity constraint in problem (\ref{equ:unaware:obj})-(\ref{equ:unaware:constraint}) and $\left(\omega,p\right)$ is feasible to problem (\ref{equ:unaware:obj})-(\ref{equ:unaware:constraint}), we can see that the solution $\left(\omega,p_{\rm I},p_{\rm II}\right)$ satisfies constraint (\ref{equ:diff:capacity}). Therefore, we can conclude that the solution $\left(\omega,p_{\rm I},p_{\rm II}\right)$ is feasible to problem (\ref{equ:diff:obj})-(\ref{equ:diff:adc2}). 

{\bf Step 2:} We prove that the value of objective (\ref{equ:diff:obj}) under $\left(\omega,p_{\rm I},p_{\rm II}\right)$ equals the value of objective (\ref{equ:unaware:obj}) under $\left(\omega,p\right)$. Since both objectives include ${\hat R}^{\rm data}\left(\omega\right)$ (i.e., the revenue from the data market) and the values of ${\hat R}^{\rm data}\left(\omega\right)$ are the same, we only need to compare the remaining terms (i.e., the ad revenues) in these two objectives. Next, we prove that $K {\hat m}_{\rm I}^*\left(\omega,p_{\rm I}\right) p_{\rm I}+K {\hat m}_{\rm II}^*\left(\omega,p_{\rm II}\right) p_{\rm II}$ equals $K {\hat m}^*\left(\omega,p\right) p$. 

According to (\ref{SM:equ:the4:a}) and (\ref{SM:equ:the4:c}), we have the following relation:
\begin{align}
\nonumber
& K {\hat m}_{\rm I}^*\left(\omega,p_{\rm I}\right) p_{\rm I}+K {\hat m}_{\rm II}^*\left(\omega,p_{\rm II}\right) p_{\rm II}\\
=& K \frac{B-p}{2A} \frac{{\mathbb E}\left[{\hat y}\right]}{{\mathbb E}\left[{\hat y}^2\right]} \left({\hat N}_{\rm I}^{\rm ad}\left(\omega\right) {\mathbb E}\left[{\hat y}_{\rm I}\right] p_{\rm I} +  {\hat N}_{\rm II}^{\rm ad}\left(\omega\right) {\mathbb E}\left[{\hat y}_{\rm II}\right] p_{\rm II}\right).\label{SM:equ:the4:d}
\end{align}
Then, we substitute the expressions of $p_{\rm I}$ and $p_{\rm II}$ in (\ref{SM:equ:the4:p1}) and (\ref{SM:equ:the4:p2}) and compute ${\hat N}_{\rm I}^{\rm ad}\left(\omega\right) {\mathbb E}\left[{\hat y}_{\rm I}\right] p_{\rm I} +  {\hat N}_{\rm II}^{\rm ad}\left(\omega\right) {\mathbb E}\left[{\hat y}_{\rm II}\right] p_{\rm II}$ in (\ref{SM:equ:the4:long}). Note that in step (a) and step (b) of (\ref{SM:equ:the4:long}), we have expanded the expressions of some terms as follows:
\begin{align}
& {\hat N}_{\rm I}^{\rm ad}\left(\omega\right)=\frac{{\hat N}^{\rm ad}\left(\omega\right) \int_{\theta_1}^{\theta_{\max}} g\left(\theta\right) d\theta }{\int_{\theta_1}^{\theta_{\max}} g\left(\theta\right) d\theta + \int_{\theta_3}^{\theta_4} g\left(\theta\right) d\theta},\\
& {\hat N}_{\rm II}^{\rm ad}\left(\omega\right)=\frac{{\hat N}^{\rm ad}\left(\omega\right) \int_{\theta_3}^{\theta_4} g\left(\theta\right) d\theta }{\int_{\theta_1}^{\theta_{\max}} g\left(\theta\right) d\theta + \int_{\theta_3}^{\theta_4} g\left(\theta\right) d\theta},\\
& {\mathbb E}\left[{\hat y}_{\rm I}\right]=\int_{\theta_1}^{\theta_{\max}} {\hat x}^*\left(\theta,\omega\right) \frac{g\left(\theta\right)}{\int_{\theta_1}^{\theta_{\max}} g\left(\theta\right)d\theta} d\theta ,\\
& {\mathbb E}\left[{\hat y}_{\rm II}\right]=\int_{\theta_3}^{\theta_4} {\hat x}^*\left(\theta,\omega\right) \frac{g\left(\theta\right)}{\int_{\theta_3}^{\theta_4} g\left(\theta\right)d\theta} d\theta,\\
& {\mathbb E}\left[{\hat y}\right]=\frac{\int_{\theta_1}^{\theta_{\max}} {\hat x}^*\left(\theta,\omega\right) g\left(\theta\right) d\theta + \int_{\theta_3}^{\theta_4} {\hat x}^*\left(\theta,\omega\right) {g\left(\theta\right)} d\theta}{\int_{\theta_1}^{\theta_{\max}} g\left(\theta\right) d\theta + \int_{\theta_3}^{\theta_4} g\left(\theta\right) d\theta},\\
& {\mathbb E}\left[{\hat y}_{\rm I}^2\right]=\int_{\theta_1}^{\theta_{\max}} \left({\hat x}^*\left(\theta,\omega\right)\right)^2 \frac{g\left(\theta\right)}{\int_{\theta_1}^{\theta_{\max}} g\left(\theta\right)d\theta} d\theta,\\
& {\mathbb E}\left[{\hat y}_{\rm II}^2\right]=\int_{\theta_3}^{\theta_4} \left({\hat x}^*\left(\theta,\omega\right)\right)^2 \frac{g\left(\theta\right)}{\int_{\theta_3}^{\theta_4} g\left(\theta\right)d\theta} d\theta,\end{align}
\begin{align}
{\mathbb E}\left[{\hat y}^2\right]=\frac{\int_{\theta_1}^{\theta_{\max}} \!\left({\hat x}^*\left(\theta,\omega\right)\right)^2 g\left(\theta\right) d\theta \!+\! \int_{\theta_3}^{\theta_4} \!{\left({\hat x}^*\left(\theta,\omega\right)\right)^2 g\left(\theta\right)} d\theta}{\int_{\theta_1}^{\theta_{\max}} g\left(\theta\right) d\theta + \int_{\theta_3}^{\theta_4} g\left(\theta\right) d\theta}.
\end{align}
These equations can be derived based on the definitions of these terms and Proposition \ref{proposition:unaware:user}. Specifically, as studied in Proposition \ref{proposition:unaware:user}, users with $\theta\in\left[\theta_1,\theta_{\max}\right]$ subscribe and watch ads, and users with $\theta\in\left[\theta_3,\theta_4\right)$ watch ads without subscription.

According to (\ref{SM:equ:the4:d}) and (\ref{SM:equ:the4:long}), we can derive the following relation:
\begin{align}
\nonumber
& K {\hat m}_{\rm I}^*\left(\omega,p_{\rm I}\right) p_{\rm I}+K {\hat m}_{\rm II}^*\left(\omega,p_{\rm II}\right) p_{\rm II}\\
\nonumber
=& K \frac{B-p}{2A} \frac{{\mathbb E}\left[{\hat y}\right]}{{\mathbb E}\left[{\hat y}^2\right]} p {\hat N}^{\rm ad}\left(\omega\right) {\mathbb E}\left[{\hat y}\right]\\
=& K {\hat m}^*\left(\omega,p\right) p.
\end{align}
Then, we can see that the value of objective (\ref{equ:diff:obj}) under $\left(\omega,p_{\rm I},p_{\rm II}\right)$ equals the value of objective (\ref{equ:unaware:obj}) under $\left(\omega,p\right)$.


Combining {\bf Step 1} and {\bf Step 2}, we can conclude that the optimal objective value of problem (\ref{equ:diff:obj})-(\ref{equ:diff:adc2}) is no less than the optimal objective value of problem (\ref{equ:unaware:obj})-(\ref{equ:unaware:constraint}). In other words, we have $\Pi^{\rm SURD}\ge \Pi^{\rm SUR}$.
\end{proof}

\section{Proof of Theorem \ref{theorem:comparison}}\label{appendix:theorem5}
\begin{proof}
{\bf Step 1:} We derive the operator's optimal total revenue under the SAR scheme when $C$ approaches infinity. 

Based on Proposition \ref{proposition:uniform:useup}, under the SAR scheme, the operator's optimal unit reward $\omega^*=D^{-1}\left(C\right)$. According to our analysis in Appendix \ref{appendix:monotonicityD}, when $\omega\in\left(\frac{\Phi u\left(Q\right)}{Fu'\left(Q\right)},\infty\right)$, the expression of $D\left(\omega\right)$ is given by
\begin{align}
\nonumber
D\left(\omega\right)=& NQ \int_{\theta_2}^{\theta_{\max}} g\left(\theta\right) d\theta \\
\nonumber
& + N  \int_{\theta_2}^{\theta_{\max}} \left(\left(u'\right)^{-1}\left(\frac{\Phi}{\omega\theta}\right)-Q\right) g\left(\theta\right)d\theta\\
=& NQ\frac{\theta_{\max}-\theta_2}{\theta_{\max}}+\frac{N}{\theta_{\max}} \int_{\theta_2}^{\theta_{\max}} \left(\frac{\omega \theta}{\Phi}-1-Q\right) d\theta.
\end{align}
We can see that $\lim_{\omega\rightarrow\infty} D\left(\omega\right)=\infty$. Hence, when $C\rightarrow \infty$, we have $\omega^* \rightarrow \infty$. 

Next, we prove that when $\omega \rightarrow \infty$, the corresponding $\theta_2 \rightarrow 0$. According to the defitition of $\theta_2$, the value of $\theta_2$ is upper bounded by $\theta_0$ and is lower bounded by $0$. In Appendix \ref{appendix:monotonicity:theta2}, we have proved that $\frac{d \theta_2}{d \omega}<0$ for any given $\omega$. Hence, when $\omega\rightarrow \infty$, the limit of $\theta_2$ exists. Let $\zeta\in\left[0,\theta_0\right]$ denote this limit. From $\theta_2$'s definition, when $u\left(z\right)=\ln\left(1+z\right)$ and $\theta\sim{\cal U}\left[0,\theta_{\max}\right]$, $\theta_2$ satisfies the following equation:
\begin{align}
\theta_2 \ln\left(\frac{\theta_2 \omega}{\Phi}\right)-F-\theta_2+\frac{\left(1+Q\right)\Phi}{\omega}=0.
\end{align}
When we take $\omega \rightarrow \infty$ on both sides of the equation, we can see that the equality holds only when $\zeta=0$. That is to say, when $\omega \rightarrow \infty$, the corresponding $\theta_2 \rightarrow 0$. 

Based on our analysis in Step 2 of Appendix \ref{appendix:sec:proposition4}, when $\omega\in\left(\frac{\Phi u\left(Q\right)}{Fu'\left(Q\right)},\infty\right)$, the operator's revenue function $R^{\rm total}\left(\omega,p^*\left(\omega\right)\right)$ is as follows:
\begin{multline}
R^{\rm total}\left(\omega,p^*\left(\omega\right)\right)=\frac{\theta_{\max}-\theta_2}{\theta_{\max}}NF\\
+p^*\left(\omega\right)\left(B-p^*\left(\omega\right)\right)\frac{3K\Phi}{8A\theta_{\max}}\frac{\left(y_h^2-y_l^2\right)^2}{y_h^3-y_l^3}N,
\end{multline}
where $y_h= \frac{\theta_{\max}-\theta_1}{\Phi}$, $y_l = \frac{\theta_2-\theta_1}{\Phi}$, and $p^*\left(\omega\right)=\max\left\{B-\frac{4A}{3K}\frac{y_h^3-y_l^3}{y_h^2-y_l^2},\frac{B}{2}\right\}$. When $\omega\rightarrow \infty$, we can verify that $\theta_1\rightarrow 0$, $y_h\rightarrow \frac{\theta_{\max}}{\Phi}$, and $y_l \rightarrow 0$. Then, we can derive the limit of the revenue function under $\omega\rightarrow \infty$ as follows:
\begin{multline}
\lim_{\omega\rightarrow \infty} R^{\rm total}\left(\omega,p^*\left(\omega\right)\right)=NF
\\
\!+\max\!\!\left\{\!B\!-\!\frac{4A}{3K}\frac{\theta_{\max}}{\Phi},\!\frac{B}{2}\!\right\}\!\! \left(\!B\!-\!\max\left\{B\!-\!\frac{4A}{3K}\!\frac{\theta_{\max}}{\Phi},\!\frac{B}{2}\!\right\} \!\right)\!\frac{3K}{8A}\!N.\label{appendix:SAR:long}
\end{multline}
When $C\rightarrow \infty$, $\omega^* \rightarrow \infty$. Hence, the operator's optimal revenue under the SAR scheme for $C\rightarrow \infty$ is characterized in (\ref{appendix:SAR:long}).

{\bf Step 2:} We discuss the maximum possible value of the operator's total revenue under the SUR scheme. 

(i) When $\omega\in\left[0,\frac{\left(1+Q\right)\Phi}{\theta_{\max}}\right]$, no user watches ads. When $\omega\in\left(\frac{\left(1+Q\right)\Phi}{\theta_{\max}},\frac{\Phi\ln\left(1+Q\right)}{F}\right]$, the value of $y$ is uniformly distributed in $\left[0,\frac{\theta_{\max}-\theta_1}{\Phi}\right]$. Hence, when $\omega\in\left[0,\frac{\Phi\ln\left(1+Q\right)}{F}\right]$, we can derive the upper bound for the operator's total revenue as follows:
\begin{align}
\nonumber
{\hat R}^{\rm total}\left(\omega,{\hat p}^*\left(\omega\right)\right)\le & \frac{\theta_{\max}-\theta_0}{\theta_{\max}}NF\\
& +{\hat p}^*\left(\omega\right) \left(B-{\hat p}^*\left(\omega\right)\right) \frac{3K}{8A} \frac{\theta_{\max}-\theta_1}{\theta_{\max}} N,
\end{align}
where ${\hat p}^*\left(\omega\right)=\max\left\{B-\frac{4A}{3K} \frac{\theta_{\max}-\theta_1}{\Phi},\frac{B}{2}\right\}$. We compare the right-hand side of the above inequality with the right-hand side of (\ref{appendix:SAR:long}). We can see that the right-hand side of (\ref{appendix:SAR:long}) is always larger.

(ii) When $\omega\in\left(\frac{\Phi\ln\left(1+Q\right)}{F},\frac{\Phi Q}{F}\right)$, we have the following relation:
\begin{multline}
{\hat R}^{\rm total}\left(\omega,{\hat p}^*\left(\omega\right)\right)=\frac{\theta_{\max}-\theta_4}{\theta_{\max}}NF\\
\!+{\hat p}^*\!\left(\omega\right)\!\left(B\!-\!{\hat p}^*\left(\omega\right)\right)\! \frac{3K}{8A} \!\frac{\left(1\!+\!{\hat \eta}^2\right)^2}{\left(1\!+\!{\hat \eta}\right)\left(1\!+\!{\hat \eta}^3\right)} \!\frac{\theta_{\max}\!-\!\theta_1\!+\!\theta_4\!-\!\theta_3}{\theta_{\max}} \!N,
\end{multline}
where ${\hat \eta} \triangleq \frac{\min\left\{\theta_{\max}-\theta_1,\theta_4-\theta_3\right\}}{\max\left\{\theta_{\max}-\theta_1,\theta_4-\theta_3\right\}}$, ${\hat y}_{\max} \triangleq \frac{\max\left\{\theta_{\max}-\theta_1,\theta_4-\theta_3\right\}}{\Phi}$, and ${\hat p}^*\left(\omega\right)=\max\left\{B-\frac{4A}{3K} \frac{1+{\hat \eta}^3}{1+{\hat \eta}^2}{\hat y}_{\max},\frac{B}{2}\right\}$. Based on $0\le \hat \eta\le1$, we can easily show that $\frac{\left(1+{\hat \eta}^2\right)^2}{\left(1+{\hat \eta}\right)\left(1+{\hat \eta}^3\right)}\le1$. Moreover, we have $\theta_{\max}>\theta_1$ and $\theta_4>\theta_3$. Therefore, the following relation holds:
\begin{align}
{\hat R}^{\rm total}\left(\omega,{\hat p}^*\left(\omega\right)\right)\!<\!\frac{\theta_{\max}\!-\!\theta_4}{\theta_{\max}}NF\!+\!{\hat p}^*\left(\omega\right)\left(B\!-\!{\hat p}^*\left(\omega\right)\right) \frac{3K}{8A} N.\label{appendix:long:2}
\end{align}
Next, we compare $\frac{\theta_{\max}}{\Phi}$ and $\frac{1+{\hat \eta}^3}{1+{\hat \eta}^2}{\hat y}_{\max}$. Since $0\le \hat \eta\le1$, we have $\frac{1+{\hat \eta}^3}{1+{\hat \eta}^2}=1-\frac{1-{\hat \eta}}{1+{\hat \eta}^2}{\hat \eta}^2 \le 1$. Moreover, ${\hat y}_{\max} = \frac{\max\left\{\theta_{\max}-\theta_1,\theta_4-\theta_3\right\}}{\Phi}<\frac{\max\left\{\theta_{\max},\theta_4\right\}}{\Phi}\le \frac{\theta_{\max}}{\Phi}$. Hence, we can see that $\frac{\theta_{\max}}{\Phi}>\frac{1+{\hat \eta}^3}{1+{\hat \eta}^2}{\hat y}_{\max}$. Based on this relation, we can verify that the following relation holds:
\begin{align}
\nonumber
& \!\!\!\max\!\left\{\!B\!-\!\frac{4A}{3K}\!\frac{\theta_{\max}}{\Phi},\frac{B}{2}\!\right\}\! \left(\!B\!-\!\max\!\left\{\!B\!-\!\frac{4A}{3K}\!\frac{\theta_{\max}}{\Phi},\frac{B}{2}\!\right\}\! \right) \!\ge\\
& \!\!\!\max\!\left\{\!B\!-\!\frac{4A}{3K}\!\frac{1\!+\!{\hat \eta}^3}{1\!+\!{\hat \eta}^2}\!{\hat y}_{\max},\!\frac{B}{2}\!\right\}\! \left(\!B\!-\!\max\!\left\{\!B\!-\!\frac{4A}{3K} \!\frac{1\!+\!{\hat \eta}^3}{1\!+\!{\hat \eta}^2}\!{\hat y}_{\max},\!\frac{B}{2}\!\right\}\!\right)\!. 
\end{align}
Therefore, we can see that the right-hand side of (\ref{appendix:long:2}) is smaller than (\ref{appendix:SAR:long}). 

(iii) When $\omega\in\left[\frac{\Phi Q}{F},\infty\right)$, we have the following relation:
\begin{align}
{\hat R}^{\rm total}\left(\omega,{\hat p}^*\left(\omega\right)\right)={\hat p}^*\left(\omega\right) \left(B-{\hat p}^*\left(\omega\right)\right) \frac{3K}{8A} \frac{\theta_{\max}-\theta_3}{\theta_{\max}} N,
\end{align}
where ${\hat p}^*\left(\omega\right)=\max\left\{B-\frac{4A}{3K} \frac{\theta_{\max}-\theta_3}{\Phi},\frac{B}{2}\right\}$. Since $\theta_3=\frac{\Phi}{\omega}$ decreases with $\omega$, we can easily prove that ${\hat R}^{\rm total}\left(\omega,{\hat p}^*\left(\omega\right)\right)$ increases with $\omega$. We can compute the limit of this revenue function under $\omega\rightarrow \infty$ as follows:
\begin{multline}
\lim_{\omega\rightarrow \infty} {\hat R}^{\rm total}\left(\omega,{\hat p}^*\left(\omega\right)\right) =\\\!\max\!\left\{\!B\!-\!\frac{4A}{3K} \!\frac{\theta_{\max}}{\Phi},\!\frac{B}{2}\!\right\} \!\left(\!B\!-\!\max\!\left\{\!B\!-\!\frac{4A}{3K}\! \frac{\theta_{\max}}{\Phi},\!\frac{B}{2}\right\} \!\right)\! \frac{3K}{8A}\!N.
\end{multline}
We compare the right-hand side of the above equation with the right-hand side of (\ref{appendix:SAR:long}). We can see that the right-hand side of (\ref{appendix:SAR:long}) is always larger.

Combining the above analysis for all cases of $\omega$ (including the case where $\omega\rightarrow\infty$), we can see that the operator's maximum possible total revenue under the SUR scheme is always smaller than the value of the right-hand side of (\ref{appendix:SAR:long}). Because the right-hand side of (\ref{appendix:SAR:long}) is the operator's optimal total revenue under the SAR when $C\rightarrow \infty$, we complete the proof.
\end{proof}

\begin{figure*}[t]
  \centering
  \subfigure[Example A.]{
    \includegraphics[scale=0.29]{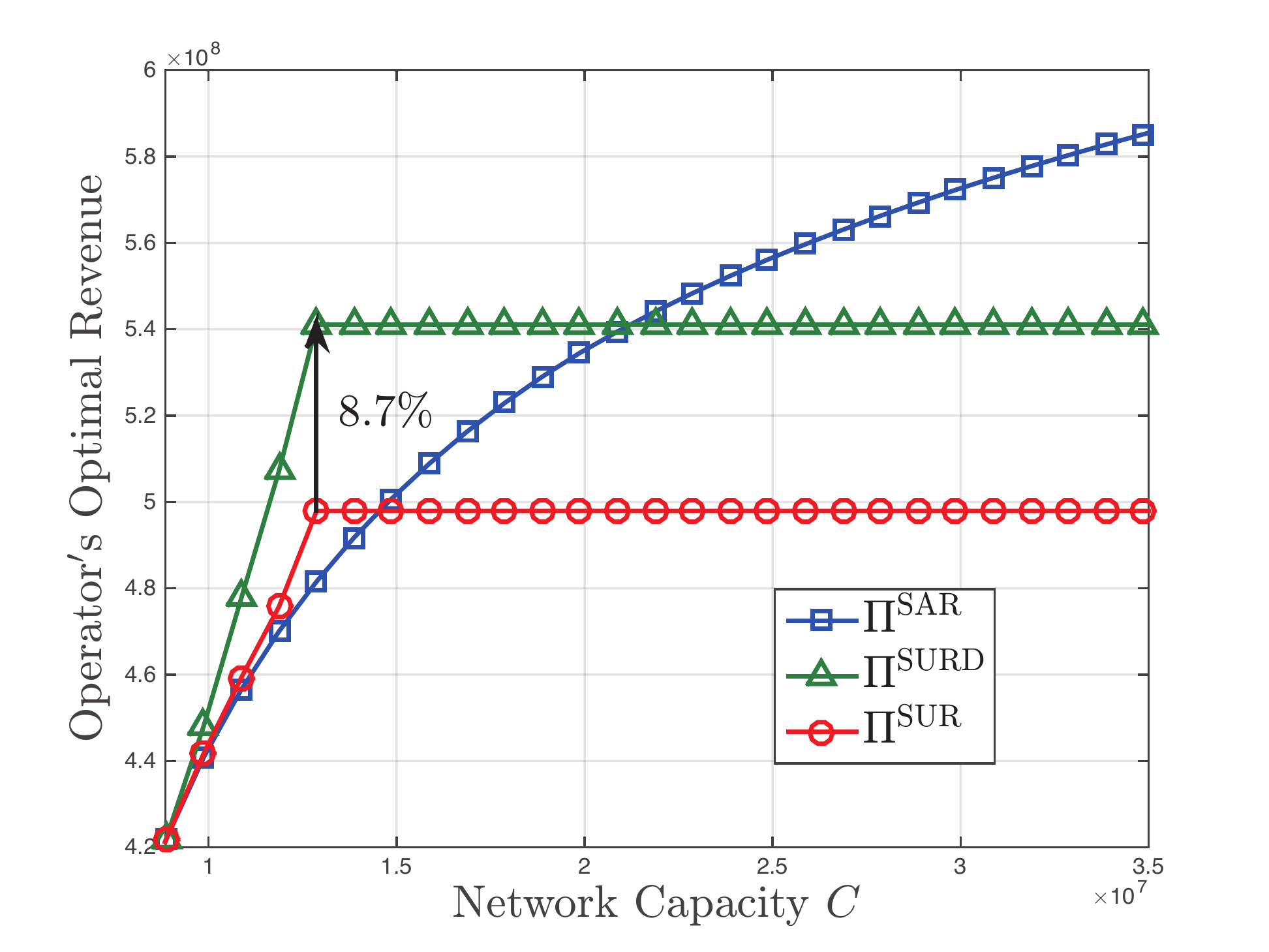}\label{appendix:fig:morepara:a}}
    \subfigure[Example B.]{
    \includegraphics[scale=0.29]{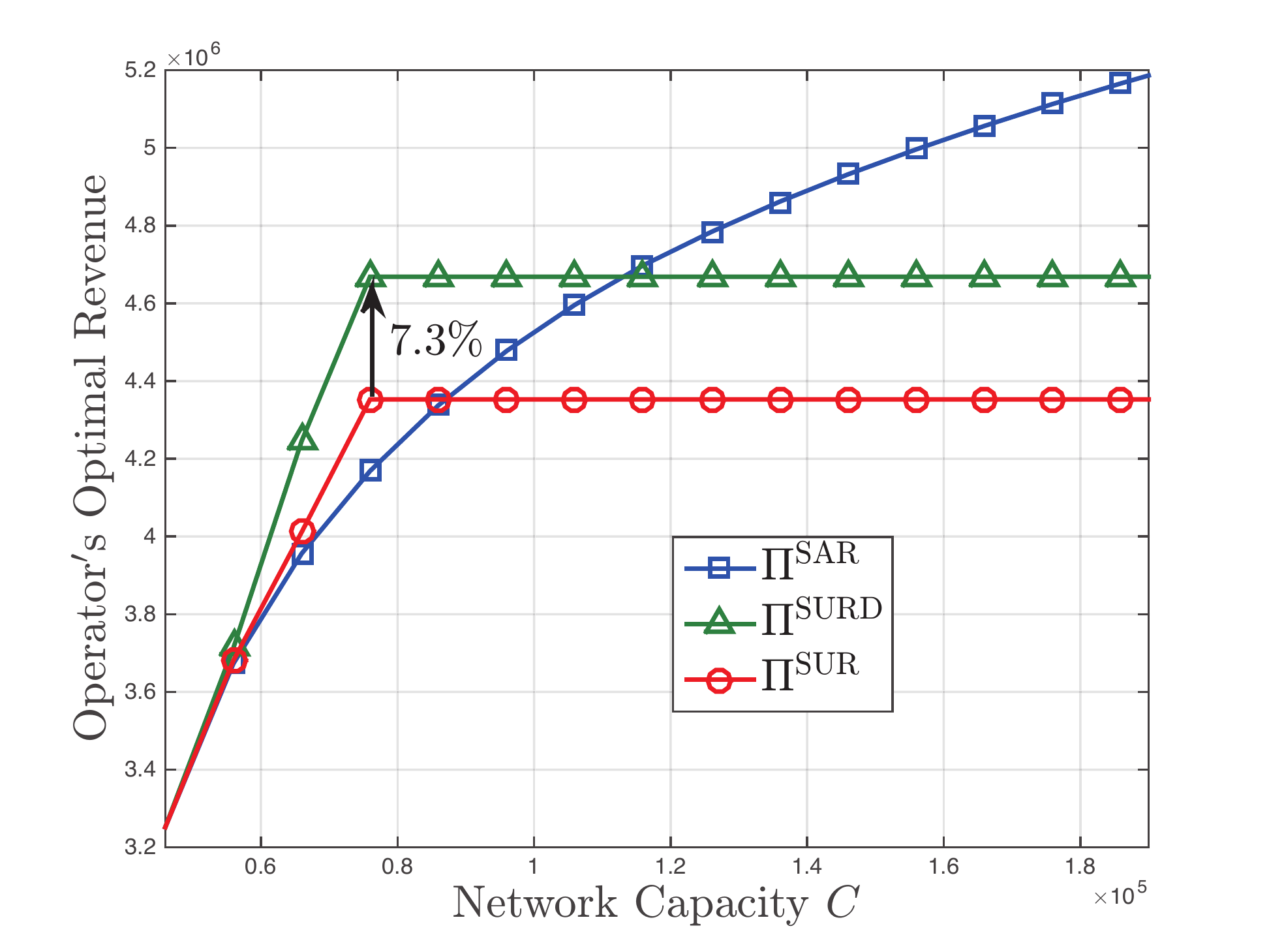}\label{appendix:fig:morepara:b}}
  \subfigure[Example C.]{
    \includegraphics[scale=0.29]{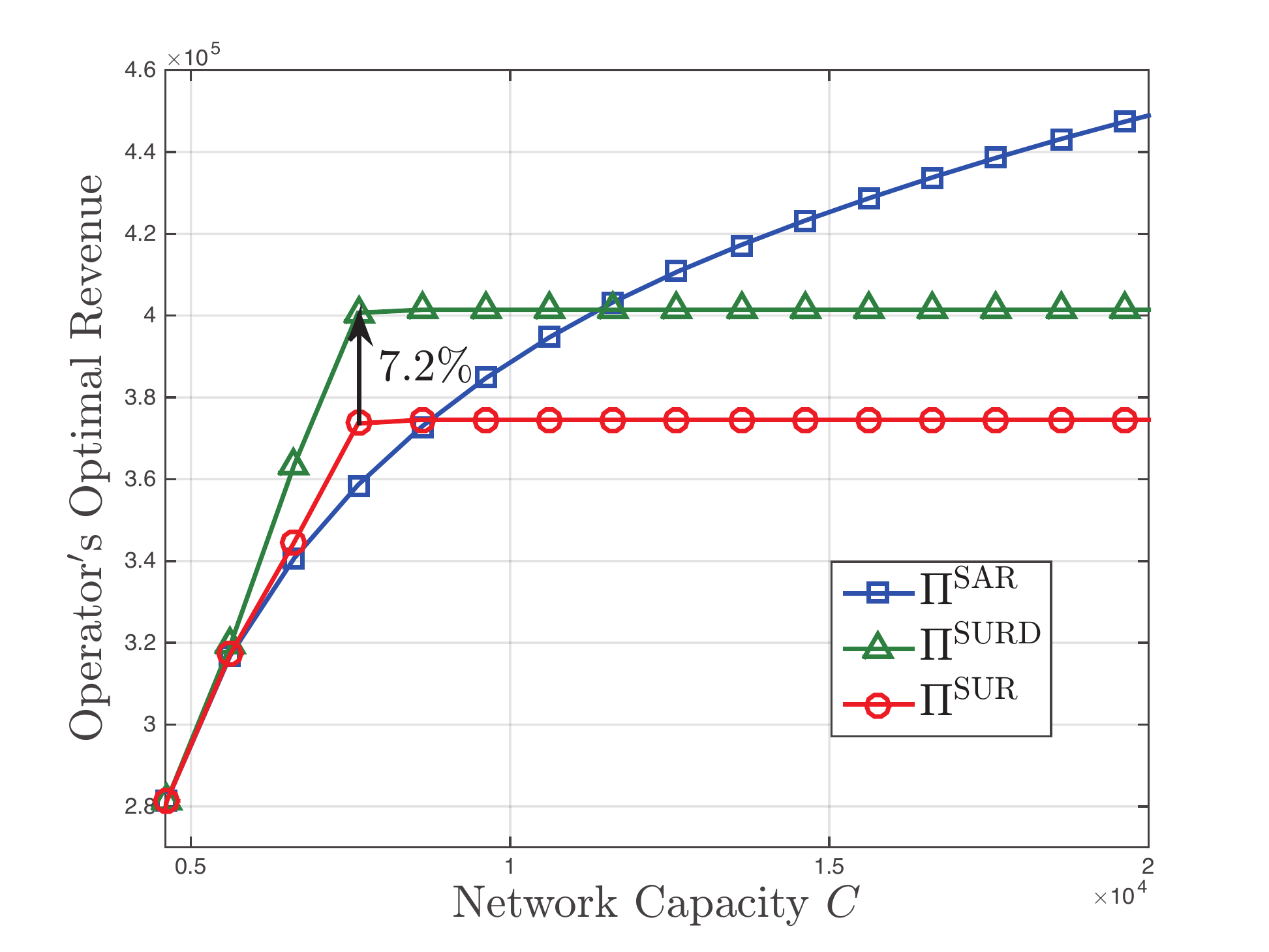}\label{appendix:fig:morepara:c}}
  \caption{$\Pi^{\rm SAR}$, $\Pi^{\rm SUR}$, and $\Pi^{\rm SURD}$ Under Different Network Capacity (Uniformly Distributed $\theta$ and Logarithmic Utility).}
  \label{appendix:fig:morepara}
\end{figure*}

\section{Numerical Results Under Different Settings}\label{appendix:morepara}
In this section, we show the numerical comparison among $\Pi^{\rm SAR}$, $\Pi^{\rm SUR}$, and $\Pi^{\rm SURD}$ under more parameter settings. Similar to Fig. \ref{fig:insimu:a}, we assume that each user's type $\theta$ follows a uniform distribution and $u\left(z\right)=\ln\left(1+z\right)$. We run experiments under three different parameter settings.

First, we choose $N=10^7$, $F=25$, $Q=0.7$, $\theta\sim{\cal U}\left[0,155\right]$, $\Phi=0.2$, $K=30$, $A=0.5$, and $B=3$. We plot $\Pi^{\rm SAR}$, $\Pi^{\rm SUR}$, and $\Pi^{\rm SURD}$ against $C$ in Fig. \ref{appendix:fig:morepara:a}.

Second, we choose $N=10^5$, $F=32$, $Q=0.6$, $\theta\sim{\cal U}\left[0,170\right]$, $\Phi=0.3$, $K=18$, $A=0.6$, and $B=4$. We plot $\Pi^{\rm SAR}$, $\Pi^{\rm SUR}$, and $\Pi^{\rm SURD}$ against $C$ in Fig. \ref{appendix:fig:morepara:b}.

Third, we choose $N=10^4$, $F=28$, $Q=0.6$, $\theta\sim{\cal U}\left[0,150\right]$, $\Phi=0.2$, $K=18$, $A=0.4$, and $B=3$. We plot $\Pi^{\rm SAR}$, $\Pi^{\rm SUR}$, and $\Pi^{\rm SURD}$ against $C$ in Fig. \ref{appendix:fig:morepara:c}.

We can see that our key observations in Fig. \ref{fig:insimu:a} also hold in Fig. \ref{appendix:fig:morepara:a}, \ref{appendix:fig:morepara:b}, and \ref{appendix:fig:morepara:c}. For example, if $C$ is small, the SUR scheme achieves a higher operator's revenue; otherwise, the SAR scheme achieves a higher operator's revenue. The ad slots' differentiation can improve the operator's revenue under the SUR scheme.

}}
\end{document}